\documentstyle [12pt,epsf,floatfig] {article}

\newcommand{\twofig}[6]
{\begin{figure}[htb!]
\vspace{-0.2cm}
 \begin{minipage}[t]{0.48\linewidth}
   \epsfxsize \linewidth
   \epsffile[10 128 550 675]{#1}
   \vspace{-0.5cm}
   \caption{#2}
   \label{#3}
 \end{minipage}
 \hfil
 \begin{minipage}[t]{0.480\linewidth}
   \epsfxsize \linewidth
   \epsffile[10 128 550 675]{#4}
   \vspace{-0.5cm}
   \caption{#5}
   \label{#6}
 \end{minipage}
\end{figure}
}

\newcommand{\onefig}[3]
{\begin{figure}[htb!]
\vspace{-0.2cm}
 \begin{minipage}[t]{0.48\linewidth}
   \epsfxsize \linewidth
   \epsffile[10 128 550 675]{#1}
   \vspace{-0.5cm}
   \caption{#2}
   \label{#3}
 \end{minipage}
 \hfil
\end{figure}
}

\newcommand{\dtfloat}[4]
{
\begin{floatingfigure}{0.48\linewidth}
\vspace*{0cm}
\hspace*{-1cm}
\epsfxsize #4\linewidth
\epsffile[30 128 550 725]{#1}
\vspace*{-0.3cm}
\caption{#2}
\label{#3}
\end{floatingfigure}
}

\def \Et {{\rm E}_{\rm T}}
\newcommand{\Etg}{${\rm E}_{\rm T}^{\gamma}$}
\newcommand{\ett}{${\rm E}_{\rm T}$}
\newcommand{\Etgt}{${\rm E}_{\rm T}^{\gamma_2}$}
\newcommand{\boxiso}{\mbox{E$^{\rm Iso}_{\rm 3x3}$}}
\newcommand{\Etggmh}{${\rm E}_{\rm T}^{\gamma}>22$~GeV}

\newcommand{\ptt}{${\rm P}_{\rm T}$}

\newcommand{\mett}{\mbox{${\rm \not\! E}_{\rm T}$}}
\newcommand{\mettx}{\mbox{${\rm \not\! E}_{\rm T}^x$}}
\newcommand{\mettmin}{\mbox{${\rm \not\! E}_{\rm T}^{\rm min}$}}
\newcommand{\mettsm}{\mbox{\scriptsize ${\rm \not\! E}_{\rm T}$}}
\newcommand{\mettgmh}{\mbox{${\rm \not\! E}_{\rm T}>35$~GeV}}
\newcommand{\mettgl}{\mbox{${\rm \not\! E}_{\rm T}>25$~GeV}}

\newcommand{\Etggl}{\mbox{${\rm E}_{\rm T}^{\gamma}>12$~GeV}}
\newcommand{\Etggh}{\mbox{${\rm E}_{\rm T}^{\gamma}>25$~GeV}}
\def\Z{{ Z^0}}
\newcommand{\zoee}{\mbox{$Z^0 \rightarrow e^+e^-$}}
\newcommand{\sumetc}{\mbox{$\Sigma {\rm E}_{\rm T}^{\rm Corrected}$}}
\newcommand{\eeggmett}{ee\gamma\gamma\mett}
\def\selectron{\tilde{e}}
\def\goes{\rightarrow}
\newcommand{\NTWO}{\mbox{$N_2$}}
\newcommand{\NONE}{\mbox{$N_1$}}
\def\Gravitino{\tilde{G}}
\def\gravitino{\tilde{G}}
\newcommand{\NTGNO}{\mbox{$\NTWO \rightarrow \gamma \NONE$}}
\newcommand{\etal}{{\em et al.}}
\def\pbarp{{\bar p}p}
\def\ppbar{{\bar p}p}

\def\roots{{\sqrt s}}
\def\degrees{^\circ}

\newcommand{\scinotn}[2]{\mbox{${#1}\times 10^{#2}$}}

\newcommand{\mettfake}{\mbox{${\rm \not\! E}_{\rm T}^{\rm Fake}$}}

\newcommand{\secorchap}{Section}
 \newcommand{\LUMTOT               }{   85}

\newcommand{\chapterone}
{
\section{INTRODUCTION}
\label{Chap-Intro}
}

\newcommand{\chaptertwo}
{
\section{Data Selection and Photon Identification}
\label{Chap-Datasets}
}

 \newcommand{\CESEFF               }{99.5\pm 0.1}
 \newcommand{\PHOEFFIDLOW          }{   68\pm    3}
 \newcommand{\PHOEFFIDHIGH         }{   84\pm    4}
 \newcommand{\CORRIDLOW            }{ 0.69\pm  0.07}
 \newcommand{\CORRIDHIGH           }{ 0.84\pm  0.08}
 \newcommand{\EFFVERTDAT           }{93.0\pm 0.6}
 \newcommand{\EFFVERTMC            }{96.4\pm 0.5}
 \newcommand{\RATIOVERT            }{0.965\pm 0.008}
 \newcommand{\PEFFETOUT            }{97.5\pm 0.4}
 \newcommand{\PTRIGLOW             }{ 96\pm  1}
 \newcommand{\PTRIGHIGH            }{  100}

 \newcommand{\NGGTOT               }{ 2239}

 \newcommand{\RESOFFSET            }{\mbox{$2.66\pm 0.34$}}
 \newcommand{\RESSLOPE             }{\mbox{$0.043\pm 0.007$}}
 \newcommand{\EEGGTOTMET           }{   55 \pm 7}
 \newcommand{\NMETEXPLOW           }{\mbox{$     0.5\pm      0.1$}}
 \newcommand{\NMETEXPHIGH          }{\mbox{$     0.5\pm      0.1$}}
 \newcommand{\NMETLOW              }{    1}
 \newcommand{\NMETHIGH             }{    2}

 \newcommand{\NIVJETEXPLOW         }{\mbox{$     1.6\pm      0.4$}}
 \newcommand{\NIVJETLOW            }{    2}
 \newcommand{\NIIIJETEXPHIGH       }{\mbox{$     1.7\pm      1.5$}}
 \newcommand{\NIIIJETHIGH          }{    0}
 \newcommand{\NCENTEORMU           }{    3}

 \newcommand{\MUMUGGMASS           }{   92 \pm 1}
 \newcommand{\EGGMASS              }{   91 \pm 2}
 
 \newcommand{\NCENTLEPDIBOSON      }{\mbox{$0.04\pm 0.04$}}
 \newcommand{\NCENTLEPFAKE         }{\mbox{$     0.2\pm      0.1$}}
 \newcommand{\NCENTEORMUEXP        }{\mbox{$     0.3\pm      0.1$}}

 \newcommand{\EGFAKERATE           }{\mbox{$     1.9\pm      0.3$}}

 \newcommand{\NTAUEXPLOW           }{\mbox{$     0.2\pm      0.1$}}
 
 \newcommand{\NCENTTAU             }{    1}

 \newcommand{\NBEXPLOW             }{\mbox{$     1.3\pm      0.7$}}
 \newcommand{\NBTAG                }{    2}

 \newcommand{\CPFAKEP              }{\scinotn{1}{ -3}}
 \newcommand{\NADDGAMMAEXP         }{\mbox{$     0.1\pm      0.1$}}
 \newcommand{\NADDGAMMA            }{    0}

 \newcommand{\PURITYLOW            }{\mbox{$   15\pm    4$}}

\newcommand{\chapterfour}
{

\section{The $\eeggmett$ Candidate event}
\label{Chap-The Event}
}

\newcommand{\metstatement}{}

 \newcommand{\EEGGPT               }{   48 \pm 2}
 \newcommand{\PEFAKEP              }{\scinotn{2}{ -3}}

\newcommand{\chapterfive}
{
\section{Estimating the Number of $\eeggmett$ Events from Standard 
Model Sources}
\label{Chap-SM Expectations}
}

\newcommand{\correctedmett}{}

\def\ovpr#1{$\frac{#1}{3\times 10^{12}}$}

 \newcommand{\NTOTW                }{58000}
 \newcommand{\NWCJEVENTS           }{ 1383}
 \newcommand{\NWPJEVENTS           }{  597}
 \newcommand{\NWCJETS              }{ 1513}
 \newcommand{\NWPJETS              }{  620}
 \newcommand{\CPFAKE               }{\scinotn{3}{ -5}}
 \newcommand{\RADGAMMA             }{\scinotn{6}{ -4}}
 \newcommand{\RADDGAMMA            }{\scinotn{6}{ -4}}
 \newcommand{\CEFAKEP              }{\scinotn{7}{ -5}}
 \newcommand{\CEFAKE               }{\scinotn{2}{ -6}}
 \newcommand{\PEFAKE               }{\scinotn{2}{ -5}}
 \newcommand{\METFAKE              }{\scinotn{4}{ -3}}
 \newcommand{\WWGGHIGHRATE         }{\scinotn{8}{ -7}}
 \newcommand{\EWWGGHIGHRATE        }{\scinotn{(8\pm 4)}{ -7}}
 \newcommand{\TOTTOPEXP            }{\scinotn{3}{ -7}}
 \newcommand{\ETOTTOPEXP           }{\scinotn{(3\pm 3)}{ -7}}
 \newcommand{\EEG                  }{\scinotn{5}{-2}}
 \newcommand{\EGG                  }{\scinotn{3}{-2}}
 \newcommand{\EEGG                 }{\scinotn{3}{-5}}

 \newcommand{\EEGMET               }{\scinotn{2}{-3}}
 \newcommand{\EGGMET               }{\scinotn{2}{-3}}
 \newcommand{\EEGFM                }{\scinotn{8}{-8}}
 \newcommand{\EEGGF                }{\scinotn{1}{-7}}
 \newcommand{\EFGGM                }{\scinotn{5}{-8}}
 \newcommand{\EEFFM                }{\scinotn{5}{-9}}
 \newcommand{\EFGFM                }{\scinotn{3}{-9}}
 \newcommand{\EEGFF                }{\scinotn{6}{-9}}
 \newcommand{\EFGGF                }{\scinotn{3}{-9}}
 \newcommand{\EFGFF                }{\scinotn{1}{-10}}
 \newcommand{\EFFFM                }{\scinotn{2}{-9}}
 \newcommand{\EEFFF                }{\scinotn{4}{-10}}

 \newcommand{\EEEGGMETTOTRATE      }{\scinotn{(1\pm 1)}{ -6}}
 \newcommand{\EFAKEGGMETTOTRATE    }{\scinotn{6}{ -8}}
 \newcommand{\EEFAKEGGMETTOTRATE   }{\scinotn{(6\pm 6)}{ -8}}
 \newcommand{\SUMEEGGMETNOWWGG     }{\scinotn{3}{-7}}
 \newcommand{\CCDY                 }{ 1660}
 \newcommand{\CPDY                 }{ 1771}

 \newcommand{\CCDYMET              }{   12}
 \newcommand{\CPDYMET              }{    7}
 \newcommand{\CCDYGAMMA            }{    0}
 \newcommand{\CPDYGAMMA            }{    2}
 \newcommand{\CCZEE                }{ 1470}
 \newcommand{\CPZEE                }{ 1613}
 \newcommand{\CCZEEMET             }{    9}
 \newcommand{\CPZEEMET             }{    3}
 \newcommand{\CCZEEGAMMA           }{    0}
 \newcommand{\CPZEEGAMMA           }{    1}
 \newcommand{\CCMEE                }{   40}
 \newcommand{\CPMEE                }{   40}
 \newcommand{\CCMEEMET             }{    1}
 \newcommand{\CPMEEMET             }{    3}
 \newcommand{\CCMEEGAMMA           }{    0}
 \newcommand{\CPMEEGAMMA           }{    1}
 \newcommand{\EG                   }{   49}
 \newcommand{\EGMET                }{    4}
 \newcommand{\NMEE                 }{   80}
 \newcommand{\NMEEMET              }{    4}
 \newcommand{\NGG                  }{  218}
 \newcommand{\NPG                  }{   22}
 \newcommand{\PGG                  }{\scinotn{1}{-2}}
 \newcommand{\EEGTOT               }{\scinotn{5}{-2}}
 \newcommand{\NGMET                }{ 3181}
 \newcommand{\WGG                  }{\scinotn{3}{-2}}
 \newcommand{\WWG                  }{\scinotn{3}{-3}}
 \newcommand{\NTOTPW               }{40000}
 \newcommand{\NPGM                 }{    4}
 \newcommand{\OVERLAPTOTVER        }{\scinotn{1}{ -9}}
 \newcommand{\OVWWGG               }{\scinotn{3}{-10}}
 \newcommand{\OVWGWG               }{\scinotn{5}{-12}}
 \newcommand{\OVPGGW               }{\scinotn{3}{-10}}
 \newcommand{\OVEEGGMET            }{\scinotn{5}{-11}}
 \newcommand{\OVWGGCOS             }{\scinotn{4}{-10}}
 \newcommand{\OVWWGCOS             }{\scinotn{3}{-12}}
 \newcommand{\OVERLAPTOT           }{\scinotn{  8}{ -9}}
 \newcommand{\EOVERLAPTOT          }{\scinotn{(8\pm 8)}{ -9}}

\newcommand{\chaptersix}
{
\section{Setting Limits with the $\gamma\gamma + X$ Analysis}
\label{Chap-Limits}
}

\newcommand{\tanbeta}{\tan\beta}
\newcommand{\sgnmu}{Sgn($\mu$)}
 \newcommand{\KANEXSECTOT          }{    11.5}
 \newcommand{\EFFETOUT             }{0.975\pm 0.004}
 \newcommand{\TRIGLOW              }{ 0.96\pm  0.01}
 \newcommand{\TRIGHIGH             }{     1.0}
 \newcommand{\EFFCORLOW            }{ 0.62\pm  0.06}
 \newcommand{\EFFCORHIGH           }{ 0.79\pm  0.08}
\newcommand{\CONE}{\mbox{$C_1$}}
 \newcommand{\KANELOWNEXPTOT       }{     2.4}
 \newcommand{\KANELOWACC           }{     5.4}
 \newcommand{\KANELOWXSEC          }{     1.1}

\newcommand{\effref}{Chapter 2}

\def\Journal#1#2#3#4{{#1} {\bf #2}, #3 (#4)}
\def\PRL{\em Phys. Rev. Lett.}
\def\PRD{{\em Phys. Rev.} D}
\def\NIMA{{\em Nucl. Instrum. Methods} A}

\def\PrePrintE#1{\mbox{hep-ex/#1}}
\def\PrePrintP#1{\mbox{hep-ph/#1}}

\begin{document}
\initfloatingfigs

\vspace*{-1.8cm}
\small
\normalsize
{\bf 
Searches for New Physics in Diphoton Events in $p{\bar p}$ collisions at
$\roots= 1.8$~TeV}


\font\eightit=cmti8
\def\r#1{\ignorespaces $^{#1}$}
\hfilneg
\begin{sloppypar}
\noindent
F.~Abe,\r {17} H.~Akimoto,\r {39}
A.~Akopian,\r {31} M.~G.~Albrow,\r 7 A.~Amadon,\r 5 S.~R.~Amendolia,\r {27} 
D.~Amidei,\r {20} J.~Antos,\r {33} S.~Aota,\r {37}
G.~Apollinari,\r {31}
% ...
\end{sloppypar}
\vskip .026in
\begin{center}
(CDF Collaboration)
\end{center}

\vskip .026in
\begin{center}
\r 1  {\eightit Argonne National Laboratory, Argonne, Illinois 60439} \\
\r 2  {\eightit Istituto Nazionale di Fisica Nucleare, University of Bologna,
I-40127 Bologna, Italy} \\
\r 3  {\eightit Brandeis University, Waltham, Massachusetts 02254} \\
\r 4  {\eightit University of California at Los Angeles, Los 
Angeles, California  90024} \\  
\r 5  {\eightit University of Chicago, Chicago, Illinois 60637} \\
\r 6  {\eightit Duke University, Durham, North Carolina  27708} \\
\r 7  {\eightit Fermi National Accelerator Laboratory, Batavia, Illinois 
60510} \\
\r 8  {\eightit University of Florida, Gainesville, FL  32611} \\
\r 9  {\eightit Laboratori Nazionali di Frascati, Istituto Nazionale di Fisica
               Nucleare, I-00044 Frascati, Italy} \\
\r {10} {\eightit University of Geneva, CH-1211 Geneva 4, Switzerland} \\
\r {11} {\eightit Harvard University, Cambridge, Massachusetts 02138} \\
\r {12} {\eightit Hiroshima University, Higashi-Hiroshima 724, Japan} \\
\r {13} {\eightit University of Illinois, Urbana, Illinois 61801} \\
\r {14} {\eightit Institute of Particle Physics, McGill University, Montreal 
H3A 2T8, and University of Toronto,\\ Toronto M5S 1A7, Canada} \\
\r {15} {\eightit The Johns Hopkins University, Baltimore, Maryland 21218} \\
\r {16} {\eightit Institut f\"{u}r Experimentelle Kernphysik, 
Universit\"{a}t Karlsruhe, 76128 Karlsruhe, Germany} \\
\r {17} {\eightit National Laboratory for High Energy Physics (KEK), Tsukuba, 
Ibaraki 305, Japan} \\
\r {18} {\eightit Ernest Orlando Lawrence Berkeley National Laboratory, 
Berkeley, California 94720} \\
\r {19} {\eightit Massachusetts Institute of Technology, Cambridge,
Massachusetts  02139} \\   
\r {20} {\eightit University of Michigan, Ann Arbor, Michigan 48109} \\
\r {21} {\eightit Michigan State University, East Lansing, Michigan  48824} \\
\r {22} {\eightit University of New Mexico, Albuquerque, New Mexico 87131} \\
\r {23} {\eightit The Ohio State University, Columbus, OH 43210} \\
\r {24} {\eightit Osaka City University, Osaka 588, Japan} \\
\r {25} {\eightit Universita di Padova, Istituto Nazionale di Fisica 
          Nucleare, Sezione di Padova, I-35131 Padova, Italy} \\
\r {26} {\eightit University of Pennsylvania, Philadelphia, 
        Pennsylvania 19104} \\   
\r {27} {\eightit Istituto Nazionale di Fisica Nucleare, University and Scuola
               Normale Superiore of Pisa, I-56100 Pisa, Italy} \\
\r {28} {\eightit University of Pittsburgh, Pittsburgh, Pennsylvania 15260} \\
\r {29} {\eightit Purdue University, West Lafayette, Indiana 47907} \\
\r {30} {\eightit University of Rochester, Rochester, New York 14627} \\
\r {31} {\eightit Rockefeller University, New York, New York 10021} \\
\r {32} {\eightit Rutgers University, Piscataway, New Jersey 08855} \\
\r {33} {\eightit Academia Sinica, Taipei, Taiwan 11530, Republic of China} \\
\r {34} {\eightit Texas A\&M University, College Station, Texas 77843} \\
\r {35} {\eightit Texas Tech University, Lubbock, Texas 79409} \\
\r {36} {\eightit Istituto Nazionale di Fisica Nucleare, University of Trieste/
Udine, Italy} \\
\r {37} {\eightit University of Tsukuba, Tsukuba, Ibaraki 315, Japan} \\
\r {38} {\eightit Tufts University, Medford, Massachusetts 02155} \\
\r {39} {\eightit Waseda University, Tokyo 169, Japan} \\
\r {40} {\eightit University of Wisconsin, Madison, Wisconsin 53706} \\
\r {41} {\eightit Yale University, New Haven, Connecticut 06520} \\
\end{center}

\clearpage

\begin{abstract}

We present a detailed description of a search 
for anomalous production of missing
E$_{\rm T}$ ($\mett$), jets, leptons ($e, \mu, \tau$), $b$-quarks,
or  additional photons in events containing two isolated,
central  \mbox{($|\eta|<1.0$)} photons with
\mbox{$\Et>12$~GeV}.  The results are consistent with standard model
expectations, with the possible exception of one event that has
in addition to
the two photons a central electron,  a
high-E$_{\rm T}$ electromagnetic
cluster, and large  $\mett$.  We set
limits using two specific SUSY
scenarios for production of diphoton events with $\mett$.

\end{abstract}
\vspace*{0.2in}
\hspace*{1.0in} PACS numbers 13.85Rm, 12.60.Jv, 13.85.Qk, 14.80.Ly
\vspace*{0.2in}

%
%
%
\clearpage

\chapterone


  In many  models  involving  physics beyond  the  standard model  
(SM)~\cite{SM Reference}, cascade
decays of  heavy new  particles  generate   $\gamma\gamma$  signatures involving
missing transverse  energy ($\mett$),   jets, leptons,  gauge bosons ($W$, $\Z$,
$\gamma$),  and   possibly  $b$-quarks. Some examples are
supersymmetry with a light gravitino~\cite{Gravitino Reference},  
radiative  decays to a higgsino-LSP~\cite{Higgsino LSP} and 
models with large symmetry groups~\cite{Non-SUSY}.
In the data taken during 1993-1995 by the CDF detector~\cite{bluebook,CDFcoo}
an `$\eeggmett$'
candidate event~\cite{Park} was recorded. 
Supersymmetric models can explain the $\eeggmett$ signature, for example, 
via the pair production and decay of selectrons via
$\selectron\goes e\NTWO\goes e\gamma\NONE$, (see Figure~\ref{LSP FEYN}) or
$\selectron\goes e\NONE\goes e\gamma\gravitino$, (see
Figure~\ref{GRAV FEYN}).

\twofig{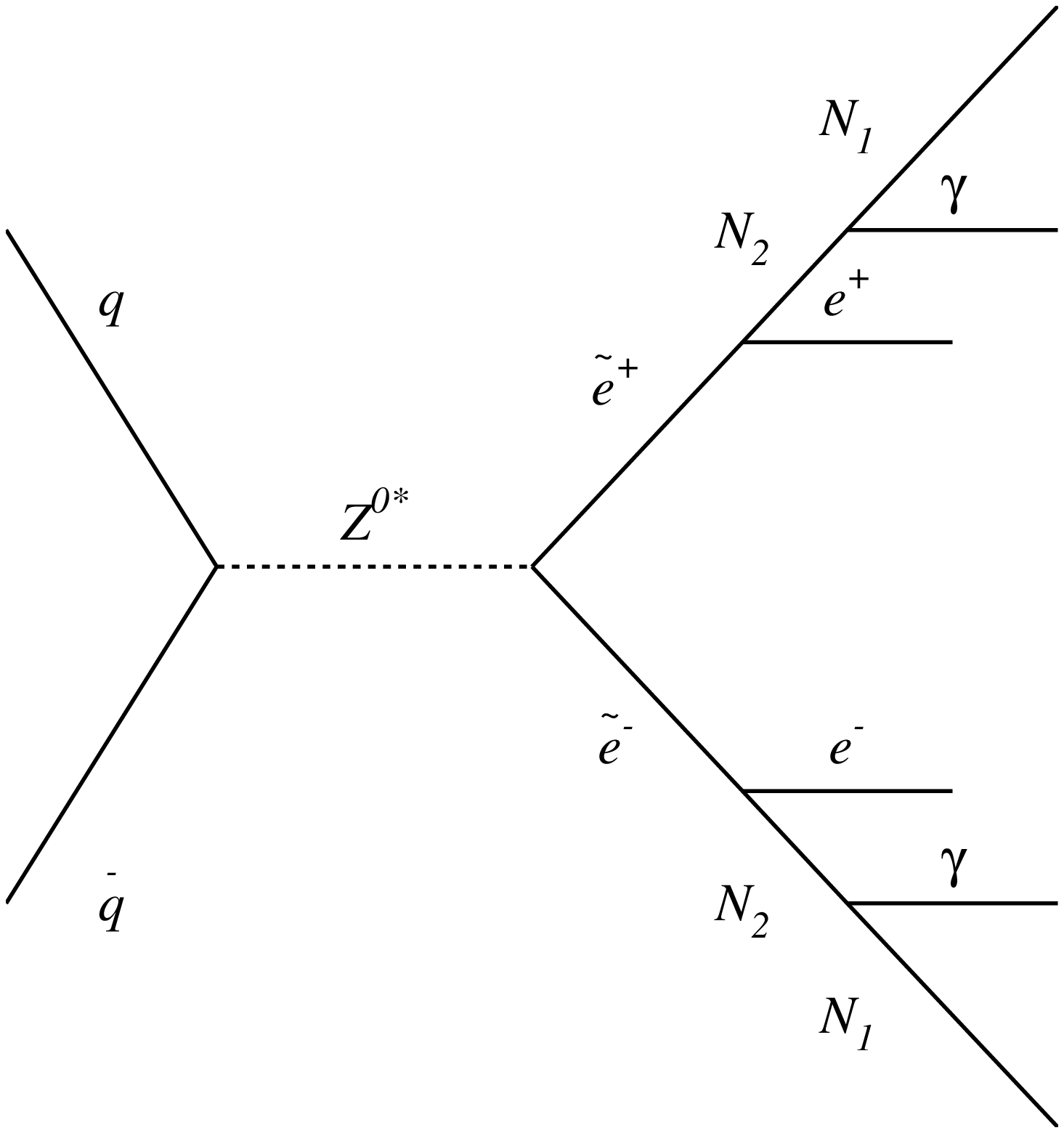}{The Feynman diagram for
$\selectron$ pair production and decay in the $\NTGNO$ scenario of Kane \etal\
Both 
selectrons decay via ${\tilde e}\rightarrow eN_2$ where $N_2$ is the 
next-to-lightest neutralino which in turn decays
via $N_2 \rightarrow \gamma\NONE$.}{LSP FEYN}
{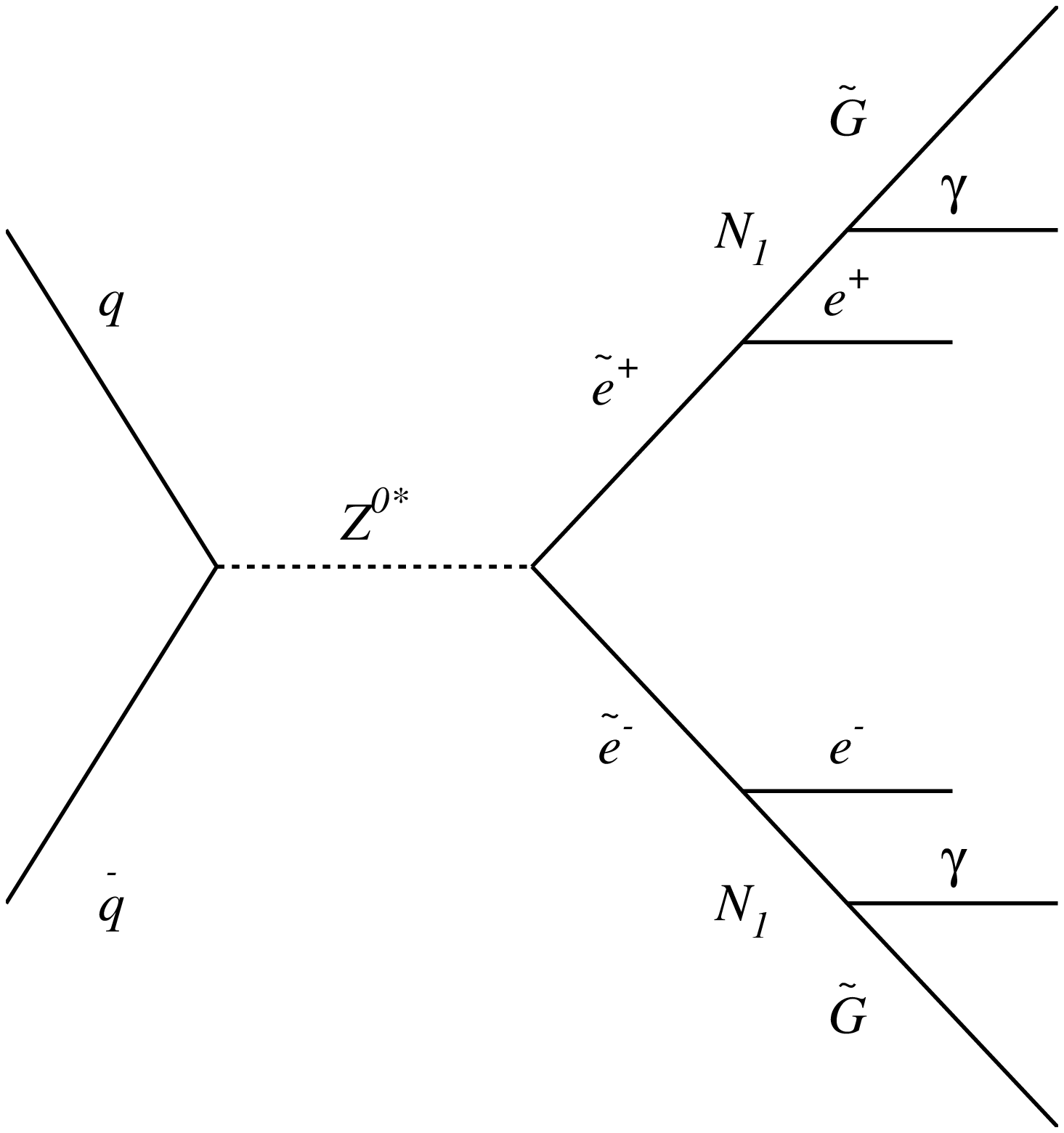}{The Feynman diagram for
$\selectron$ pair production and decay in the light
gravitino scenario. Both 
selectrons decay via ${\tilde e}\rightarrow eN_1$ where $N_1$ is the lightest
neutralino which in turn decays via $N_1 \rightarrow \gamma\gravitino$.}{GRAV FEYN}


This paper describes a systematic 
search for other anomalous $\gamma\gamma$ events
by examining
events with two isolated,
central  \mbox{($|\eta|<1.0$)} photons with
\mbox{$\Et>12$~GeV} which contain 
$\mett$, jets, leptons ($e, \mu, \tau$), $b$-quarks,
or  additional photons~\cite{Toback}. The search is based on
85~pb$^{-1}$ of data from $\pbarp$
collisions at \mbox{$\roots = 1.8$~TeV} collected with  the CDF
detector. 

The remainder of \secorchap\ 1 is devoted to a description of the 
detector.  
\secorchap\ 2 discusses the
diphoton event selection, the 
efficiencies of the selection criteria, and the purity
of the sample. \secorchap\ 3 discusses a search for anomalous events in the
sample.  \secorchap\ 4 discusses the 
$\eeggmett$ candidate event.
\secorchap\ 5 discusses the possible standard
model sources for the $\eeggmett$ signature and estimates the number of events
expected from each. \secorchap\ 6 discusses the possible
interpretations of this event and places 
limits on some of the models which have
risen to explain it. \secorchap\ 7 contains the conclusions.  


\subsection{Overview of the CDF detector}

\dtfloat{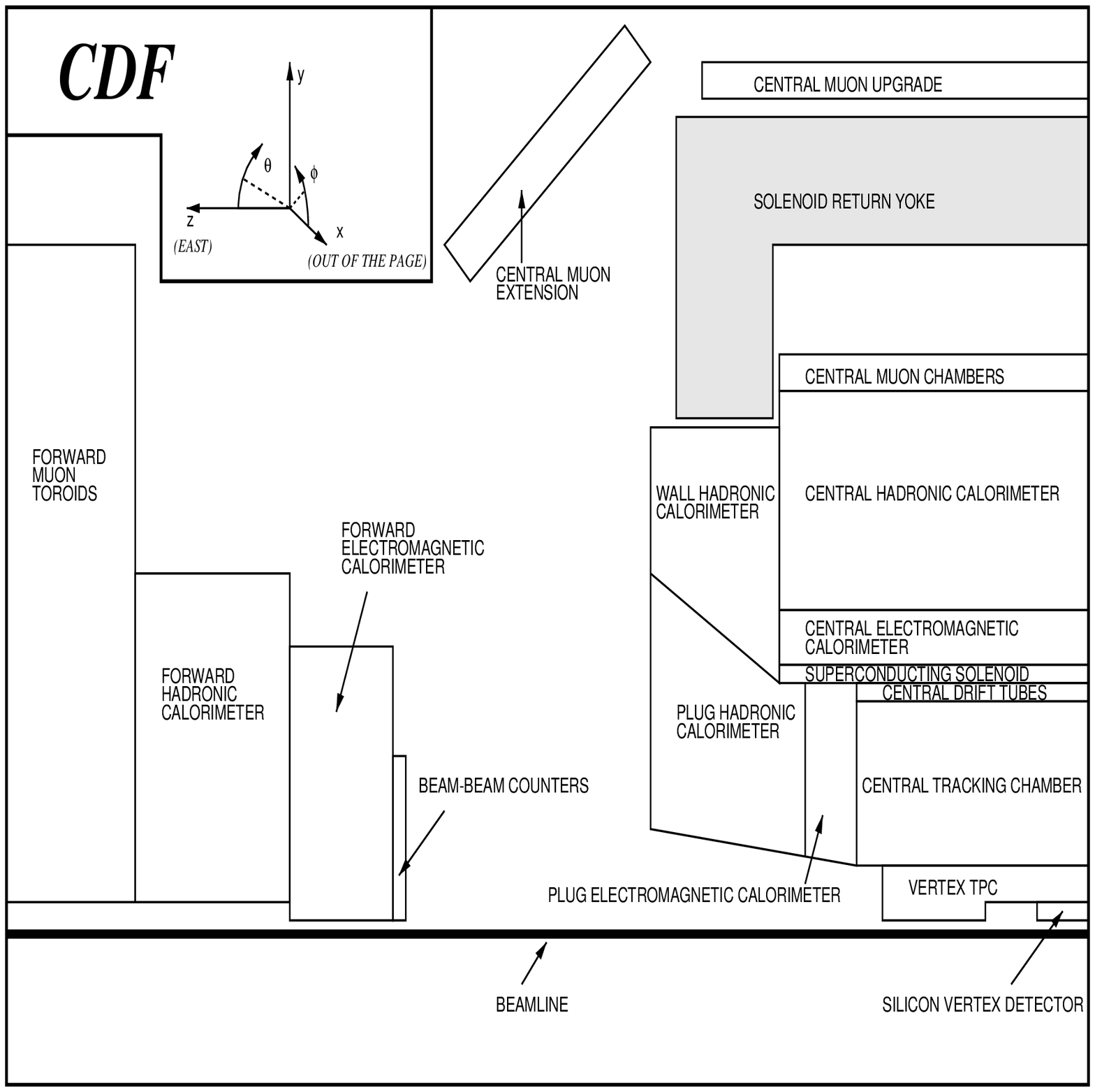}
{A schematic drawing of one-quarter of the CDF detector}
{CDF Picture}{0.48}

\begin{sloppypar}
The CDF detector is an azimuthally and forward-backward symmetric magnetic
detector designed to study $\pbarp$ collisions at the Fermilab
Tevatron. A schematic drawing of the major detector components is shown
in Figure~\ref{CDF Picture}. A more detailed description can be found 
in~\cite{bluebook}; recent detector upgrades are described 
in~\cite{svxnim}.
The magnetic spectrometer consists of tracking devices inside a
\mbox{3-m} diameter, 5-m long superconducting solenoidal magnet which operates
at 1.4~T. A four-layer silicon microstrip vertex detector (SVX)~\cite{svxnim}
makes measurements between the radii of 2.8~cm and 7.9~cm,   and
is used to identify $b$ hadron decays. A set of vertex time projection chambers
(VTX) surrounding the SVX provides  measurements in the $r$-$z$ plane  up to a
radius of 22~cm,  and is used to find the $z$ position of the $\pbarp$
interaction (${\rm z_{vertex}}$).  The 3.5-m long central tracking chamber
(CTC), which provides up to 84 measurements between the radii of 31.0~cm and
132.5~cm,  is used to measure the momentum of charged particles with momentum
resolution  \mbox{${\sigma_p}/{p}<0.001 p$} ($p$ in GeV/$c$). The calorimeter,
constructed of projective electromagnetic and   hadronic towers,   is divided
into a central barrel which surrounds the solenoid coil (\mbox{$|\eta|<1.1$}),
`end-plugs' (\mbox{$1.1<|\eta|<2.4$}),   and  forward/backward modules
(\mbox{$2.4<|\eta|<4.2$}).   Wire chambers with cathode strip readout   give
2-dimensional  profiles of electromagnetic showers in the central  and plug
regions (CES and PES systems, respectively).  A system of drift chambers (CPR)
outside the solenoid and in front of the electromagnetic calorimeters uses the
1-radiation-length thick magnet coil as a `preradiator',   allowing
photon/$\pi^0$ discrimination on a statistical basis by measuring the conversion
probability~\cite{photons}. Muons are identified with the central muon chambers,
situated outside the calorimeters in the region \mbox{$|\eta|<1.1$}.
\end{sloppypar}

To ensure that events are well measured, only events in which 
both photon candidates fall within the fiducial volume of the 
central electromagnetic calorimeter (CEM) are selected. 
The CEM is made of shower counters arranged in
a projective tower geometry, with each tower composed of 
absorber sheets interspersed with
scintillator. 
The towers are constructed in 48 wedges, each consisting of 10 towers in
$\eta$ by one tower in $\phi$. 
The position and transverse profile of a photon shower, within a tower,
is measured using the CES which is embedded near shower
maximum at approximately 6 radiation lengths. These chambers 
have wires in the $r-\phi$ view and
cathode strips in the $z$ view.  To be in the fiducial region, 
the shower position is required to lie
within 21~cm of the tower center ($|X_{\rm wire}|<21.0$~cm) in the $r-\phi$ view
so that the shower is fully contained in the active region.  
The region $|\eta|<0.05$, where the two halves of the detector
meet, is excluded. The region 0.77$< \eta < 1.0, 75^{\circ} <\phi <90^{\circ}$
is uninstrumented because it is the penetration for the
cryogenic connections to the solenoidal magnet. In addition, the region 1.0$<
|\eta|<1.1$ is excluded because of the smaller depth of the electromagnetic
calorimeter in the region. Within the angular region 
0.05$<|\eta|<1.0$ and 0$< \phi< 2\pi$ the fiducial coverage is roughly 
87\% per photon. For low-E$_{\rm T}$
photons (\Etgt $\leq$ 22~GeV)
the fiducial region is reduced to $|X_{\rm wire}|<17.5$~cm to be
consistent with the trigger requirements. The tight fiducial region coverage 
is 
approximately 73\% for low-E$_{\rm T}$ photons.

The CDF detector is a relatively 
well-understood measuring instrument and there
exist standard  identification selection criteria 
for electrons, muons, taus, $b$-quarks, and jets that were 
developed for, among other things, the discovery of the top quark and 
studies of its properties. Descriptions for these criteria
can be found 
in Refs.~\cite{WMass,R,top,Marcus,Doug G}. 
Photon identification~\cite{photons} is 
described in more detail in \secorchap\ 2.  The \mett\ calculation used in this
search has been
customized for this analysis and is described in \secorchap\ 3.

\chaptertwo

Events are selected based on the identification of two photon candidates
in the central region of the CDF detector,
\mbox{$|\eta|<1.0$}.  The
final selection criteria are listed in Table~\ref{Event Cuts} and are described
below. The central region contains calorimeters and tracking chambers, and 
since the interaction of a high energy photon with the detector
is similar to that of
an electron, many of the same techniques for
identifying electrons are used to identify isolated photons~\cite{photons,R}. 
The calorimeters are 
used to measure the 4-momentum of the photon as well as to
distinguish between photons produced directly in the 
$\ppbar$ collision and those which
are produced in the decay of hadrons, such as $\pi^0\rightarrow \gamma\gamma$.
The tracking chambers are 
used to provide additional rejection against jets of hadrons as well as 
electrons. 

\begin{table}[htb]
\centering
\begin{tabular}{ll} \hline
\multicolumn{2}{c}{Photon Identification and Isolation Cuts} \\ \hline
\multicolumn{2}{l}{Central $(|\eta| < 1.0)$}  \\
\multicolumn{2}{l}{$\le$1 3D tracks pointing at the cluster,
        P$_{\rm T}$ $<$  1 GeV}\\
\multicolumn{2}{l}{$\chi^2_{\rm CES} < 10.0$}\\
\multicolumn{2}{l}{$|\sigma_{\rm CES}|<$  2.0} \\
\multicolumn{2}{l}{E$^{\rm 2nd~cluster} \leq
-0.00945+0.144\times {\rm E}_{\rm T}^{\rm \gamma}$
 (${\rm E}_{\rm T}^{\rm \gamma}< 17.88$ GeV)}\\
\multicolumn{2}{l}{E$^{\rm 2nd~cluster} \leq
    2.39+0.01 \times {\rm E}_{\rm T}^{\rm \gamma} \;\;\;\;\;\;\;\;\;\;\;
 ({\rm E}_{\rm T}^{\rm \gamma}> 17.88$ GeV)}\\
\hline
\multicolumn{1}{c|}{12~GeV $<$ \Etgt $<$ 22~GeV} &
\multicolumn{1}{c}{\Etgt $\geq$ 22~GeV} \\
\hline
\multicolumn{1}{l|}{Low-Threshold Trigger}  & High-Threshold Trigger \\
\multicolumn{1}{l|}{Tight Fiducial}  & Loose Fiducial \\
\multicolumn{1}{l|}{\boxiso $\le$ 4 GeV} & Had$/$EM   $<$  0.055 + 0.00045E.\\
\multicolumn{1}{l|}{}                   &
    ${\rm Iso}^{\rm Cal} <$ 0.10\\
\multicolumn{1}{l|}{}                   &
P$_{\rm T}^{\rm Tracks}$ in a cone of 0.4  $<$  5 GeV\\
\hline\hline
\multicolumn{2}{c}{Global Event Cuts} \\
\hline
\multicolumn{2}{l}{
                $|{\rm z\/}_{\rm Vertex}| <$  60.0 cm  }         \\
\multicolumn{2}{l}{E$_{\rm T}$ out-of-time = 0 GeV}\\
\hline
\end{tabular}
\caption[The selection criteria used to identify $\gamma\gamma$ candidate
events]{The selection criteria used to identify diphoton candidate
events.}
\label{Event Cuts}
\end{table}

The initial data sample for the search consists of events with two
photon candidates selected by the three-level trigger~\cite{trigger}. At Level
1, (L1), 
events are required to have two central electromagnetic calorimeter (CEM)
trigger towers~\cite{Trigger Towers} which measure more than 
4~GeV.  At Level 2 (L2), two triggers, 
one optimized for good background rejection at
low $\Et$ and the other for high efficiency at high $\Et$, are
`OR'd.  The
low-threshold diphoton trigger requires two
electromagnetic clusters~\cite{WMass} with $\Et > 10$~GeV and
less than 4~GeV in a 3-by-3 array of trigger
towers around the cluster (\boxiso); the high-threshold (16~GeV) trigger
has no isolation requirement.  Corresponding Level 3 (L3) 
triggers require cluster energies calculated with the offline photon
algorithm~\cite{photons} to be above the 10 and 16~GeV thresholds respectively.
The low-threshold trigger also requires the
clusters be in a restricted fiducial region of the calorimeter 
to ensure a good cluster measurement in the strip
chambers. 

\subsection{Photon Identification}

Photon candidates are 
identified as electromagnetic clusters of energy 
deposited in the central electromagnetic calorimeter and
are required 
to be consistent with being produced from a single prompt photon 
shower~\cite{photons}.
To reject against backgrounds from electrons and hadronic jets, each 
candidate is required to pass the identification 
selection criteria of Table~\ref{Event Cuts}. Electrons, which have 
shower characteristics similar to those of photons, 
can be removed by identifying the associated track. Hadronic jets, which
can contain photons from neutral meson decays, 
can be removed since they typically contain
multiple particles that can be identified by the calorimeter and/or 
tracking chamber. 

Electrons and charged hadrons
can be rejected by the presence of their tracks 
pointing at a photon candidate.   Each photon candidate 
is required to have no charged track pointing at it. However, 
to reduce the inefficiency due to 
unrelated particles, a single track is
allowed to point at the cluster if the track 
has  a measured P$_{\rm T} \le$ 1~GeV.

The ratio Had/EM of the energy in the hadronic towers of the photon cluster
(Had) to the energy in the electromagnetic towers in the photon cluster (EM) 
is used to reject hadronic backgrounds~\cite{R}.   Electromagnetic showers 
deposit most (typically $>95\%$) of their energy in the 
electromagnetic calorimeters, 
while hadron showers in general deposit
energy in both the hadronic and electromagnetic compartments.  For
events with both photons with \Etg$>$22~GeV, each photon is required to have  
Had$/$EM   $<$  0.055 + 0.00045$\times$E$^{\gamma}$.  

The shower shape measured in the CES is used to
distinguish between single photon
and the remaining hadronic backgrounds.  A
$\chi^2$ test is used to separately compare the energy
deposited in the $z$ view and in the $r-\phi$ view
to that expected from 
test beam data~\cite{photons}.  The average of the 
two measurements, $\chi^2_{\rm CES}$, is required to be below 10. 
To reject cosmic rays, the measured 
shower shape
for each candidate in the CES is fitted to that 
expected 
from the measured CEM energy and vertex 
position.  The result of the comparison, $\sigma_{\rm CES}$, 
is required to be within 2 standard deviations from expectations.

\subsection{Photon Isolation}

Photons from the radiative decays of heavy new particles
are, in general,  expected to be ``isolated,''  
that is, they are
not expected to be produced in association with other nearby particles. 
A number of different isolation variables help 
reduce hadronic jet backgrounds (see Table~\ref{Event Cuts}).

The energy in a 3x3 trigger 
tower array~\cite{Trigger Towers} around the 
primary tower (in both the 
hadronic and electromagnetic calorimeters, but not including the primary
electromagnetic tower) is summed and is 
referred to as \boxiso. A requirement of 
\boxiso$<$4~GeV is imposed both at the trigger level and
offline on each photon 
if either photon candidate has \Etg$<$22~GeV.
For high energy photons, the leakage of the shower into the
hadronic compartments makes this requirement inefficient, and it is removed if 
both photons have \Etggmh.

\begin{sloppypar}
The cluster isolation, ${\rm Iso^{Cal}}$, 
is similar to the trigger tower isolation, but is more
efficient for higher energy photons since it scales with the photon energy.  The
${\rm Iso^{Cal}}$ variable is defined as 
\begin{eqnarray}
{\rm Iso^{Cal}} & = &
\frac{
{\rm E_T^{cone}} - {\rm E_T^{cluster}} 
}  
{
{\rm E_T^{cluster}}
},
\end{eqnarray}
where ${\rm E_T^{cone}}$ is the sum of the electromagnetic and hadronic 
transverse energies in
all of the towers (including those in the 
photon cluster) in a cone of 
\mbox{$R = \sqrt{(\Delta\eta)^2 + (\Delta\phi)^2} = 0.4$}
centered around the photon cluster, and ${\rm E_T^{cluster}}$ is the
electromagnetic transverse energy in the photon cluster. 
For events with both photons with \Etg$>$22~GeV, each photon is
required to have ${\rm Iso}^{\rm Cal} <$0.1.
\end{sloppypar}

While there may be no track pointing directly at the cluster, 
tracks near the cluster may indicate that the
cluster is due to a jet. The track isolation is defined 
as the scalar sum 
of the transverse momenta of all tracks in a cone of radius
$R$=0.4 in $\eta-\phi$ space centered on the photon. 
For
events with both photons with \Etg$>$22~GeV, each photon 
is required to have the sum be less than
5.0~GeV.

To remove photons from $\pi^0 \rightarrow \gamma\gamma$ production, 
photon candidates 
which have a second electromagnetic cluster, as measured by the 
strip chambers, are rejected. 
To maintain a constant efficiency for all photon
energies separate requirements for low energy and high
energy photon candidates are made~\cite{Wgamma}: 
\begin{eqnarray}
{\rm E^{2nd~cluster}}  & \leq & 
-0.00945+0.144\times {\rm E_T^{\gamma}} 
\end{eqnarray}
for ${\rm E_T^{\gamma}}< 17.88~{\rm GeV}$, and 
\begin{eqnarray}
{\rm E^{2nd~cluster}}  & \leq &
 2.39+0.01    \times {\rm E_T^{\gamma}} 
\end{eqnarray}
for ${\rm E_T^{\gamma}}> 17.88~{\rm GeV}$.

\subsection{Additional Event Requirements}\label{Add Event Req}

In addition to the
photon identification and isolation requirements, 
there are cuts on the primary vertex and on the time of the
energy deposited in the calorimeter to
ensure that events are well measured and are not due to cosmic ray sources.

To maintain the projective geometry of the detector, only events in which the
primary vertex occur near the center of the detector are selected. 
The position, in $z$, of the primary event vertex, $z_{\rm vertex}$,
is measured by the vertex
tracking chamber (VTX). The 
$z$ positions of the interactions are distributed around the nominal 
interaction in an approximately Gaussian distribution with 
$\sigma\approx 30$~cm. 
A requirement of $|z_{\rm vertex}| <$  60~cm is used.

To reduce 
cosmic ray interaction backgrounds which might occur during an
event, requirements are made on the time of arrival of energy in the hadronic
calorimeter. The typical time-of-flight for relativistic
particles to travel from the interaction point at the center of the detector 
to the calorimeter is approximately 7~nsec. 
Since every tower in the central hadronic calorimeter has timing information 
associated with the energy deposited~\cite{bluebook}, 
all energy deposited at time $t$ must
occur within a 28~nsec
window around the nominal collision time, $t_0$, and corrected for 
the time of flight to be considered
`in time' with the collision. The window is
defined by 
\begin{equation}
-20~{\rm nsec} < t-t_0 < 35~{\rm nsec}.
\end{equation}
The event
is rejected if any  tower has more than 1~GeV
deposited outside the timing window.

\subsection{Final Offline Selection}

The final offline event selection criteria are
listed in Table~\ref{Event Cuts}. 
The two different sets of selection criteria 
correspond to the two trigger paths 
and allow the efficiencies to be well
measured. The low-threshold criteria require both photons to have ${\rm
E_T^{\gamma}} >12$~GeV (where the 10~GeV trigger becomes $>$~98\% efficient)
while the high-threshold criteria are used if 
both photons have \mbox{\Etg$>$ 22~GeV} 
(where the 16~GeV trigger becomes $>$~98\%
efficient). The final data set consists 
of $\NGGTOT$ events.

\subsection{Efficiency of the Selection Criteria}

The 
efficiencies of the selection criteria listed in 
Table~\ref{Event Cuts} are measured using electrons.
Two samples of electrons from $Z^0/\gamma^* \rightarrow e^+e^-$ events are used:
one from  the data 
and one generated using the PYTHIA Monte Carlo 
generator~\cite{PYTHIA} and a detector simulation. 
Each sample is composed of events with one electron 
candidate in the fiducial region which passes tight identification and 
isolation criteria, a second candidate in the
fiducial region with
$\Et>20$~GeV and a matching track with P$_{\rm T}>13$~GeV.  As 
shown in Ref.~\cite{R}, an additional requirement on the invariant 
pair mass, M$_{e^+e^-}$,
within 10~GeV of the $Z^0$ mass produces 
a fairly pure and unbiased sample of electrons which can be used to measure the
efficiency of the selection criteria. As a check, 
the efficiency for each requirement is also calculated using a 
second sample of events with M$_{\rm e^+e^-}>30$~GeV. 
Differences between data and Monte Carlo are quantified as corrections, 
$C_i$, to the data. 

The efficiencies of the identification and isolation selection criteria 
are shown in 
Figures~\ref{EFF Plot Data}, \ref{EFF Plot MC} and
\ref{EFF Plot Ratio} as a
function of \Etg. Figure~\ref{EFF Plot Ratio} 
shows that the ratio of efficiencies
is fairly flat as a function of E$_{\rm T}$.

\twofig{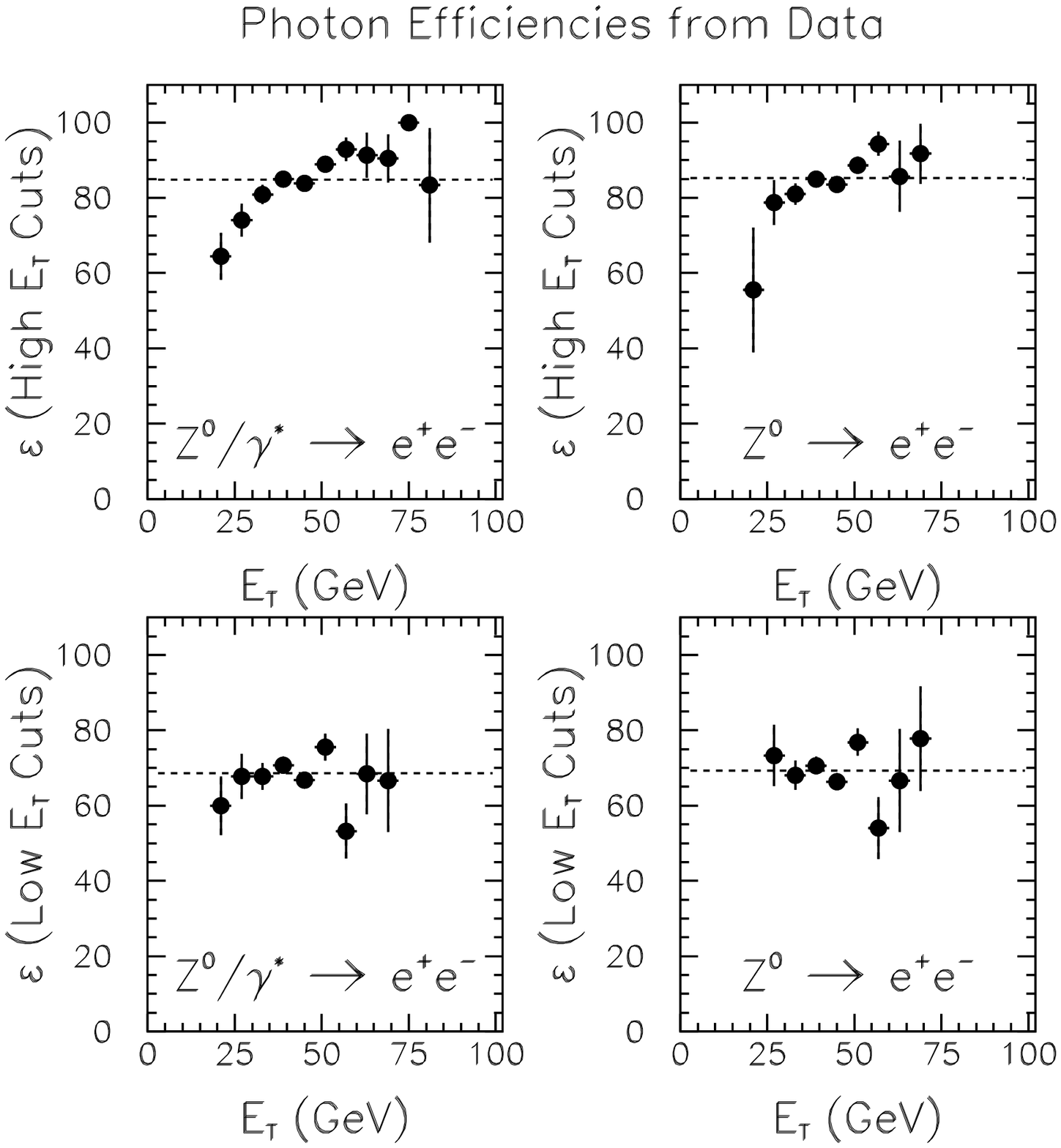}
{The efficiency of the photon identification 
and isolation selection criteria as a function of E$_{\rm
T}$ as measured from a sample of 
$e^+e^-$ events in the data. 
The left-hand plots show the results for events 
with M$_{\rm e^+e^-}>30$~GeV, the right-hand plots
for \mbox{81~GeV $< {\rm M}_{\rm e^+e^-}<$ 101~GeV}. 
The upper plots show the results
for the high \Etg\ threshold selections, 
the lower plots for the lower \Etg\ threshold
selections. The dashed line is the average efficiency.}
{EFF Plot Data}
{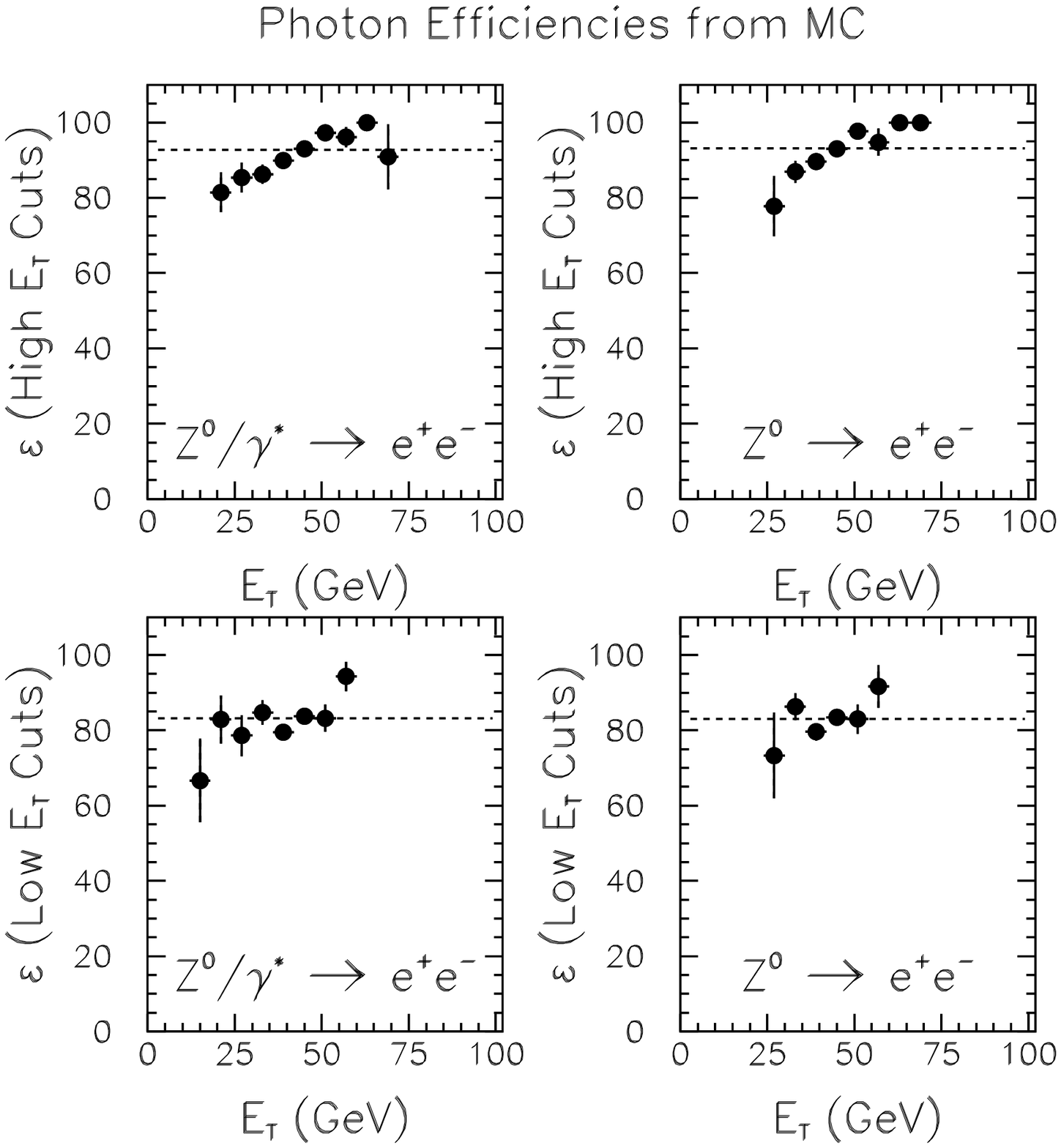}
{The efficiency of the photon identification 
and isolation selection criteria as a function of E$_{\rm
T}$ as measured from the detector simulation. All the criteria are the same
as in Figure~\protect\ref{EFF Plot Data}.}
{EFF Plot MC}

The  distribution in the shower shape variable for rejecting cosmic rays,
$\sigma_{\rm CES}$,
is different for electrons and photons
and is not well modeled in the detector simulation. 
The efficiency is estimated from the data and taken to be 
\mbox{$\epsilon_{\sigma_{\rm CES}} = (\CESEFF)\%$.} 
The $\sigma_{\rm CES}$ requirement is
not used in Monte Carlo simulations, so no correction is made.

An additional 
systematic uncertainty is estimated as half the range of 
efficiencies as a function of \Etg\ (5\%)
and is added in quadrature with the 
statistical uncertainties.
The total photon identification (ID) and isolation (Iso) efficiency is
$\epsilon_{(\rm ID~and~Iso)} = \epsilon_{\rm raw}\times
\epsilon_{\sigma_{\rm CES}}$ and is measured to be 
$\epsilon^{\rm Low  \; Threshold}_{(\rm ID~and~Iso)} = (\PHOEFFIDLOW)\%$  
and 
$\epsilon^{\rm High \; Threshold}_{(\rm ID~and~Iso)} = (\PHOEFFIDHIGH)\%$.
The correction to the efficiency for detecting both photons,
$C_{(\rm Id~and~Iso)}$, to be used in \secorchap~6, 
is determined using:
\begin{equation}
C_{(\rm Id~and~Iso)} = 
(\frac{\epsilon^{\rm Data}_{(\rm ID~and~Iso)}}
{\epsilon^{\rm MC}_{(\rm ID~and~Iso)}})^2 .
\end{equation}
The measured values are $C_{(\rm Id~and~Iso)} = \CORRIDLOW
\; {\rm and}\;  \CORRIDHIGH$ for the low-threshold  and high-threshold
selection criteria,
respectively.

\onefig{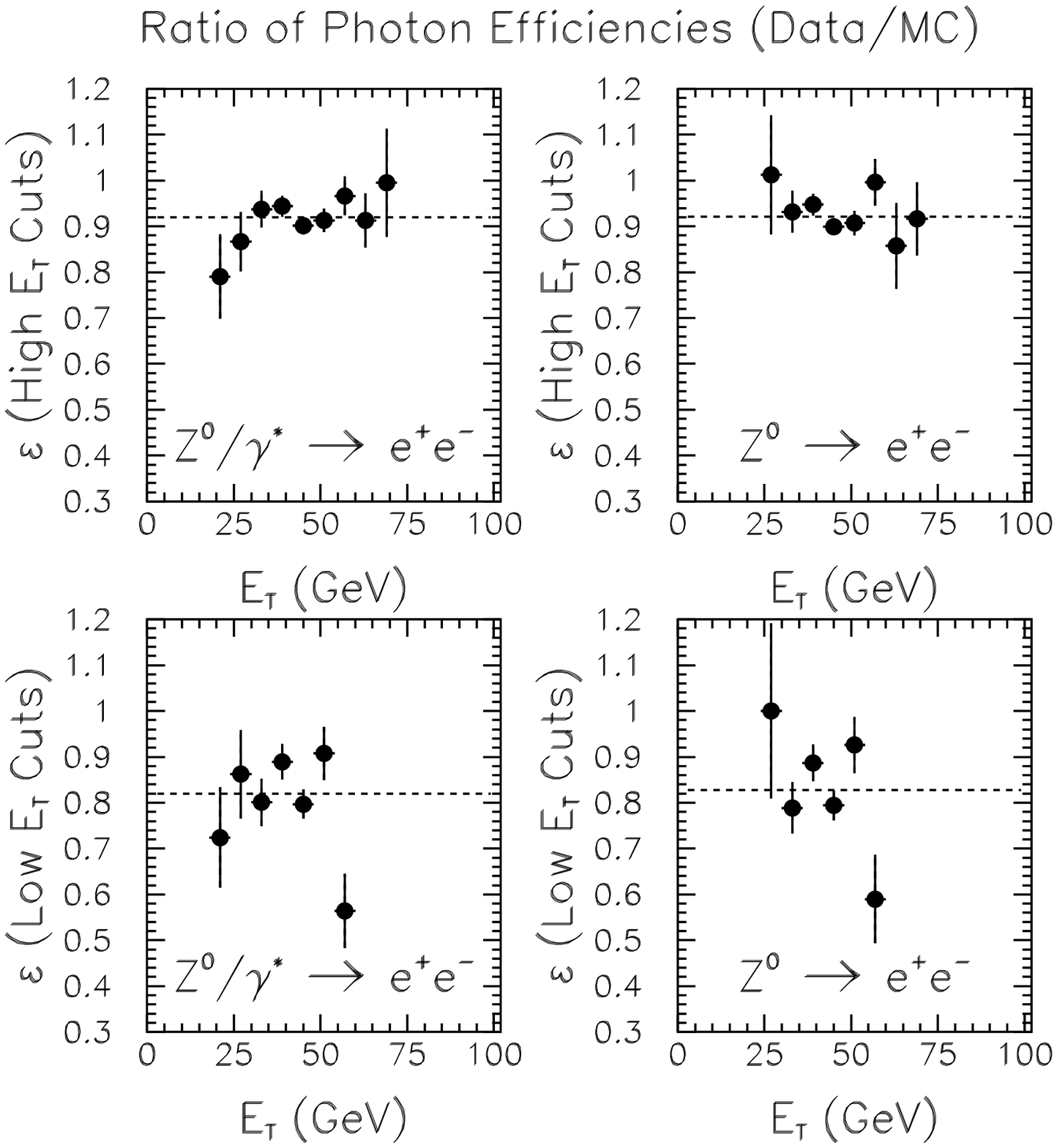}{
The same as Figure~\protect\ref{EFF Plot Data} except that the
   points are the ratio of efficiencies of the 
photon identification and isolation selection criteria as measured from the 
data and the detector simulation.
The dashed line is the average correction factor.}{EFF Plot Ratio}

\begin{sloppypar}
Additional corrections are made for data and Monte Carlo
differences in the 
vertex and Energy-Out-of-Time (ETOUT) distributions. 
The efficiency of the vertex requirements are estimated to be 
\mbox{$\epsilon^{\rm Data}_{\rm z \;
vertex} = (\EFFVERTDAT)\%$}, and 
\mbox{$\epsilon^{\rm MC}_{\rm z \; vertex} = (\EFFVERTMC)\%$} which gives
\mbox{$C_{\rm z \; vertex} = \frac{\epsilon_{\rm Data}}{\epsilon_{\rm MC}} 
= \RATIOVERT$.} 
Since the Energy-Out-of-Time (ETOUT) distribution 
is not simulated in the Monte Carlo, 
the efficiency and correction are taken to be 
$C_{\rm ETOUT} = \epsilon_{\rm ETOUT} = (\PEFFETOUT)\%$.
\end{sloppypar}

\begin{sloppypar}
The efficiencies of the diphoton 
triggers are measured using independent triggers. 
Figure~\ref{Low ET Trigger Plot}a shows the efficiency of the 
L2 low-threshold trigger as a function of the isolation energy 
in a 3-by-3 array of trigger towers around the cluster, \boxiso.
Figure~\ref{Low ET Trigger Plot}b 
shows the trigger efficiency as a function of \Etg; the efficiency is
flat as a function of \Etg\ above 12~GeV. Above 12~GeV 
the trigger efficiency is taken to be 
$\epsilon^{\rm Low\; Threshold}_{\rm Trigger} = (\PTRIGLOW)$\%. 
The trigger efficiencies
for the L2 trigger and the 
L2-L3 high-threshold trigger path are shown  in
Figure~\ref{High ET Trigger Plot}. The efficiency is
flat as a function of \Etg\ above 22~GeV.
Above 22~GeV the trigger efficiency is taken to be 
$\epsilon^{\rm High\; Threshold}_{\rm Trigger} = \PTRIGHIGH$\%. 
\end{sloppypar}

\twofig{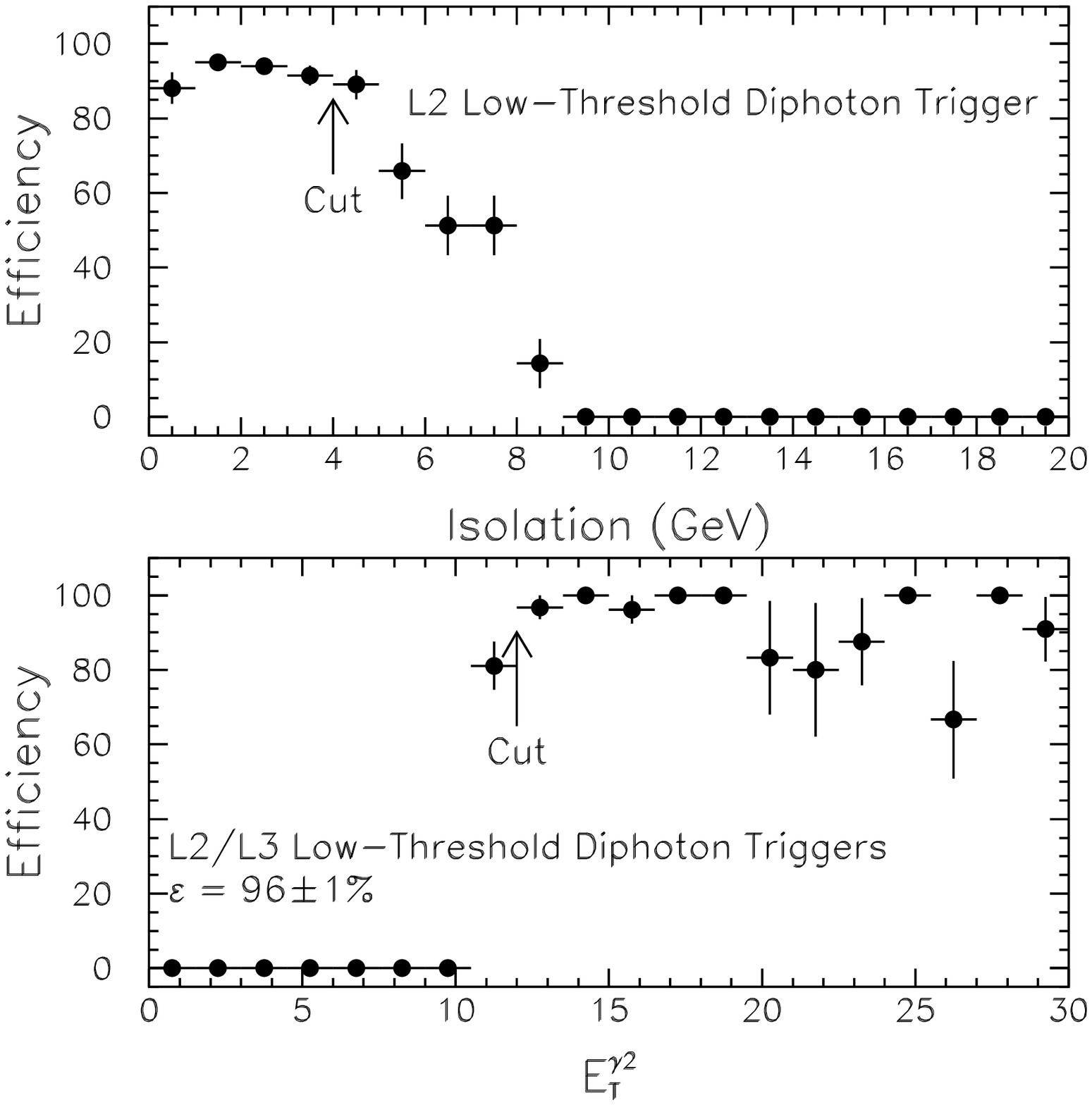}
{The top plot shows the efficiency of the L2 low-threshold trigger 
as a function of the \boxiso\ selection. The lower plot shows the efficiency of
the L2/L3 low-threshold 
trigger path as a function of \Etgt\, the softer of the two photons.
 The trigger is fully efficient for \Etggl\
and has an efficiency of ($\PTRIGLOW)$\%.}
{Low ET Trigger Plot}
{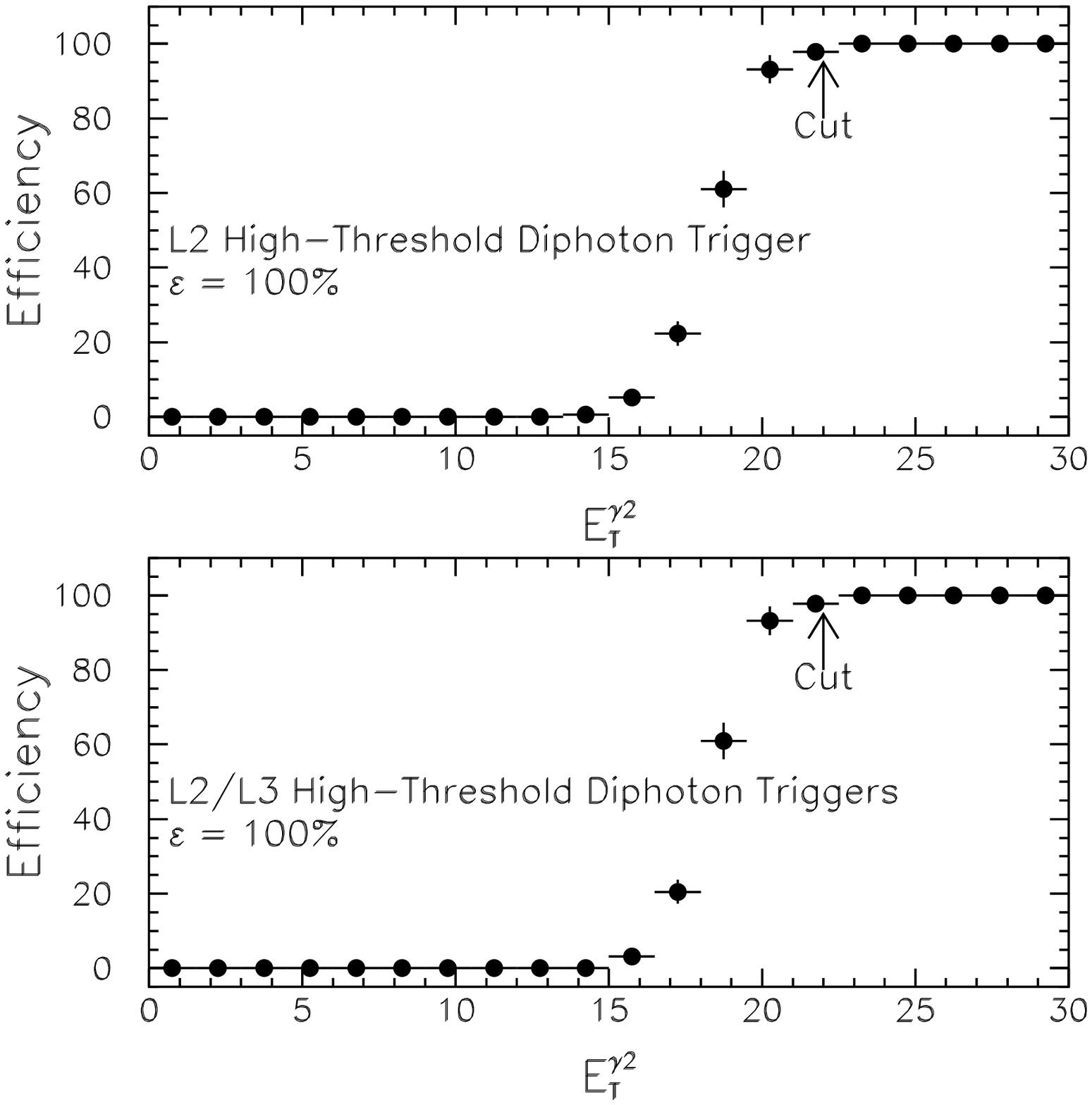}
{The efficiency of the L2 high-threshold and 
the L2/L3 high-threshold triggers as a function of \Etgt\, the softer of the 
two photons.
The trigger is fully efficient for \Etggmh\
and has an efficiency of $\PTRIGHIGH$\%.}
{High ET Trigger Plot}


\subsection{Purity of the Diphoton Sample}

Since the purity of the sample is of less importance than the efficiency for
searches for anomalous events, the selection criteria have been chosen 
to have high efficiency. Even 
after requiring each photon to pass all of the 
selection criteria, there are still a substantial number of background events in
the sample.  The 
backgrounds are primarily due to hadronic jets which contain pions, 
kaons or etas, each of which can decay to multiple photons. 

To estimate the photon backgrounds, each photon candidate is compared to the
single photon hypothesis and the background
hypothesis in
a manner similar to that in Ref.~\cite{photons}.
For candidates with ${\rm E_T} < 35$~GeV, the strip chamber 
system  can distinguish the difference between a single 
$\gamma$ and 
$\pi^0\rightarrow \gamma\gamma$. For higher energies, 
${\rm E_T} > 35$~GeV, the two photons cannot be resolved in the CES. Instead, the
central preradiator system (CPR) 
is used to measure the conversion probability 
in the magnet coil. 
In both cases, it is not
possible to separate prompt photons and backgrounds 
on an event-by-event basis.  However, standard techniques 
allow the extraction of purity information on a statistical basis in large
samples.  Using these techniques on 
both candidates 
the average purity of the photon sample is estimated to be 
(\PURITYLOW)\% prompt diphoton events.
            
\section{Searches for Deviations from Standard Model Predictions}
\label{Chap-Searches}

Each of the $\NGGTOT$ events in the diphoton sample
is searched for the presence of 
$\mett$, jets,  electrons, muons, taus, \mbox{$b$-quarks}, or additional
photons.
Deviations from standard model predictions
are searched for using two values of the photon $\Et$  thresholds: \mbox{$\Et
=12$~GeV} and \mbox{$\Et = 25$~GeV}.  The $\Et$=12~GeV threshold
has better acceptance for low-$Q^2$  decays to photons, but
has more background. The $\Et =25$~GeV threshold 
accepts many fewer standard model
events and so has better discrimination for high-$Q^2$ decays.


\subsection{Missing E$_{\rm T}$}\label{Resolution Section}

The standard method for inferring the presence of particles that do not interact
in the calorimeter, 
such as neutrinos, is measuring the missing transverse energy 
(\mett) in the event~\cite{WMass}. 
The \mett\ is corrected for the measured detector response to jets 
and takes into account cracks between
detector components and nonlinear calorimeter 
response~\cite{top,jets}.  In addition, 
the \mett\ is corrected for the presence of
muons, which do not deposit their total 
energy in the calorimeter~\cite{top}.   

\dtfloat{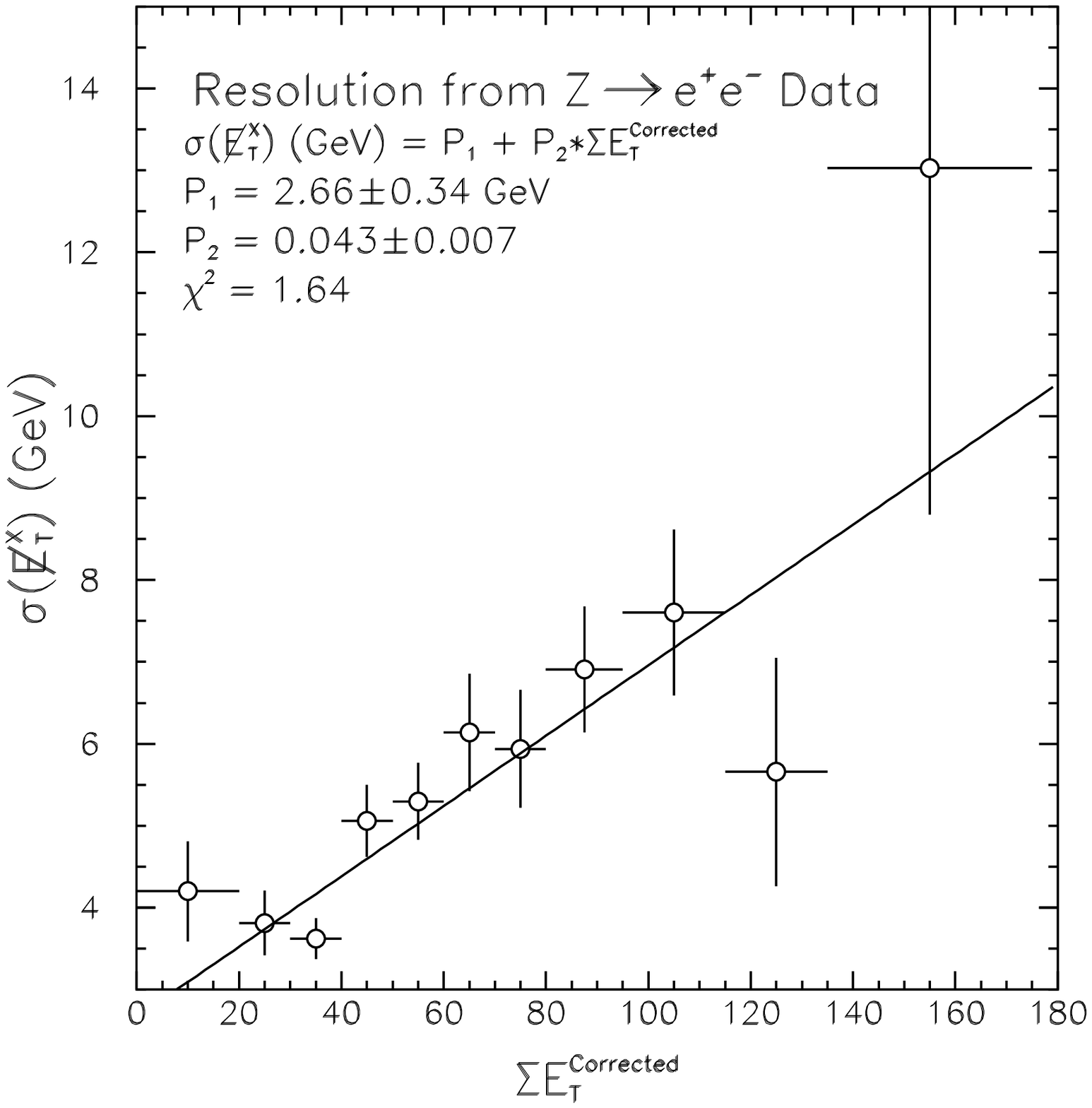}
{The resolution on one component of the \mett\ (\mettx) 
as determined from a sample of $\zoee$ 
events. }
{Met Resolution Plot}{0.48}

\begin{sloppypar}
While the corrections improve the \mett\ resolution on average, some events
still have a substantially mismeasured \mett.
Many of these events can be removed by rejecting 
events which have a jet with ${\rm E_T}>10$~GeV pointing within
10$^{\circ}$ in azimuth of the \mett. Since this requirement introduces an
unnecessary inefficiency and a possible bias when 
searching for leptons, bosons, or jets, it is only imposed 
when searching for the presence of \mett\ and in making 
all \mett\ plots. The requirement 
removes only 48 of the \NGGTOT\ events in the sample. 

The \mett\ resolution is measured using a fairly pure sample of 
$\zoee$ events. Events are selected if they have two electrons, 
each passing the standard requirements, and 
 M$_{ee}$ within 10~GeV of the mass of the $Z^0$~\cite{No Met}. 
The resolution is 
plotted in Figure~\ref{Met Resolution Plot} 
as a function of \sumetc\ where \mbox{\sumetc $= 
\Sigma {\rm E_T} - {\rm E}_{\rm T}^{e_1} - {\rm E}_{\rm T}^{e_2}$. }
In the region $\Sigma {\rm E}_{\rm T}^{\rm Corrected} <150~{\rm GeV}$
the
distribution is  well-parameterized by 
\begin{eqnarray}\label{Resolution Equation}
\sigma(\mettx) & = & (\RESOFFSET \; {\rm~GeV}) 
+ (\RESSLOPE) \times \Sigma 
{\rm E}_{\rm T}^{\rm Corrected}. 
\end{eqnarray}
\end{sloppypar}

Standard model 
diphoton events have no intrinsic \mett; thus the expected \mett\ distribution
can be predicted from the resolution alone. This has the advantage that the
estimate is determined from the data.
The expected \mett\ distribution is estimated
by smearing the 
X and Y components of the true \mett\ (assumed to be zero) by the 
resolution (estimated from $\Sigma {\rm E}_{\rm T}^{\rm Corrected}$ on an 
event-by-event basis). 
The systematic uncertainty on the distribution
is found by varying the resolution within its uncertainty.

The data are shown along with the expectations from the resolution simulation
in Figures~\ref{Diphoton Met Data 12} and \ref{Diphoton Met Data 25}. With 
the exception of one event on the tail on the
distribution, the `$\eeggmett$' candidate 
event~\cite{Park} \mbox{($\mett = \EEGGTOTMET$~GeV)},  the data agree
well with the expectations.  For a photon E$_{\rm T}$ threshold of 12~GeV 
one event with $\mett>35$~GeV is observed, with a expectation of
\NMETEXPLOW\ events. For a photon threshold  E$_{\rm T}$ of 25~GeV,  
two events are observed with $\mett>25$~GeV, with \NMETEXPHIGH\ events
expected.   The $\eeggmett$ candidate event will be discussed in more
detail in \secorchap~4.  The other event has both photons above 25~GeV and
$\mett = 34~$GeV. However, on close inspection,  it appears
to be due to two mis-measurements. The event contains an energetic jet 
($\Et$ = 44~GeV) 
which points 
directly at the region
between the plug and forward calorimeters and near the \mett\ in $\phi$ and is
therefore likely to be significantly mismeasured. 
Moreover, one of the photons is at the
edge of the fiducial region and may be undermeasured~\cite{Why Under}, 
causing the $\phi$ position of the \mett\ 
to be just far enough away from the jet to pass the
$\Delta\phi_{\rm \mettsm-jet} >10^{\circ}$ requirement.  The 4-vectors of the
event are given in Table~\ref{Met Event}.

\twofig{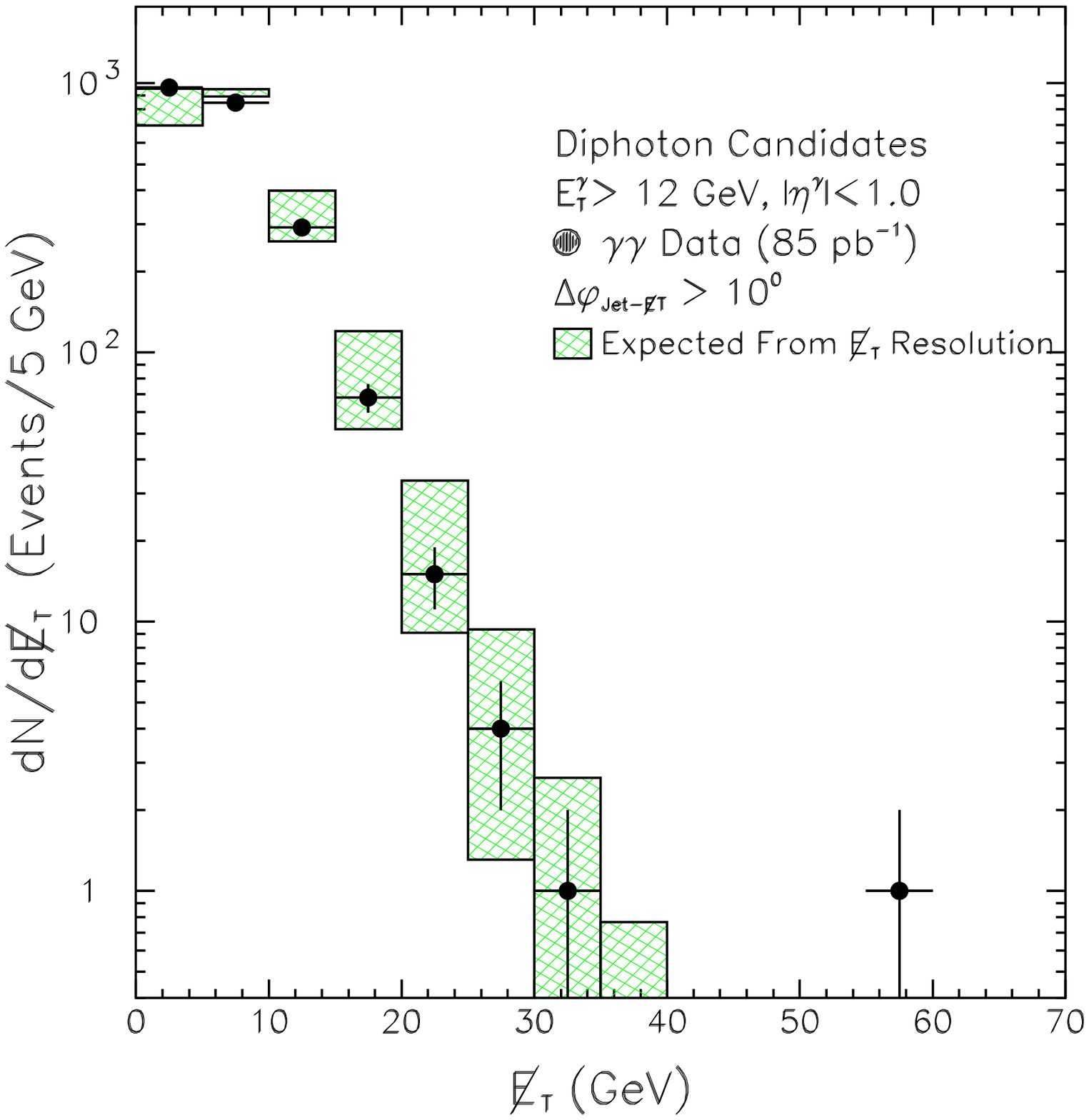}
{The \mett\ spectrum for 
diphoton events with \Etggl\ in the data. The boxes indicate
the range of the values of the 
\mett\ distribution predicted from detector resolution. 
The one event on
the tail is the $\eeggmett$ candidate 
event, described in detail in \secorchap~4.}
{Diphoton Met Data 12}
{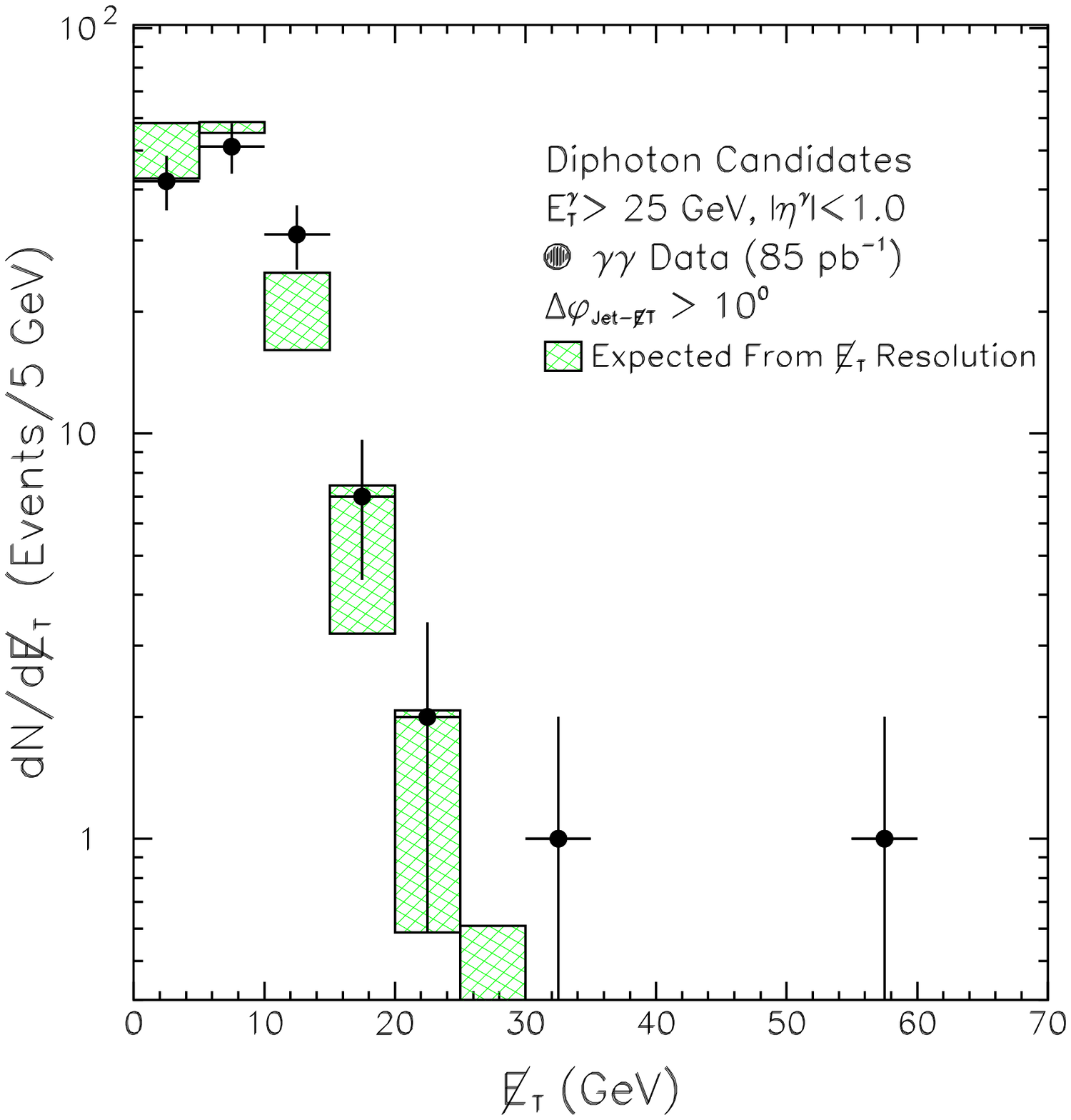}
{The \mett\ spectrum for 
diphoton events with \Etggh\ in the data. The boxes indicate
the range of the values of 
the \mett\ distribution predicted from detector resolution.  
The one event on
the tail is the $\eeggmett$ candidate 
event, described in detail in \secorchap~4.}
{Diphoton Met Data 25}

\begin{table}[htb]
\centering
\begin{tabular}{l|c|c|c|c|c} 
\multicolumn{6}{c}{Run 67397, Event 47088} \\ \hline
  & P$_{\rm x} $  & P$_{\rm y}$  & P$_{\rm z}$ & E     & E$_{\rm T}$  \\ \hline
  & (GeV/c)       & (GeV/c)      & (GeV/c)     & (GeV) & (GeV)        \\
\hline
$\gamma_1$ & -85.8& 1.6          &   63.4      & 106.7 & 85.8 \\
$\gamma_2$ & 30.8 & -15.9        &    6.4      & 35.3  & 34.7 \\ 
$j_1$      & 40.1 & 18.8         & 237         & 242   & 44.4 \\ 
\mett      & 33.6 & -5.5         & --          & --    & 34.1 \\
\end{tabular}
\caption[The 4-vectors of the $\gamma\gamma+\mett$ candidate event]
{The 4-vectors of the $\gamma\gamma+\mett$ candidate event. This event may
be due to mis-measurement as the  
$\Delta\phi$ between the jet and the \mett\ is 34$^\circ$, the
jet points at the region between the plug and forward calorimeters and the 
second photon, $\gamma_2$, is at the edge of the fiducial region of the
central calorimeter and may be undermeasured~\protect\cite{Why Under}.}
\label{Met Event}
\end{table}


\subsection{Jets}\label{Jets Section}

To search for anomalous production of quarks and gluons, the number of jets, 
N$_{\rm Jet}$, is counted in a manner identical to that used in the top-quark
discovery in the dilepton channel~\cite{top,jets}. 
Each jet is required to have 
uncorrected E$_{\rm T} > 10$~GeV and 
$|\eta|< 2.0$. The distributions in the number of jets are shown in 
Figures~\ref{Diphoton NJET Data 12} and \ref{Diphoton NJET Data 25} for photon
E$_{\rm T}$ thresholds of 12~GeV and 25~GeV respectively.

\twofig{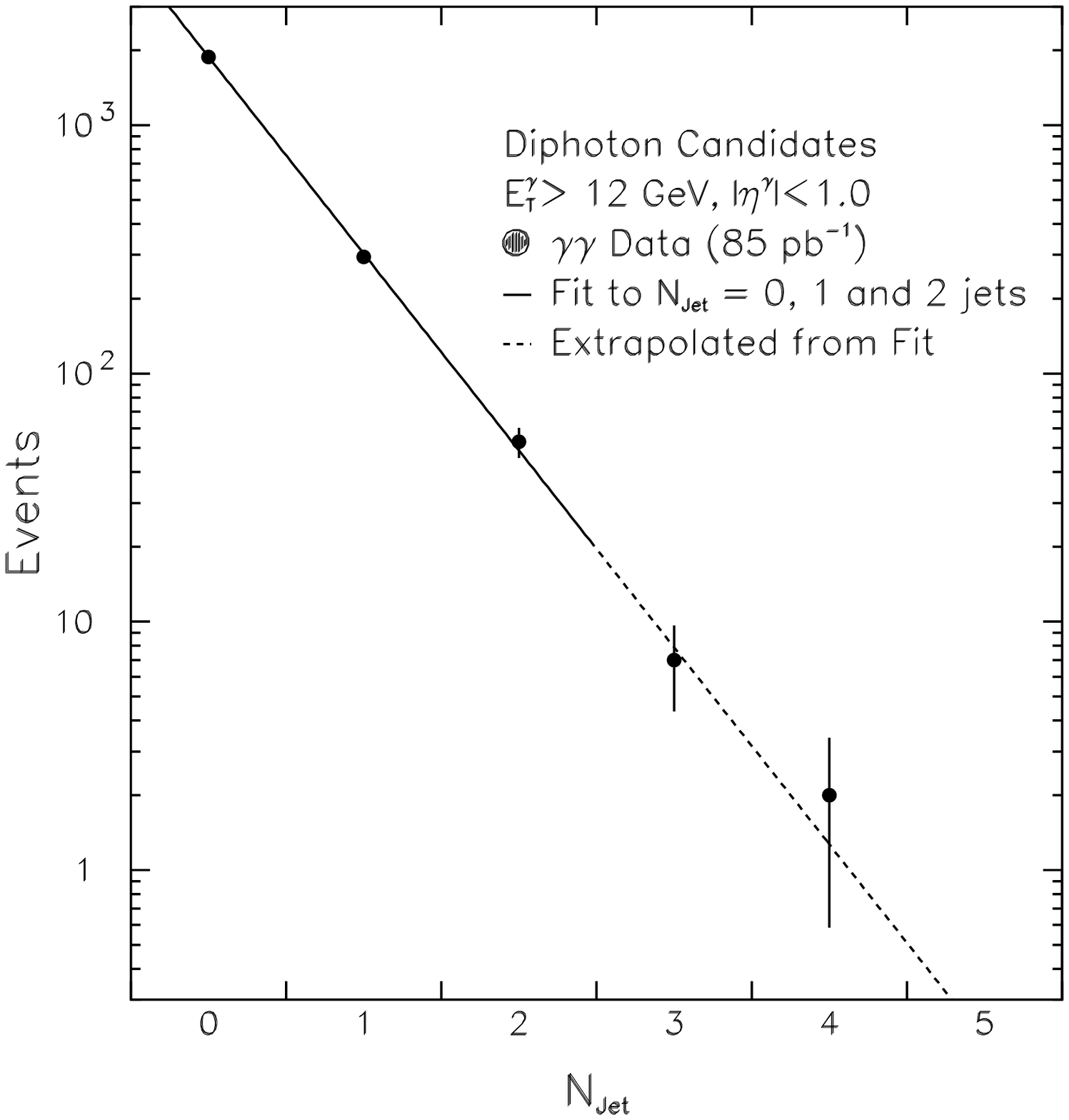}
{The number of jets, N$_{\rm Jet}$, produced in association with 
diphoton pairs with \Etggl.
The line is an exponential fit to the data with N$_{\rm Jet} \le 3$, and is
extrapolated to N$_{\rm Jet} \ge 4$.}
{Diphoton NJET Data 12}
{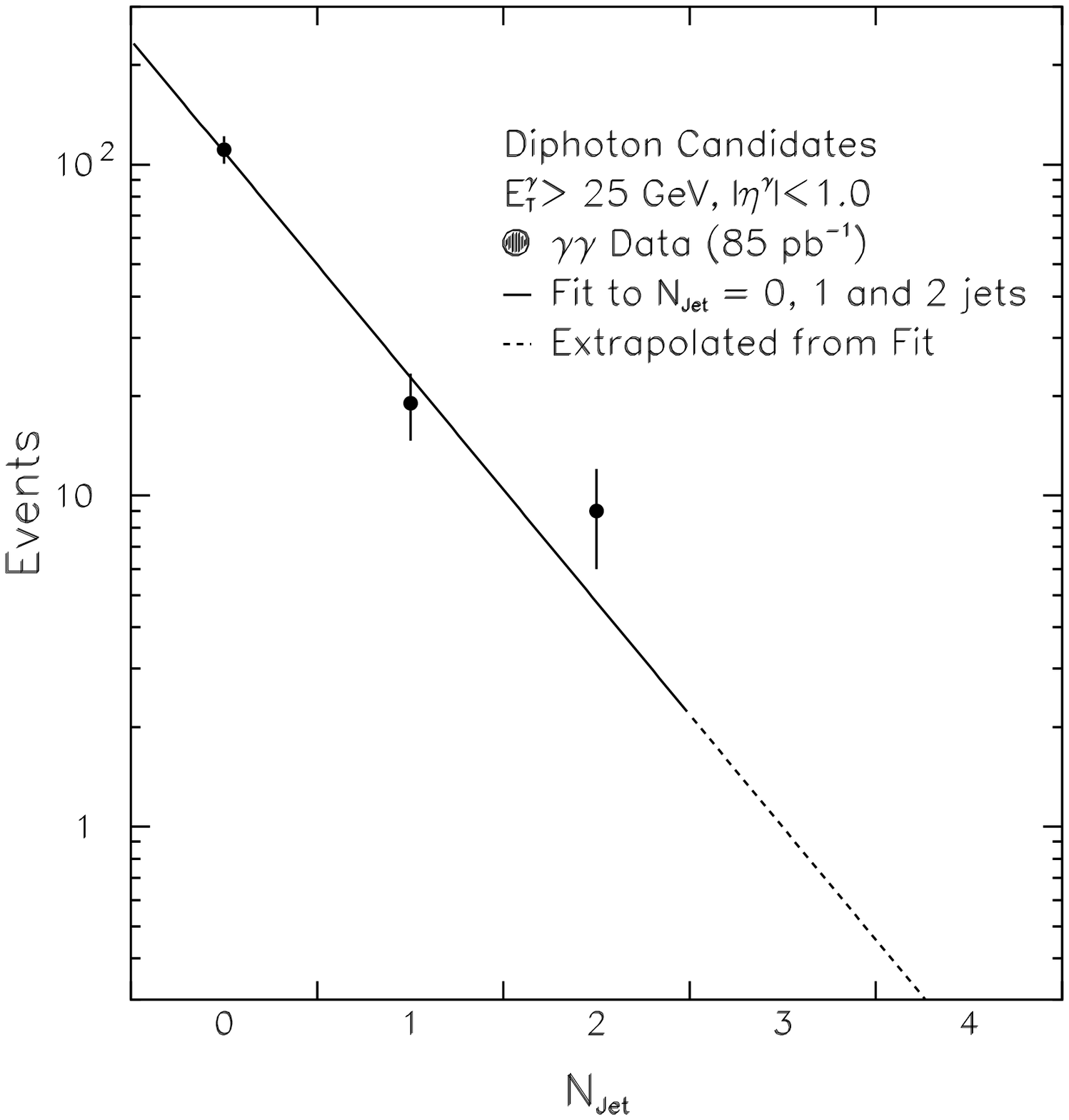}
{The number of jets, N$_{\rm Jet}$, produced in association with 
diphoton pairs with \Etggh.
The line is an exponential fit to the data with N$_{\rm Jet} \le 2$, and is
extrapolated to N$_{\rm Jet} \ge 3$.}
{Diphoton NJET Data 25}

While there are cross section predictions for $\gamma\gamma$, 
$\gamma\gamma + 1$ jet and $\gamma\gamma + 2$ jet production~\cite{Huston}, 
there is currently no theoretical 
prediction for higher jet multiplicities.  However,
it has been known for some time that the ratio 
between $n$-jet and \mbox{$(n-1)$-jet} 
cross sections for $W$ and $Z^0$ production can be approximated 
by a constant~\cite{Ellis},  
\begin{equation}
R_n = \frac{\sigma(V\; + \; n\; {\rm jets})}{\sigma(V\; + \; (n-1)\; 
{\rm jets})}
\end{equation}
where $V$ is either a $W$ or a $Z^0$.
This has been confirmed within resolution in the 
CDF data~\cite{Hauger} and is expected to hold for most processes
since additional jets are typically due to initial-state 
and final-state radiation. 

To look for anomalous
N$_{\rm Jet}$ production in the $\gamma\gamma$ data, an exponential fit 
for small values of N$_{\rm Jet}$ is used to 
extrapolate to the large N$_{\rm Jet}$ region. 
For diphoton events with \Etggl\  and  N$_{\rm Jet} \le 3$ the 
parameterization predicts \NIVJETEXPLOW\ 
events with 4 or more jets; \NIVJETLOW\ events are observed. For diphoton events
with \mbox{\Etggh}  and  N$_{\rm Jet} \le 2$, 
the parameterization predicts
$\NIIIJETEXPHIGH$ events  with 3 or more jets; $\NIIIJETHIGH$ events 
are observed.


\subsection{Electrons and Muons}\label{LGG Section}

Electrons and  muons produced in association with photon pairs 
are required to be isolated, have 
E$_{\rm T}>25$~GeV, and be in the central part of the detector 
($|\eta|<1.0$). They are identified with the same 
identification and isolation selection requirements 
used in the top-quark discovery~\cite{top}.  A total of 
\NCENTEORMU\   events with a central electron or muon 
are found in the data. The first event 
has two muons and two photons. This event (see Table~\ref{Mu Mu Event}) 
is consistent with a double-radiative $Z^0$ decay, 
$\ppbar \rightarrow Z^0 \rightarrow
\mu^+\mu^-\gamma\gamma$, since the 4-body invariant mass is 
$m_{\mu^+\mu^-\gamma\gamma} = \MUMUGGMASS$~GeV/c$^2$.
The second event (see Table~\ref{EGG Event}) has a single electron. 
This event is also likely to be due to 
the decay of a $Z^0$ boson because \mbox
{$m_{e^+\gamma\gamma} = \EGGMASS$~GeV/c$^2$} and 
there is some indication in the SVX that 
there is  a charged particle traveling in the direction of $\gamma_2$.
The third event is the `$\eeggmett$' candidate event and will be discussed
further in \secorchap~4. Only the `$\eeggmett$' candidate event passes the 
photon threshold of \Etggh.

\begin{table}
\centering
\begin{tabular}{l|c|c|c|c|c}
\multicolumn{6}{c}{Run 69571, Event 769815} \\
\hline\
  & P$_{\rm x} $  & P$_{\rm y}$  & P$_{\rm z}$ & E     & E$_{\rm T}$  \\
  & (GeV/c)       & (GeV/c)      & (GeV/c)     & (GeV) & (GeV)        \\
\hline
$\gamma_1$ & 12.4 & 5.3          & -1.4        & 13.5  & 13.4         \\ 
$\gamma_2$ & -16.0& -5.3         & -3.2        & 17.1  & 16.8         \\
$\mu_1$    & 28.2 & 9.5          & -33.4       & 44.7  & 29.7           \\
$\mu_2$    & -26.5& -4.9        & -14.9         & 30.8 & 27.0           \\
\mett      & 7.1  & -0.3        & --          & --    & 7.1 \\
\end{tabular}
\caption[The 4-vectors of the $\gamma\gamma+\mu\mu$ candidate event]
{The 4-vectors of the $\gamma\gamma+\mu\mu$ candidate event. This event 
is consistent with a double-radiative $Z^0$ decay, 
$\ppbar \rightarrow Z^0 \rightarrow
\mu^+\mu^-\gamma\gamma$ 
\mbox{($m_{\mu^+\mu^-\gamma\gamma} = \MUMUGGMASS$~GeV/c$^2$).}}
\label{Mu Mu Event}
\end{table}

\begin{table}
\centering
\begin{tabular}{l|c|c|c|c|c} 
\multicolumn{6}{c}{Run 63541, Event 304680} \\
\hline
  & P$_{\rm x} $  & P$_{\rm y}$  & P$_{\rm z}$ & E     & E$_{\rm T}$  \\
  & (GeV/c)       & (GeV/c)      & (GeV/c)     & (GeV) & (GeV)        \\
\hline
$\gamma_1$ & -13.7 & -19.9      & -28.8         & 37.6 & 24.2 \\
$\gamma_2$ & -9.8  & -13.1      & -10.1         & 19.2 & 16.3 \\
$e_1$      & 19.7 & 35.0         & 5.0         & 40.4 & 40.1  \\ 
\mett      & -4.3 & 0.4        & --          & --    & 4.4 \\
\end{tabular}
\caption[The 4-vectors of the $\gamma\gamma+e$ candidate event]
{The 4-vectors of the $\gamma\gamma+e$ candidate event. 
The invariant mass of the two photons
and the electron indicates that this may be a $Z^0$ 
\mbox{($m_{e^+\gamma\gamma} = \EGGMASS$~GeV/c$^2$)} where one of the electrons
was identified as a photon or the electron emitted all its energy in a photon
via bremsstrahlung.}
\label{EGG Event}
\end{table}

The dominant standard model sources of extra leptons in $\gamma\gamma$ events
is inclusive $W$ and $Z^0$ production and decay. Diagrams include
$W \rightarrow \ell\nu\gamma\gamma, W\gamma \rightarrow \ell\nu\gamma\gamma,
 W\gamma\gamma \rightarrow \ell\nu\gamma\gamma, 
Z^0 \rightarrow \ell\ell\gamma\gamma, 
Z^0\gamma \rightarrow \ell\ell\gamma\gamma$ and 
$Z^0\gamma\gamma \rightarrow \ell\ell\gamma\gamma$.  These processes,
where $\ell$ is an electron, muon or tau, 
are simulated using the PYTHIA~\cite{PYTHIA} 
Monte Carlo and a detector simulation and checked using the 
MADGRAPH~\cite{Mrenna Lgamma,MADGRAPH} Monte Carlo.  
The Monte Carlo  estimates 
$\NCENTLEPDIBOSON$ $\ell\gamma\gamma + X$ events in the data.

A source of $e\gamma\gamma$ events  
which is not correctly simulated with the Monte Carlo is
$Z^0\gamma \rightarrow e^+e^-\gamma$ where one of the electrons 
is identified as a photon. This can occur if the electron emits a 
photon via bremsstrahlung (the photon 
carries away most of the energy and the electron is lost in the detector) or 
the track of the electron is not found by the central tracking chamber. 
The rate at which electrons are
misidentified as a photon is determined from a 
sample of $\zoee$ events from the data and is estimated 
to be (\EGFAKERATE)\% per
electron. The total number of $e\gamma\gamma$
events expected from this source is estimated from $ee\gamma$ data
to be
\NCENTLEPFAKE. 

Summing the above sources
gives an expectation of 
\NCENTEORMUEXP\ $\ell\gamma\gamma+X$ events 
in the \Etggl\ data.  Similarly, for the photon threshold of \Etggh\ these 
methods
predict a total of 0.1$\pm$0.1 events, dominated by events in which electrons
fake photons.
The dominant mechanism for producing $\ell\ell\gamma\gamma$ events is 
dominated by inclusive $Z^0$ production and decay. 
The PYTHIA Monte Carlo predicts a
total of $\NCENTLEPDIBOSON$ events to be
observed
in the data. 


\subsection{Taus}

Hadronic decays of a $\tau$ lepton produced in association with diphotons 
are identified using standard
identification criteria~\cite{Marcus} and are required to have 
E$_{\rm T}>$ 25~GeV
and \mbox{$|\eta|<1.2$}.
One $\tau\gamma\gamma$ candidate (see Table~\ref{Tau Event}) is observed 
in the data with 
\Etggl; none with
\Etggh. The dominant source of SM $\tau\gamma\gamma$ candidate events
is from hadronic jets produced in
association with diphoton pairs which fake
the hadronic $\tau$ decay signature.  This rate 
is estimated using the methods of Ref.~\cite{Marcus}.  
Figure~\ref{Tau Back} shows the E$_{\rm T}$ spectrum for $\tau$ leptons measured
in the data as well
as for backgrounds from fake $\tau$'s.   A total of 
\NTAUEXPLOW\ events where a jet fakes a $\tau$ are expected in the data for
\Etggl, and 0.03$\pm$0.03 events for \Etggh; both are
consistent with observation. 

\begin{table}
\centering
\begin{tabular}{l|c|c|c|c|c} 
\multicolumn{6}{c}{Run 66392, Event 23895} \\
\hline
  & P$_{\rm x} $  & P$_{\rm y}$  & P$_{\rm z}$ & E     & E$_{\rm T}$  \\
  & (GeV/c)       & (GeV/c)      & (GeV/c)     & (GeV) & (GeV)        \\
\hline
$\gamma_1$ & 3.6  & 23.3        & 3.9           & 23.9 & 23.6 \\
$\gamma_2$ & -11.5& 15.4        & 5.6           & 20.0  & 19.2 \\
$\tau$     &  14.6& -20.7       & 26.3          & 36.6  & 25.4 \\
$j_1$      & -13.5& -9.2        & 11.3          & 20.2  & 16.6 \\
$j_2$      & 19.1 & 6.4         & 33.5          & 39.4  & 20.3 \\
\mett      &-12.7 & -7.0        & --            & --    & 14.5\\ 
\end{tabular}
\caption[The 4-vectors of the $\gamma\gamma+\tau$ candidate event]
{The 4-vectors of the $\gamma\gamma+\tau$ candidate event. 
The \mett\ in the event is small (14.5~GeV) and the first jet, $j_1$,
is only $5.5^{\circ}$ in $\phi$ away from the \mett.}
\label{Tau Event}
\end{table}


\subsection{$b$-Quarks}

Jets from $b$-quarks are identified using the 
$b$-tagging jet algorithm (SECVTX) developed for 
the top-quark \mbox{discovery~\cite{top,Doug G}.}
Two $b\gamma\gamma$ candidate events are observed in the data with \Etggl; none
with \Etggh. Quark and gluon jets produced in
association with diphoton pairs are real 
and fake sources of $b$-jets. 
The number of $b\gamma\gamma$ events from these sources is estimated
using the same methods as developed for the top-quark 
discovery~\cite{top,Doug G,Method 1}. 
Figure~\ref{BTAG Back} shows the E$_{\rm T}$ spectrum of the 
$b$-tagged jets 
and the expectations from the background prediction.
A total of \NBEXPLOW\ $b\gamma\gamma$ events are expected to be 
in the sample due to real and fake sources of $\gamma\gamma + b$ for \Etggl; 
0.1$\pm$0.1 events are expected for \Etggh. 
The 4-vectors of the objects in the two $b\gamma\gamma$ events 
are given in Tables~\ref{Btag1 Event}
and \ref{Btag2 Event}.

\twofig{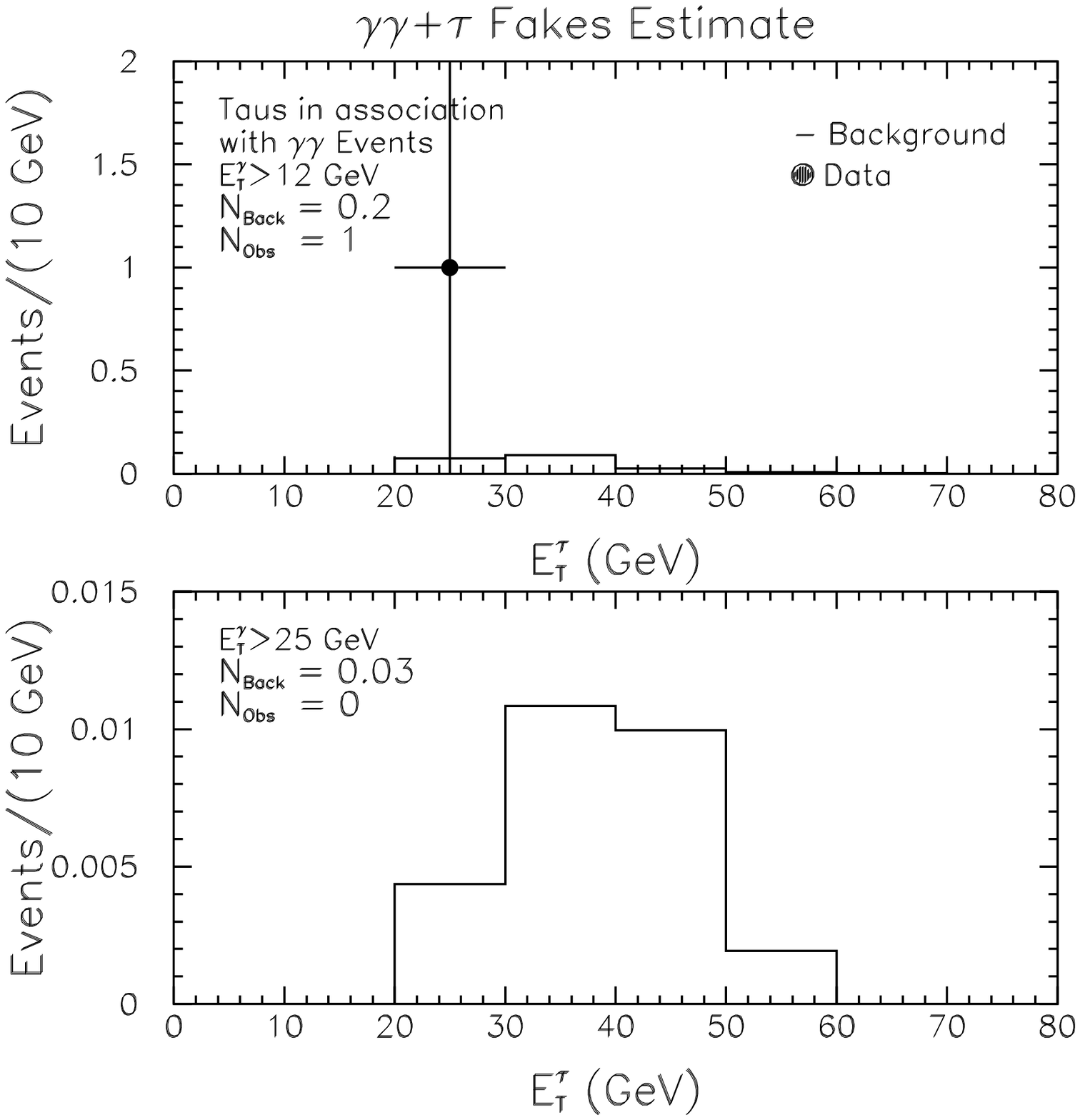}
{The E$_{\rm T}$ spectrum 
of $\tau$ candidates produced in association
with diphoton pairs. Only the hadronic decays of the $\tau$ are included.
The upper plot is for diphoton events in which both photons have
\Etggl. There are no events with a $\tau$ candidate in the data for 
\Etggh, as shown in the lower plot. 
The point represents the one event in the data; 
the histogram is the expectation from fake $\tau$'s.}
{Tau Back}
{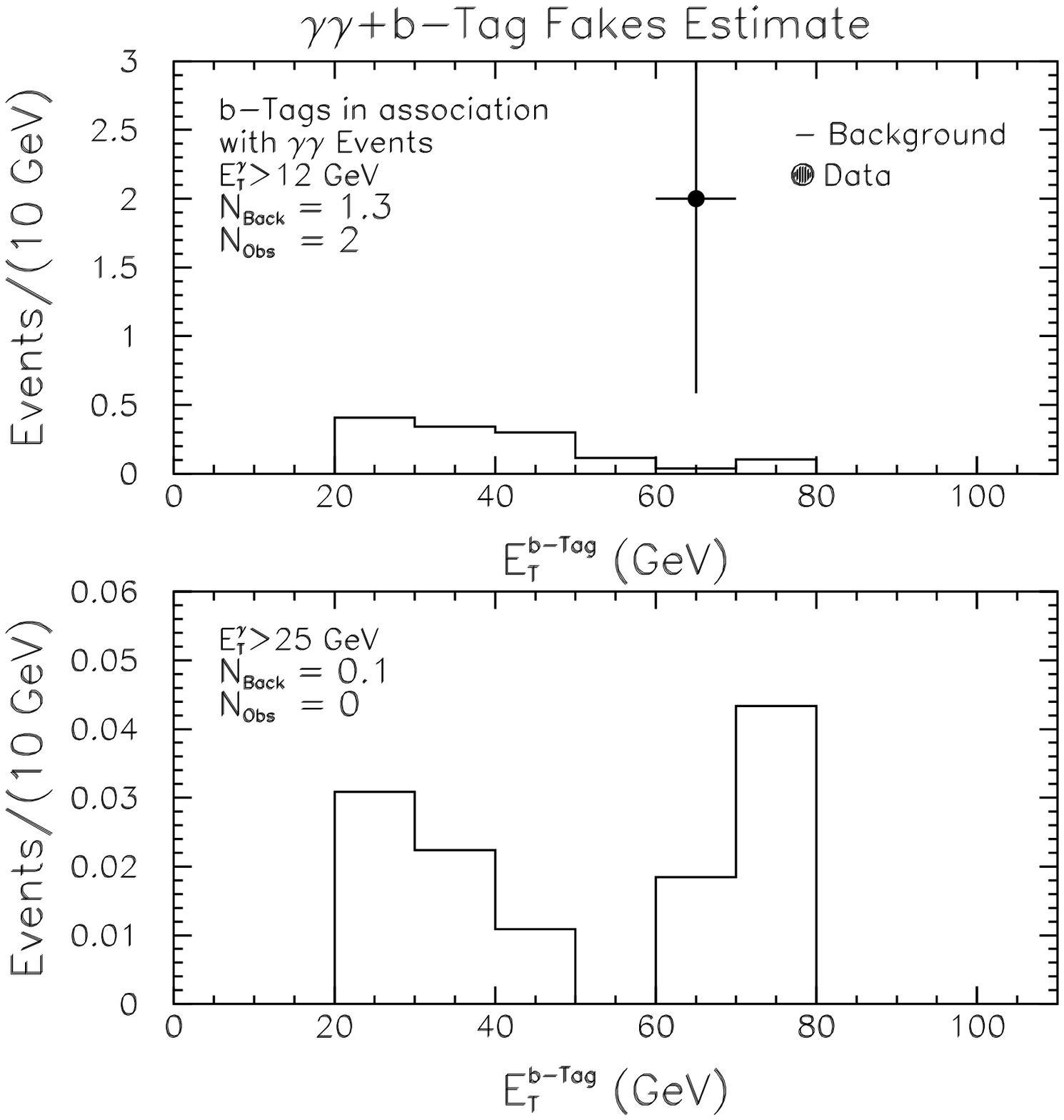}
{The E$_{\rm T}$ spectrum 
of $b$-tags produced in association
with diphoton pairs. 
The upper plot is for diphoton events in which both photons have
\Etggl. The lower plot is for diphoton events in which both photons have
\Etggh.  The point represents the data; the histogram is the expectation
from real and fake sources of $b$-tags. No events have $b$-tags in the \Etggh\
sample.}
{BTAG Back}

\begin{table}[htb]
\centering
\begin{tabular}{l|c|c|c|c|c} 
\multicolumn{6}{c}{Run 63033, Event 337739} \\
\hline
  & P$_{\rm x} $  & P$_{\rm y}$  & P$_{\rm z}$ & E     & E$_{\rm T}$  \\
  & (GeV/c)       & (GeV/c)      & (GeV/c)     & (GeV) & (GeV)        \\
\hline
$\gamma_1$ & -21.6& -8.2         & -16.7       & 28.5   & 23.1          \\
$\gamma_2$ & -14.3& -22.1        & -12.8       & 29.3   & 26.4          \\
$b$-jet    & 44.8 & 40.8         & 37.6         & 71.8  & 61.0          \\ 
$j_2$      & 4.9  & 13.0         & 12.4         & 18.6  & 13.9          \\
\mett      & -4.0 & -4.5         & --           & --   & 6.1 \\
\end{tabular}
\caption[The 4-vectors of the first $\gamma\gamma+b$ candidate event]
{The 4-vectors of the first $\gamma\gamma+b$ candidate event.}
\label{Btag1 Event}
\end{table}

\begin{table}[htb]
\centering
\begin{tabular}{l|c|c|c|c|c} 
\multicolumn{6}{c}{Run 64811, Event  62109} \\
\hline
  & P$_{\rm x} $  & P$_{\rm y}$  & P$_{\rm z}$ & E     & E$_{\rm T}$  \\
  & (GeV/c)       & (GeV/c)      & (GeV/c)     & (GeV) & (GeV)        \\
\hline
$\gamma_1$ & -7.1 & 20.2         & -11.1        & 24.1 & 21.4           \\
$\gamma_2$ & -23.6& 9.5          & -10.8        & 27.6 & 25.4           \\
$b$-jet    & 37.6 & 48.6         & -4.1         & 62.2 & 62.0           \\
$j_2$      & 4.1  & -67.5        & 17.5         & 70.4 & 68.1           \\
$j_3$      & -8.8 & -5.6         & -16.6        & 19.6 & 10.4           \\
\mett      & -0.9 & -12.8       &  --           & --    & 12.9 \\
\end{tabular}
\caption[The 4-vectors of the second $\gamma\gamma+b$ candidate event]
{The 4-vectors of the second $\gamma\gamma+b$ candidate event. 
While the \mett\ in the event is 12.9~GeV, the second jet, $j_2$,
is only $7.4^{\circ}$ in $\phi$ away from the \mett.}
\label{Btag2 Event}
\end{table}

\clearpage

\subsection{Additional Photons}

To search for events with additional photons with E$_{\rm T}>25$~GeV, 
events are required to have three photons which pass the selection criteria in 
Table~\ref{Event Cuts}. One photon must have \mbox{E$_{\rm T}>25$~GeV} and pass
the high-threshold requirements, any two other
photon candidates
 in the event must both pass the same selection criteria (low-threshold
or high-threshold) so as to trigger the event.
No events are observed with more than two photons.  The expected rate is 
dominated by 
jets which fake the photon
signature and is estimated using a method similar to that in used in 
Ref.~\cite{Marcus}. 
The average 
rate at which jets fake the photon signature 
is approximately \CPFAKEP/jet and is 
essentially flat as a function of E$_{\rm T}$ 
for \mbox{E$_{\rm T}>25~$GeV.} A total of 
\NADDGAMMAEXP\ events are estimated to be in 
the sample due to $\gamma\gamma$ + fake $\gamma$ for \Etggl; 0.01$\pm$0.01 for
\Etggh. The E$_{\rm T}$ 
spectrum for photon background sources in which jets fake additional photons
is shown in Figure~\ref{PHOTON Back}. 


\subsection{Summary of the Searches}

Table~\ref{found} summarizes the observed and expected numbers of events  with
\mett, N$_{\rm Jets}$, additional leptons, $b$-tags or photons. With the one
possible exception of the $\eeggmett$ candidate, the data appear to be well
predicted by the background expectations~\cite{Diboson Back}.  
The $\eeggmett$ candidate event is
discussed in the next \secorchap.

\onefig{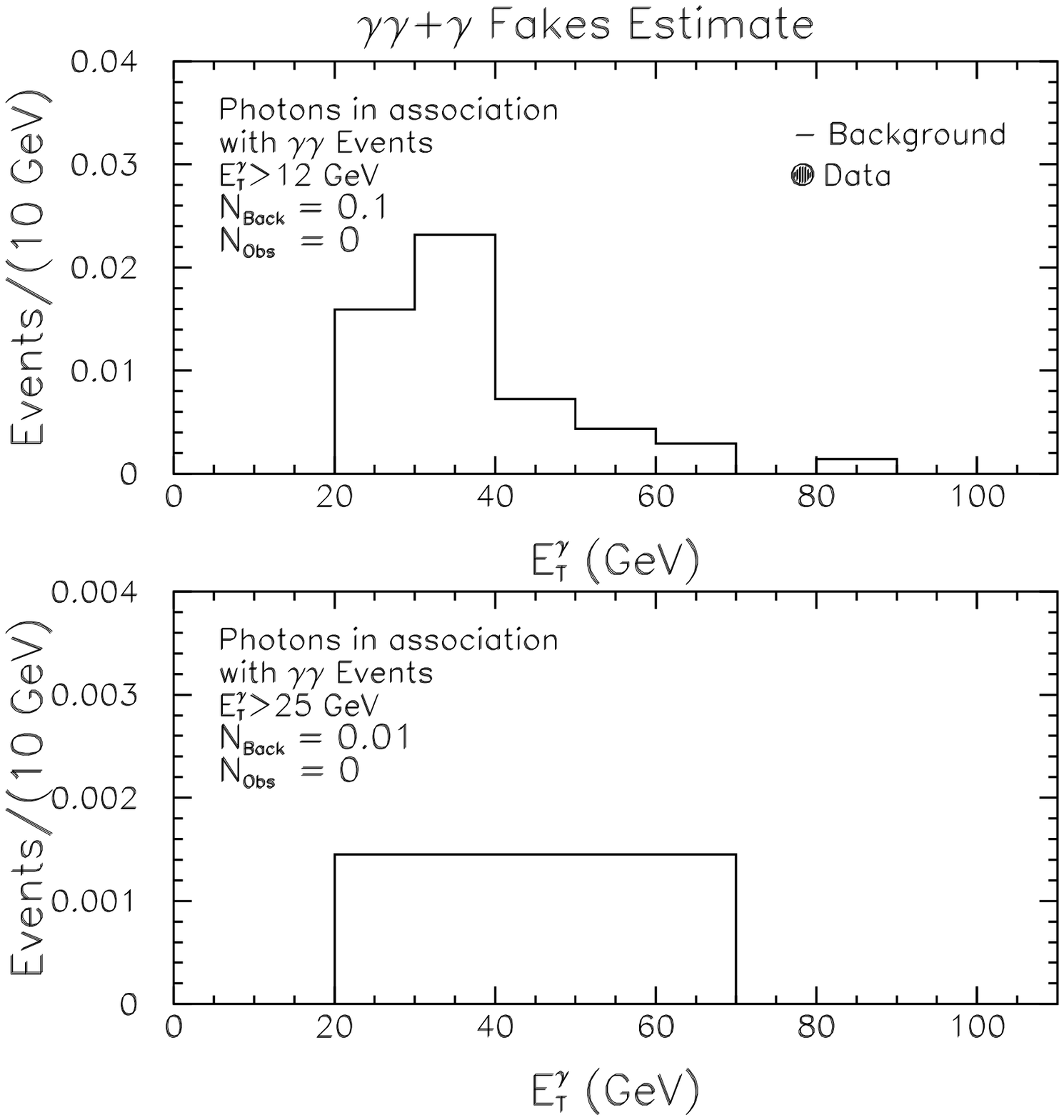}
{The E$_{\rm T}$ spectrum 
of additional photons produced in association
with diphoton pairs.   There are no events in the data with an additional
photon. The upper and lower plots show the expectation
from fake $\gamma$'s for thresholds of 
\Etggl\ and \Etggh\ respectively.}
{PHOTON Back}

\begin{table}
\centering
\begin{tabular}{lccc}
\multicolumn{4}{c}{\Etggl\ Threshold} \\
\hline
Object & Obs. & Exp. & Ref. \\
\hline
\mettgmh, $|\Delta\phi_{\mettsm-{\rm jet}}|>10^\circ$
   & \NMETLOW      & \NMETEXPLOW           & -- \\
N$_{\rm Jet} \ge 4$, ${\rm E}_{\rm T}^{\rm Jet} >10$~GeV, 
$|\eta^{\rm Jet}| < 2.0$     & \NIVJETLOW    & \NIVJETEXPLOW         & -- \\
Central $e$ or $\mu$, ${\rm E}_{\rm T}^{e~{\rm or}~\mu} >25$~GeV
    & \NCENTEORMU   & \NCENTEORMUEXP        &\cite{top}\\
Central $\tau$, ${\rm E}_{\rm T}^{\tau} >25$~GeV
          & \NCENTTAU     & \NTAUEXPLOW           &\cite{Marcus}\\
$b$-tag, ${\rm E}_{\rm T}^{b}>25$~GeV
                  & \NBTAG        & \NBEXPLOW             & \cite{top} \\
Central $\gamma$, ${\rm E}_{\rm T}^{\gamma_3}>25$~GeV
                        & \NADDGAMMA    & \NADDGAMMAEXP         & -- \\
\hline
\hline
\multicolumn{4}{c}{\Etggh\ Threshold} \\
\hline
Object & Obs. & Exp. & Ref. \\
\hline
\mettgl, $|\Delta\phi_{\mettsm-{\rm jet}}|>10^\circ$ 
         & \NMETHIGH     & \NMETEXPHIGH          & -- \\
N$_{\rm Jet} \ge 3$, ${\rm E}_{\rm T}^{\rm Jet} >10$~GeV,
$|\eta^{\rm Jet}| < 2.0$ & \NIIIJETHIGH  & \NIIIJETEXPHIGH       & -- \\ 

Central $e$ or $\mu$, ${\rm E}_{\rm T}^{e~{\rm or}~\mu} >25$~GeV
    & 1   & 0.1 $\pm$ 0.1 &\cite{top}\\
Central $\tau$, ${\rm E}_{\rm T}^{\tau} >25$~GeV
          & 0     & 0.03 $\pm$ 0.03           &\cite{Marcus}\\
$b$-tag, ${\rm E}_{\rm T}^{b}>25$~GeV
                  & 0        & 0.1 $\pm$ 0.1             & \cite{top} \\
Central $\gamma$, ${\rm E}_{\rm T}^{\gamma_3}>25$~GeV
                        & 0    & 0.01 $\pm$ 0.01         & -- \\
\hline
\end{tabular}
\caption[The number of observed and expected 
$\gamma\gamma$ events with additional objects in \LUMTOT~pb$^{-1}$]
{The number of observed and expected 
$\gamma\gamma$ events with additional objects in \LUMTOT~pb$^{-1}$. 
Note that the $\eeggmett$ candidate event appears in multiple
categories.}
\label{found}
\end{table}

\chapterfour

The `$\eeggmett$' candidate event~\cite{Park}, 
shown in Figure~\ref{Pretty EEGG LEGO}, 
consists of 
two high-$\Et$ photons, a central electron, an electromagnetic cluster
in the plug calorimeter with \mbox{\ett = 63~GeV} which passes the electron
selection criteria 
used for $Z^0$ identification~\cite{R}, and the largest  $\mett$ ($\mett =
\EEGGTOTMET$~GeV) in  the diphoton sample.  
While the event is unexpected from the standard model, 
it could also be due to one or more 
detection pathologies~\cite{Pathologies}. 
In addition to a detailed study of 
the event, its properties are compared to 
a control sample of 1009 
well-measured  $\zoee$ events.

\subsection{The Interaction Vertex} 

The primary vertex,
determined using the track from the central electron, 
is situated at \mbox{$z=20.4$~cm}. 
The scalar sum of the transverse momentum of the 
14 tracks associated with the vertex is 40.6~GeV and 
includes 31.8~GeV due to the electron in the central calorimeter.  
Since there is no track associated with a photon and the calorimeter has no
pointing capabilities, the $z$ position of the vertex for the photons cannot be
determined. 
Similarly, since there is no CTC
track for the cluster in the plug calorimeter~\cite{CTC Info}, its vertex cannot
be determined.

There are three other vertices in the event which 
are typical of soft $\ppbar$ collisions~\cite{Other Argument} 
and are described in Table~\ref{Vertices}. The
instantaneous luminosity, ${\cal L}$, during this particular 
part of the run was measured to be
\mbox{${\cal L} = 1.43\times 10^{31}/{\rm cm^2\cdot sec}$};
at this luminosity there should be, on average, 2.5 primary vertices.
There is no indication that the electron candidates, photon candidates
or the missing transverse energy are
due to anything other than the single $\ppbar$ collision which occurred at
\mbox{$z=20.4$~cm}.

\begin{table}[htb]
\centering
\begin{tabular}{r|c} 
\multicolumn{1}{c|}{$z_{\rm vertex}$} & $\Sigma{\rm P_T}$ of tracks \\
                        & associated with the vertex \\ \hline
20.4~cm            & 40.6~GeV \\
-8.9~cm            & 1.3~GeV \\
-38.9~cm           & 5.0~GeV \\
-33.7~cm           & 5.4~GeV \\
\end{tabular}
\caption[The vertices in the $\eeggmett$ candidate event]{The vertices in the
$\eeggmett$ candidate event. $\Sigma{\rm P_T}$ is the scalar 
sum of the transverse
momentum of tracks associated with the vertex. The primary vertex at 20.4~cm 
has \mbox{$\Sigma{\rm P_T}$ = 40.6~GeV} 
which includes the ${\rm P_T}$ of the central
electron.  The other vertices are typical of soft $\ppbar$
collisions~\protect\cite{Other Argument}. 
The last two vertices
are not completely independent as they share tracks with a total of 2.03~GeV of
$\Sigma{\rm P_T}$. }
\label{Vertices}
\end{table}

\subsection{Timing Information}

As described in Section~\ref{Add Event Req}, 
every tower in the central hadronic calorimeter has timing information 
associated with any energy deposited.  
Any tower with energy deposited out of time with the collision might
indicate the presence of a cosmic ray interaction in the event. No tower in the
$\eeggmett$ candidate event has
more than 1~GeV of energy deposited outside the timing window.
Timing information for clusters in the central electromagnetic 
calorimeter can be found
if the shower also deposits energy in the 
hadronic calorimeters.  The 
electron arrival time distribution of Figure~\ref{Timing Figure} and has a
resolution of $\approx 4$~nsec. In contrast, 
cosmic rays~\cite{Cosmic Rays} have an arrival rate which 
is flat in time and extends to
large times (see Figure~\ref{Timing Figure}b).
In the $\eeggmett$ candidate event only the central
electron and one of the photons ($\gamma_1$ in Table~\ref{Kinematics Table})
have associated timing information~\cite{Kuhlmann Note}. 
The arrival times of the clusters 
are measured to be 15~nsec and 18~nsec after the nominal
collision time 
respectively, well within expectations, and consistent with each other. 
There is no indication that any of the energy deposited in the event 
is due to a
cosmic ray interaction in the detector.

\twofig{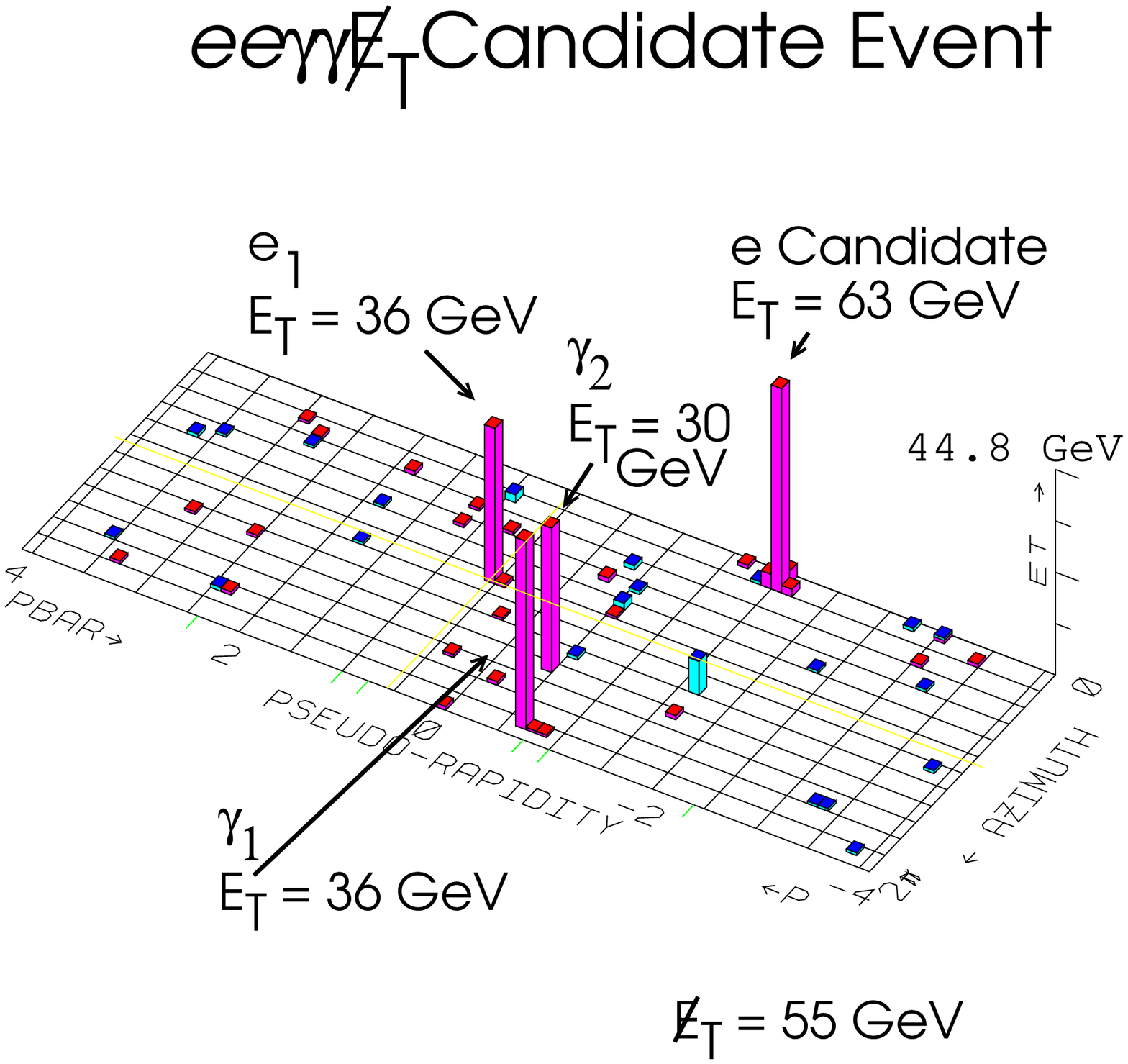}
{An event display for the $\eeggmett$ candidate event.}
{Pretty EEGG LEGO}
{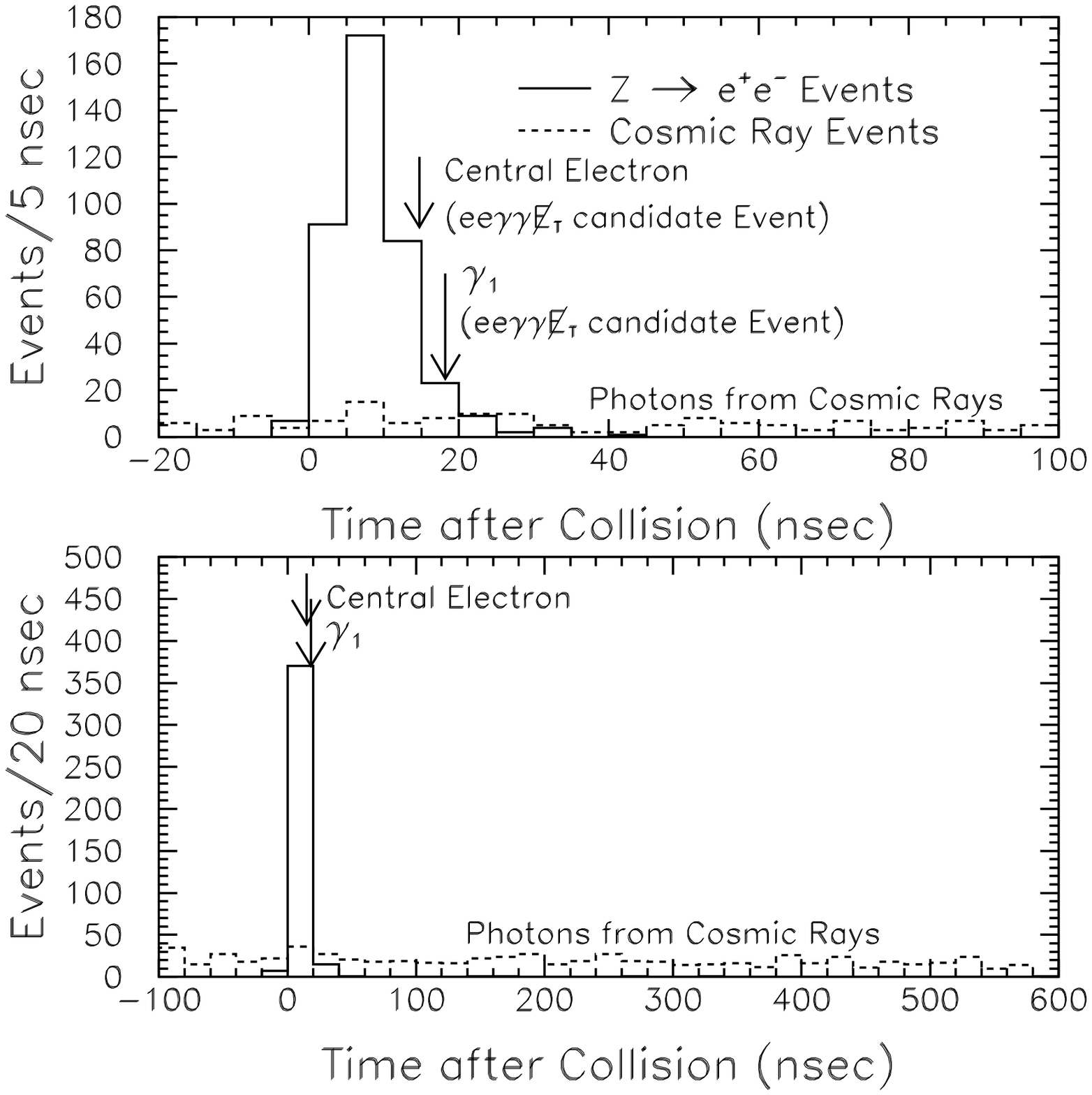}
{The arrival times of electrons and photons
at the central hadronic calorimeter
from $\zoee$ events and from a sample of photons from cosmic rays.
In the $\eeggmett$
candidate event only the central
electron and one of the photons ($\gamma_1$ in 
Table~\protect\ref{Kinematics Table})
have associated timing information and are indicated by the arrows. 
There is no
indication that any of the energy deposited in the event 
is due to a
cosmic ray interaction in the detector.}
{Timing Figure}

\subsection{The Central Electron}

The electron in the central calorimeter passes all the standard 
electron identification and isolation requirements used in
top-quark studies~\cite{top}. The measured values of the identification 
variables as well as the selection criteria are
given in Table~\ref{CEM Electron}. 

\begin{table}[htb!]
\centering
\begin{tabular}{l|l} 
\multicolumn{1}{c|}{Requirement} & \multicolumn{1}{c}{Value } \\ \hline 

 E$_{\rm T} >$ 25~GeV           & E$_{\rm T}$ =    36.4~GeV \\
 E/P $<$ 1.8                    & E/P=     1.15 \\
 Had/EM $<$ 0.05                & Had/EM = 0.026 \\
 Lshr $<$ 0.2                   & Lshr=    -0.007 \\
 $\chi^2_{\rm strip} <$ 10.0    & $\chi^2_{\rm strip}$ = 2.13 \\
$|\Delta x_{\rm track-shower}| <$ 1.5~cm & $\Delta x$ = 0.02~cm \\
$|\Delta z_{\rm track-shower}| <$ 3.0~cm & $\Delta z$ = -0.50~cm \\
 $|\Delta z_{\rm vertex-track}|<$ 5.0~cm 
                                & $\Delta z$ = 1.31~cm \\
 $|z_{\rm vertex}|$ $<$ 60.0~cm & $z_{\rm vertex}$ =    20.4~cm \\
 Fiducial                       &  Yes   \\
 ${\rm E}_{\rm T}^{\rm Iso}/{\rm E}_{\rm T} <$ 0.1 
        & ${\rm E}_{\rm T}^{\rm Iso}/{\rm E}_{\rm T}$= 0.02 \\
\end{tabular}
\caption[The measured values of the variables used to identify the 
central electron in the $\eeggmett$ candidate event]
{The measured values of the variables used to identify the 
central electron in the $\eeggmett$ candidate event.
The selection criteria are those
used to identify electrons in the top-quark analyses.
For a full description of these variables see Refs.\protect\cite{R,top}.}
\label{CEM Electron}
\end{table}

\subsection{The Central Photons}

Both photon candidates in the event 
pass all the selection requirements in Table~\ref{Event Cuts}. 
The values of the variables used in the selection,
as well as the selection criteria, are
shown in Table~\ref{Photons}.  While it is true that the purity of the 
sample is low ((\PURITYLOW)\%), it is not possible to 
determine if these photons are directly produced or are from a $\pi^0
\rightarrow \gamma\gamma$ decay except on a statistical basis.   
The fact that the showers pass the selection criteria (in particular the 
$\chi^2_{\rm CES}$ and $\sigma_{\rm CES}$ requirements) implies
that the showers are consistent with coming from the interaction region.

\begin{table}[htb!]
\centering
\begin{tabular}{l|l} 
\multicolumn{2}{c}{Photon 1} \\ \hline 
\multicolumn{1}{c|}{Requirements} & \multicolumn{1}{c}{Value } \\ \hline 
 E$_{\rm T} >$ 22~GeV           & E$_{\rm T}$ =    36~GeV \\
$\le$1 3D tracks, 
        P$_{\rm T}$ $<$  1~GeV  & \# 3D Tracks = 0      \\
 $\chi^2_{\rm CES} < 10$        & $\chi^2_{\rm CES} = 1.9$        \\
 $|\sigma_{\rm CES}|<$  2.0     & $\sigma_{\rm CES}$ = -0.29 \\
E$^{\rm 2nd~cluster} \leq
    2.39+0.01 \times {\rm E}_{\rm T}^{\rm \gamma}$  & 
     E$_{\rm 2nd~Cluster}$ = 1.4~GeV \\
\hspace*{1.75cm} = 2.92~GeV & \\
Fiducial & Yes \\

 Had$/$EM   $<$  0.055 + 0.00045E & Had/EM  = 0.012       \\
\hspace*{1.6cm} = 0.079 & \\
 ${\rm E}_{\rm T}^{\rm Iso}/{\rm E}_{\rm T}  < 0.10$    & 
    ${\rm E}_{\rm T}^{\rm Iso}/{\rm E}_{\rm T} $=     0.050 \\
 $\Sigma {\rm P}_{\rm T} (\Delta R=0.4) <$  5.0       & 
        $\Sigma {\rm P}_{\rm T} (\Delta R=0.4)$=     0.39 \\
\hline \hline
\multicolumn{2}{c}{Photon 2} \\ \hline 
\multicolumn{1}{c|}{Requirements} & \multicolumn{1}{c}{Value } \\ \hline 
 E$_{\rm T} >$ 22~GeV           & E$_{\rm T}$ =    32~GeV \\
$\le$1 3D tracks,
        P$_{\rm T}$ $<$  1~GeV  & \# 3D Tracks = 0      \\
 $\chi^2_{\rm CES} < 10$        & $\chi^2_{\rm CES} = 3.9$        \\
 $|\sigma_{\rm CES}|<$  2.0     & $\sigma_{\rm CES}$ = -1.6 \\
E$^{\rm 2nd~cluster} \leq
    2.39+0.01 \times {\rm E}_{\rm T}^{\rm \gamma}$ &
E$_{\rm 2nd~Cluster}$ = 1.2~GeV \\
\hspace*{1.75cm} = 2.76~GeV & \\
Fiducial & Yes \\
 Had$/$EM   $<$  0.055 + 0.00045E
                                & Had/EM=     0.012     \\

\hspace*{1.6cm} = 0.072 & \\ 
 ${\rm E}_{\rm T}^{\rm Iso}/{\rm E}_{\rm T}  < 0.10$    & 
           ${\rm E}_{\rm T}^{\rm Iso}/{\rm E}_{\rm T} $ =     0.015 \\
 $\Sigma {\rm P}_{\rm T} (\Delta R=0.4) <$  5.0  & 
             $\Sigma {\rm P}_{\rm T} (\Delta R=0.4)$=     1.7 \\
\end{tabular}
\caption[The measured values of the variables 
used to identify the central photons in the $\eeggmett$
candidate event]{The measured values of the variables used to 
identify the central photons in the $\eeggmett$
candidate event.}
\label{Photons}
\end{table}


\subsection{The Missing Transverse Energy}

The missing transverse energy in the $\eeggmett$ candidate event is measured to
be 55~GeV. 
The scalar sum of the transverse energy deposited in the calorimeters 
is measured to be \mbox{$\Sigma {\rm E}_{\rm T}$ = 268~GeV}.
The majority of the transverse energy ($> 60\%$) 
is deposited in the four clusters in the electromagnetic calorimeters where
the energy resolution is good~\cite{Resolutions}; The 
rest of the energy in the calorimeter is unclustered. 
To use the \mett\ resolution method 
of Section~\ref{Resolution Section},
the $\Sigma {\rm E}_{\rm T}$ is corrected by 
subtracting off all the electromagnetic clusters giving 
\mbox{$\sumetc =$ 100~GeV.}
Using Equation~\ref{Resolution Equation} yields 
$\sigma(\mett) =$ 7~GeV 
for a final result of \mbox{\mett = $\EEGGTOTMET$~GeV}. 
As a check, 
the total \ptt\ of the 4-cluster system (which is well measured) is
$\EEGGPT$~GeV, opposite to the \mett\
and in good agreement with the measured magnitude. This is a further indication
that the imbalance is not caused by spurious energy elsewhere in the detector. 
There is no indication
that the \mett\ is the result of a measurement pathology or due to a cosmic ray
interaction.

\subsection{The Electron Candidate in the Plug Calorimeter}

The cluster in the plug electromagnetic calorimeter
passes all the standard electron identification and isolation 
selection criteria used for $\zoee$ identification~\cite{R}.
In addition it passes all the
requirements used to identify electrons in the region of the plug calorimeter 
used in the 
top-quark discovery~\cite{top,top exception}. 
Table~\ref{Plug EM Cluster} and
Figure~\ref{Plug EM Variables} show a comparison of 
the values of the measured variables for the cluster to 
those of electrons from the \mbox{$\zoee$} control sample. 
The fact that the shower passes the selection criteria (in particular the 
\mbox{$\chi^2_{\rm 3x3}$ = 1.3} and 
\mbox{$\chi^2_{\rm Depth}$ = 0.43} requirements) implies
 that the shower is consistent with being from a single, isolated
 electron emanating from the interaction point.
 However, a closer inspection reveals a possible
discrepancy with the electron hypothesis.

\begin{table}[htb]
\centering
\begin{tabular}{l|l} 
\multicolumn{1}{c|}{Requirement} & \multicolumn{1}{|c}{Value } \\ \hline 
 $\Et >$ 25~GeV                 & $\Et$  = 63~GeV       \\
 $\chi^2_{\rm 3x3} <$ 3.0       & $\chi^2_{\rm 3x3}$ = 1.3  \\
 Had/EM $<$ 0.05                & Had/EM=     0.03      \\
 E$_{\rm T}^{\rm Iso}/\Et  <$ 0.1 
                                & E$_{\rm T}^{\rm Iso}/\Et  =$0.05 \\
 VTX occupancy $>$ 50\%         & VTX occupancy =     1.0      \\
 Fiducial                       & Yes                   \\ 
\hline \hline
\multicolumn{2}{c}{Additional 
Selection Criteria used in the top-quark Analysis} \\ \hline
 $\chi^2_{\rm Depth} <$ 15.0    & $\chi^2_{\rm Depth}$ = 0.43 \\
 3D Track through    & \# Tracks = 0 \\
\hspace*{0.5cm}3 CTC axial superlayers & \\
\end{tabular}
\caption[The measured values of the variables used to compare the cluster 
in the plug
calorimeter in the $\eeggmett$ candidate event to electrons]
{The measured values of the variables used to compare the cluster 
in the plug
calorimeter in the $\eeggmett$ candidate event to electrons. 
The requirements are those used to identify
electrons  
from  $\zoee$ events in the plug calorimeter and are described in 
Ref.~\protect\cite{R}. The additional
selection criteria are those used to identify electrons in the top-quark
analysis~\protect\cite{top}.
}
\label{Plug EM Cluster}
\end{table}

\subsection{A Problem with the Electron Interpretation}

The tracking information along the 
trajectory between the primary vertex at \mbox{$z=20.4$~cm} and the cluster
in the calorimeter indicates that the cluster is not due to an 
electron.
Figure~\ref{Trajectory} shows the expected path of the particle
as it passes through the SVX, VTX and CTC tracking chambers. 
The standard
electron identification selection criteria only use the information from the CTC
and VTX detectors.  However, in 
this particular event there is no expectation of finding
a track in the central tracking chamber because the trajectory only passes
through the innermost layers~\cite{CTC Info}. 

\twofig{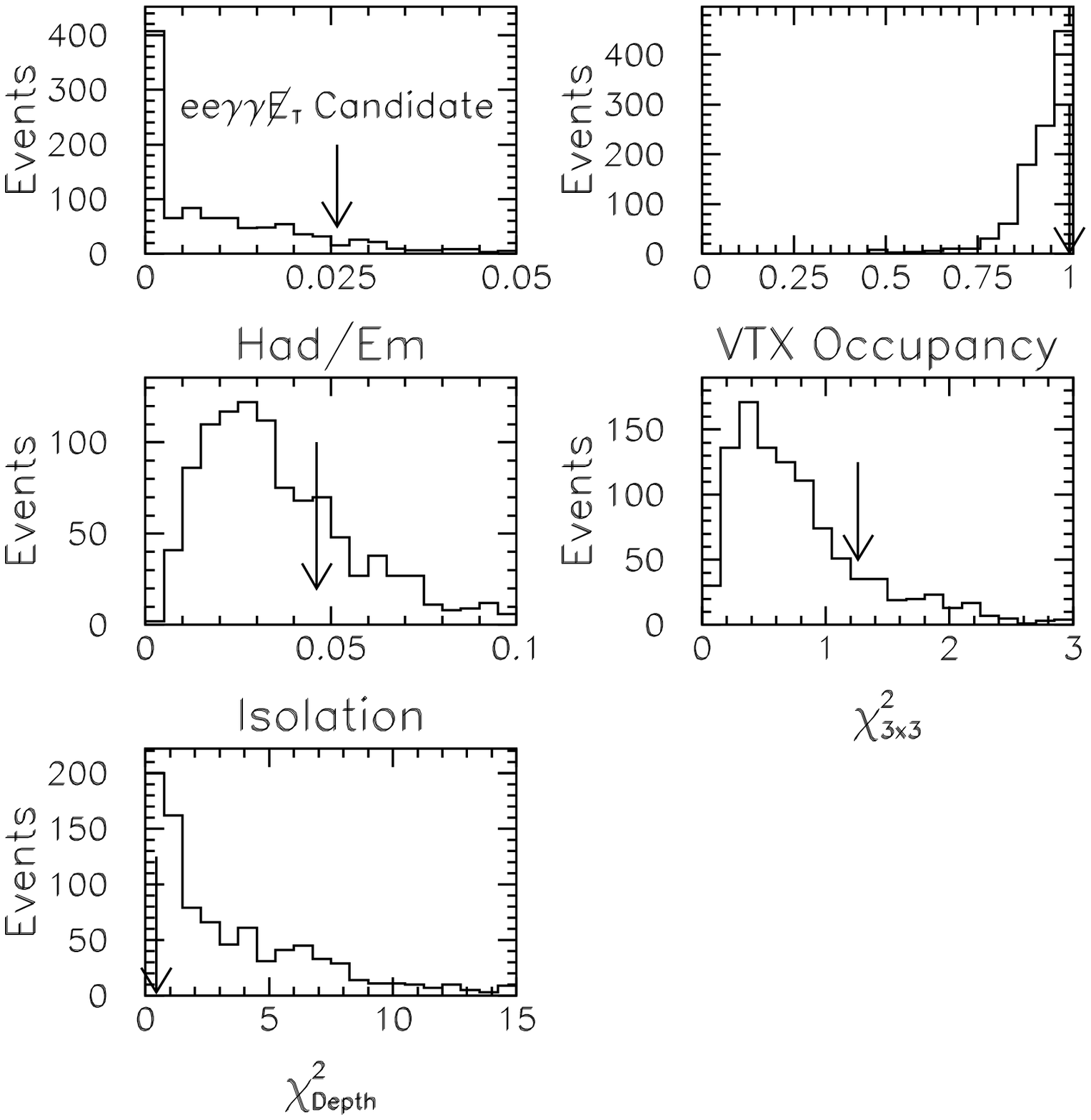}
{The values of the identification variables for electrons in the plug
calorimeter from a sample of $\zoee$ events. 
The arrows represent the measurement for 
the cluster in the $\eeggmett$ candidate event.}
{Plug EM Variables}
{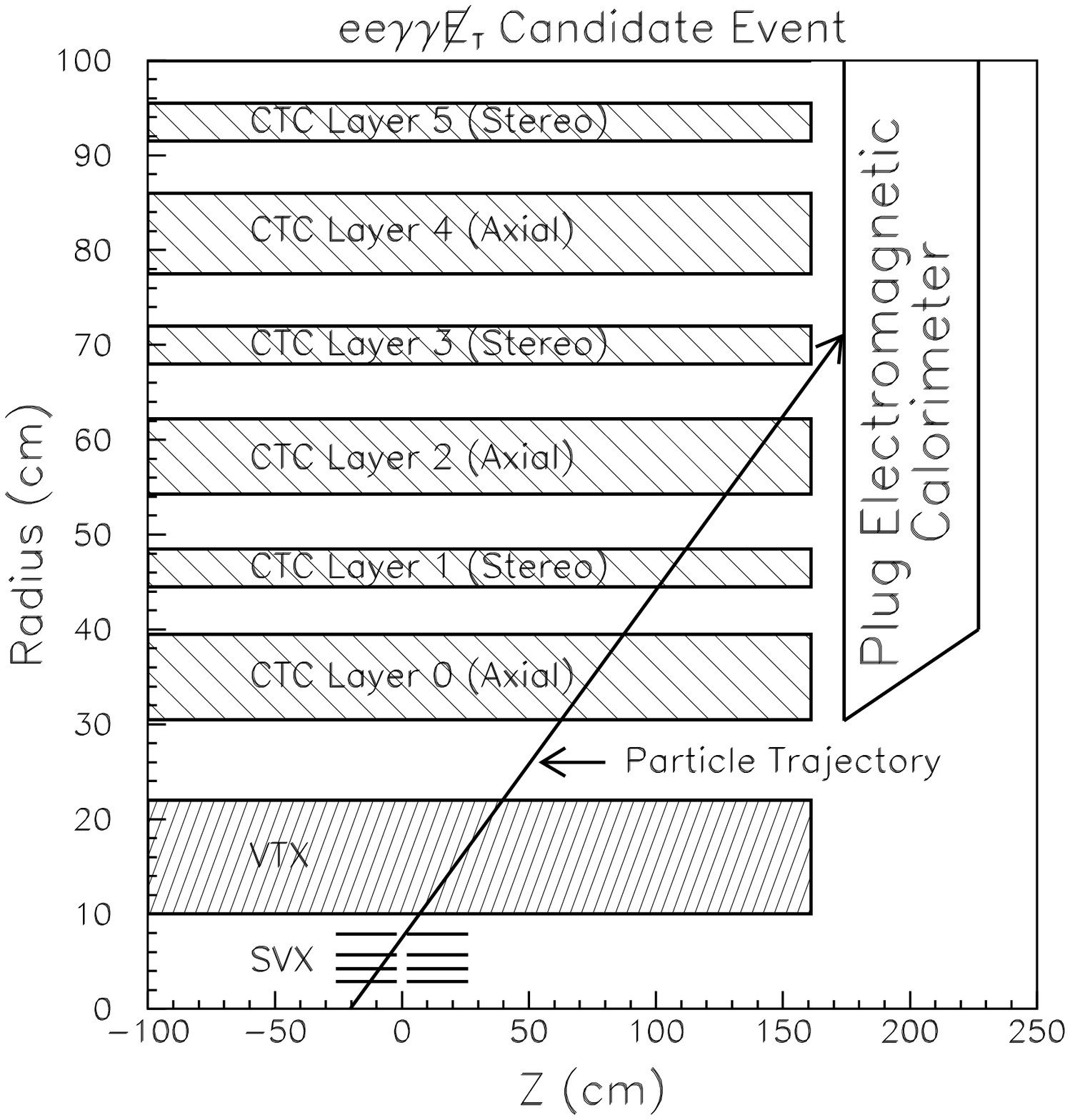}
{The expected trajectory for the cluster 
in the plug calorimeter as it passes through the SVX, VTX and
CTC tracking chambers in the $\eeggmett$ candidate event.}
{Trajectory}

\subsubsection{VTX Tracking}

The VTX is a system of eight octagonal 
time projection modules 
surrounding the beam pipe and mounted end-to-end along the 
beam direction. For every event $r-z$ tracking, 
with some $\phi$ resolution, provides a measurement 
of the vertex position as well as additional tracking information
for individual charged particles. The standard electron identification 
requirements use a VTX occupancy measurement which is defined to be 
the ratio
of the number of layers in the VTX in which the electron deposits charge 
divided by the  number of layers in the VTX expected to be
traversed by the electron, given the electron's trajectory.  
The  VTX does not provide a precision measurement of the trajectory and cannot 
distinguish between single and multiple particles. 
For more information on the VTX and electron identification 
see Refs.~\cite{bluebook,R}. 

The expected particle trajectory, from the vertex at $z=20.4$~cm to the cluster
position at $\phi\approx 0.3$~rad, $\eta\approx -1.7$, 
passes through the fiducial part of the VTX. A total of 
7 hits are recorded with 7 hits expected for an occupancy 
of 100\%.
Figure~\ref{VTXOCC vs. Etaphi} shows the VTX occupancy as a
function of $\eta$ for $\phi=0.3$~rad, and as a 
function of $\phi$ for $\eta$ = -1.72.
There 
appears to be at least one 
charged particle trajectory at the $\eta$ and $\phi$ 
of the cluster. The VTX
information is 
completely consistent with the interpretation of the cluster as an electron.

\subsubsection{SVX Tracking}

The standard electron identification selection criteria do not use the SVX
because the detector covers only the region
\mbox{$|z|<30$~cm.}
However, for interactions which occur within $|z|<30$~cm, the SVX can often 
be used to provide precision tracking for electrons~\cite{Wasymm}. 
For more details on the SVX as 
well as the `stub'-finding algorithms see Appendix A.

Figure~\ref{Final Deltaphi} shows the $\Delta\phi$ between
the measured $\phi$ position in the strip chambers 
(CES and PES respectively) and from 
SVX stubs found for electrons 
from the $\zoee$ control sample. 
The two peaks correspond to the bending of positively and negatively charged
electrons in the magnetic field.

\twofig{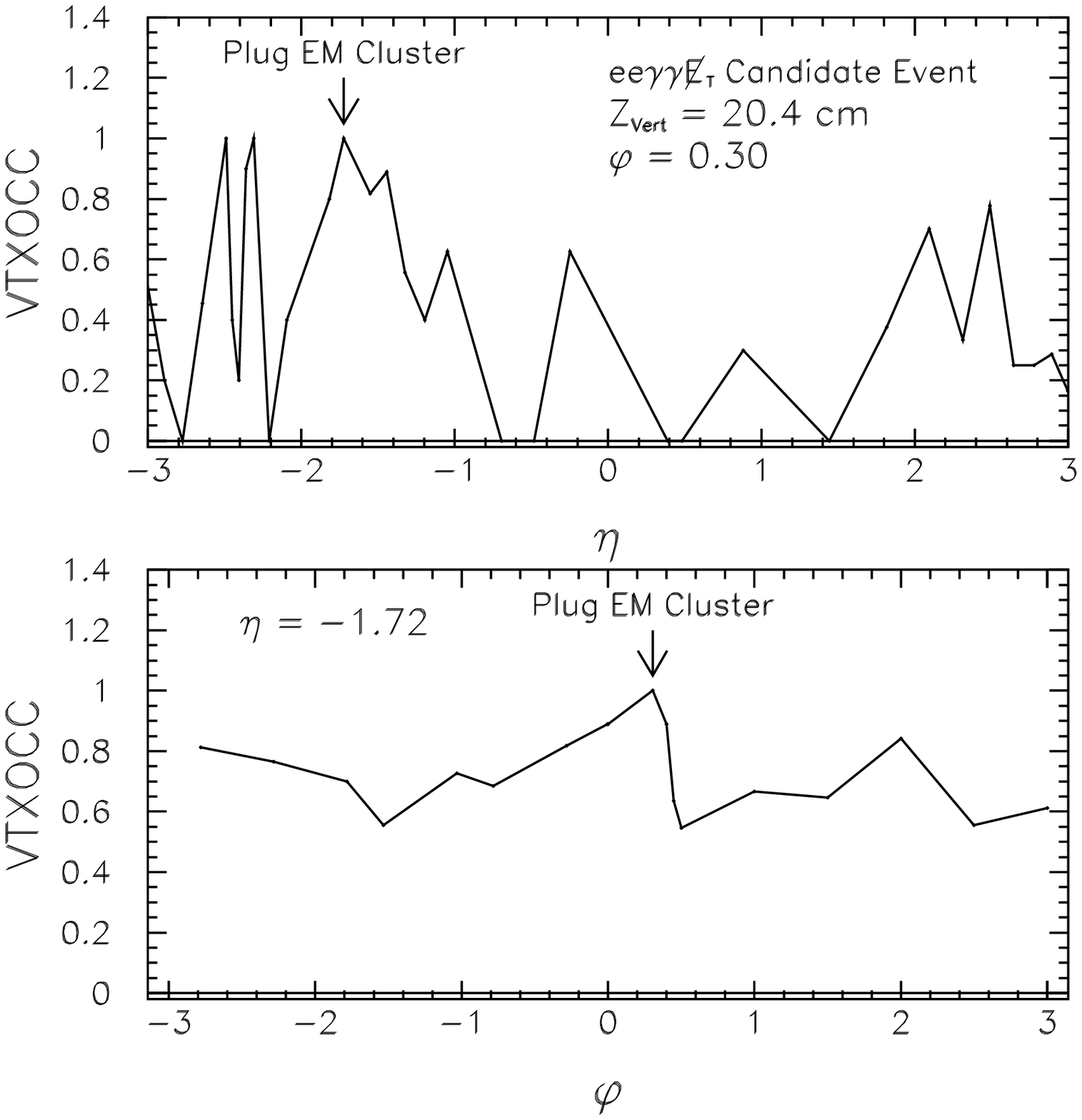}
{The VTX occupancy as measured in the $\eeggmett$ candidate event. The
trajectory is assumed to come from the vertex at \mbox{$z$=20.4~cm.}}
{VTXOCC vs. Etaphi}
{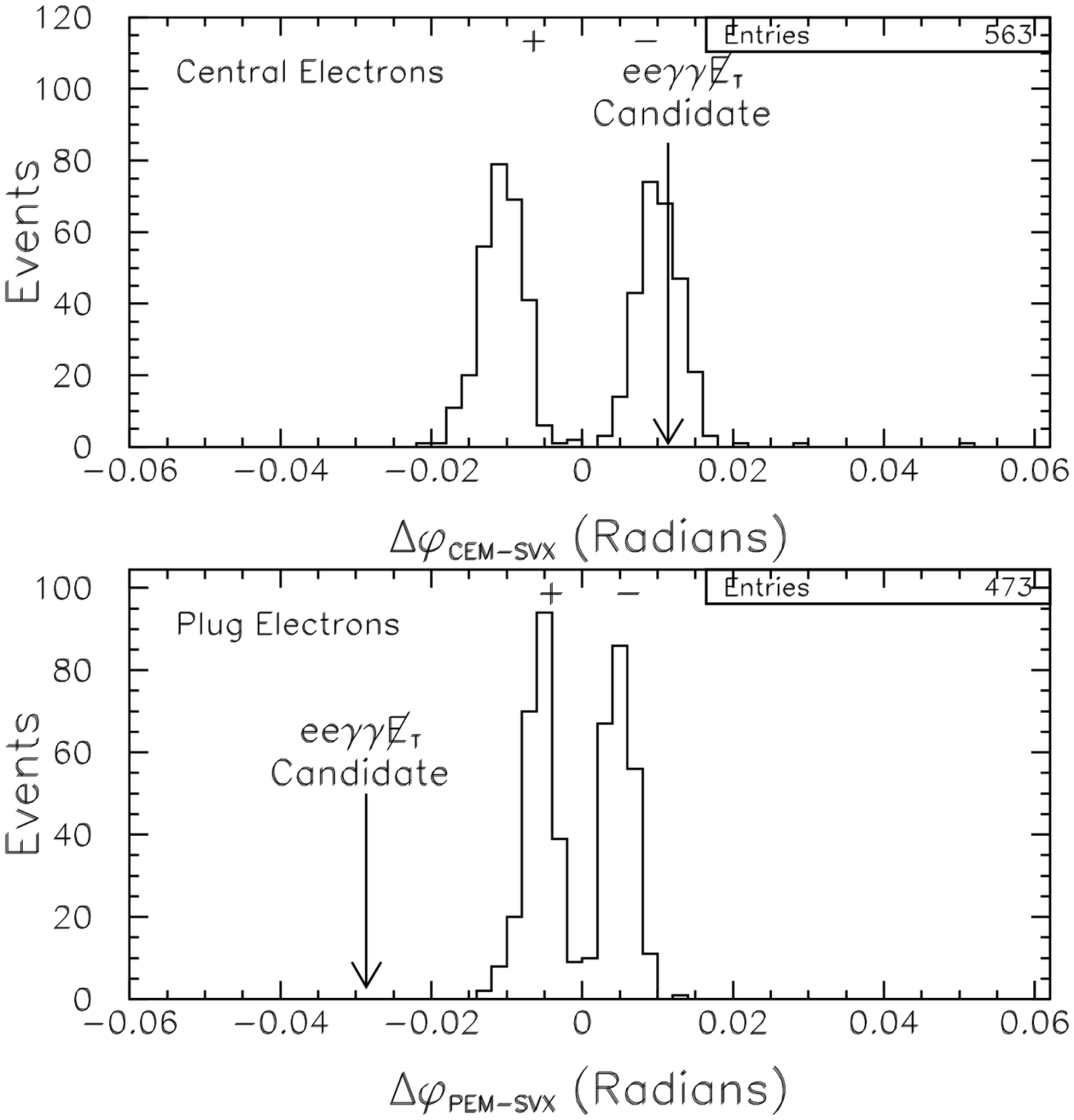}
{The $\Delta\phi$ between the measured electron position from the strip chambers
(CES and PES) 
and the $\phi$ from the SVX tracker for electrons in the $\zoee$
control sample. The 
two peaks  correspond to the bending of positively and negatively charged
electrons in the magnetic field.
}
{Final Deltaphi}

In the $\eeggmett$ candidate event, only the central electron and the 
electromagnetic cluster in the plug have stubs.
For the
electron in the central calorimeter, the 
$\Delta\phi$ measurement 
is consistent with the negative charge determination from the track in the
central tracking chamber as
shown in Figure~\ref{Final Deltaphi}. 
The stub for the cluster in the plug calorimeter is inconsistent
with the interpretation of the cluster as an electron. 
The expected $\Delta\phi$ between the SVX stub and the 
measured position in the PES, due
to bending in the magnetic field, 
is expected to be -2.6~mrad for a 63~GeV positron. 
The position, in $\phi$, of the electromagnetic cluster 
as measured by the strip chambers in the plug
calorimeter (PES) is 
\mbox{$\phi_{\rm PES} = 0.294$~rad}, but  
there are no SVX clusters in the region 
\mbox{0.29~rad $< \phi< $ 0.30~rad} 
in either SVX barrel as seen in Figure~\ref{SVX Hits}.
However, the algorithm does pick up a three-cluster stub near the expected path
which is in the barrel with $z<0$ (as would be expected for the
trajectory). The
stub appears to be well measured, but has $\phi_{\rm SVX} = 0.265$~rad 
($\Delta\phi = -29$~mrad); 
again see Figure~\ref{Final Deltaphi}. 

The non-observation of an SVX stub with the correct $\Delta\phi$ is unusual
for an electron~\cite{Large DPHI}.
There is no indication of a $\phi$ mismeasurement in
the calorimeter~\cite{Other Problem}. In the SVX
there is a bad strip in the innermost layer (layer~0) at 
$\phi=0.296\pm 0.002$ which may be along the trajectory.  While this could 
cause the loss of a cluster, an 
electron typically deposits energy in multiple strips.
The trajectory
passes near a gap in layer 1 between
silicon crystals at \mbox{$z$ = 9.6~cm}. 
These could possibly account for two of the
three missing clusters. The SVX cluster-finding efficiency is $\approx 95\%$ due
almost entirely to dead strips and gaps between crystals. With that
efficiency, the average probability
to miss all three clusters is \scinotn{1.4}{-4}; however if the true trajectory
passes through  the two bad regions, the probability of losing the
third cluster is only 
less than about 1\%~\cite{Stuart Private}. 

Prompt electrons should have an impact parameter, in the $x-y$ direction, 
with respect to
the position of the collision consistent
with zero. The distribution in the 
impact parameter of stubs associated with
central and plug electrons from $\zoee$ events 
is shown in Figure~\ref{Impact Parameter} along with
the results for the electron candidates in the 
$\eeggmett$ candidate event. While the stub associated with
the central
electron has a small impact parameter \mbox{(46 $\pm 45)~\mu$m,} the 
stub at \mbox{$\phi=0.265$~rad} has a large value,
\mbox{90$\pm$45 $\mu$m,} 
which is on the tail of the distribution but not inconsistent
with the prompt hypothesis.  However, the impact parameter measurement is
dependent on the $\phi$ position and energy 
information from the calorimeter. 
If the stub is unrelated to the cluster, removing the constraints
of calorimeter information 
from the tracking algorithm changes the impact parameter measurement to
be $D_0 = 233\pm180\mu$m. 

\twofig{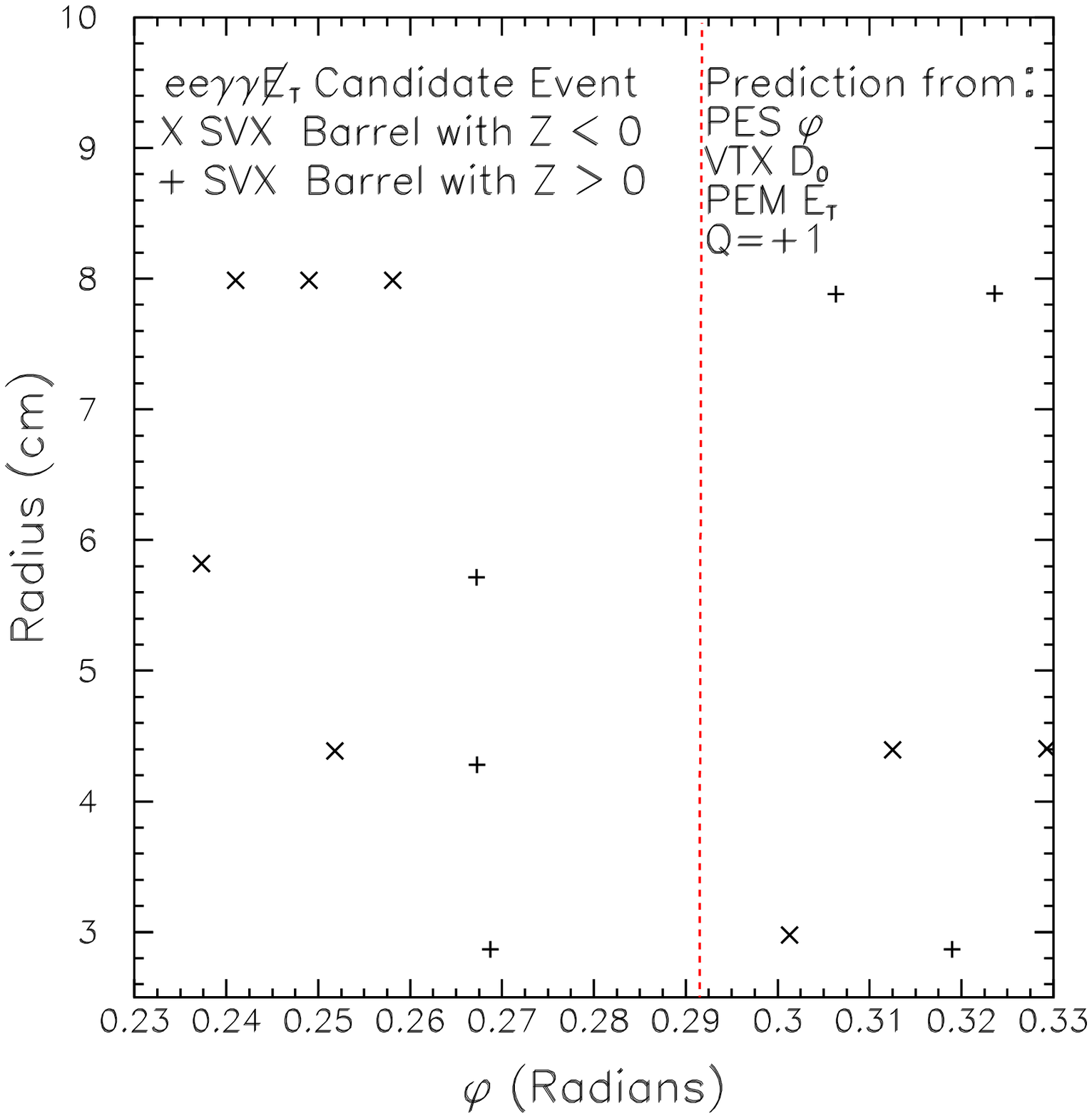}
{The positions of the
SVX clusters with 0.23~rad $<\phi< 0.33$~rad. The dashed line is the
expected trajectory from the primary vertex to the 
cluster in the plug calorimeter using the 
measured cluster position and E$_{\rm T}$.}
{SVX Hits}
{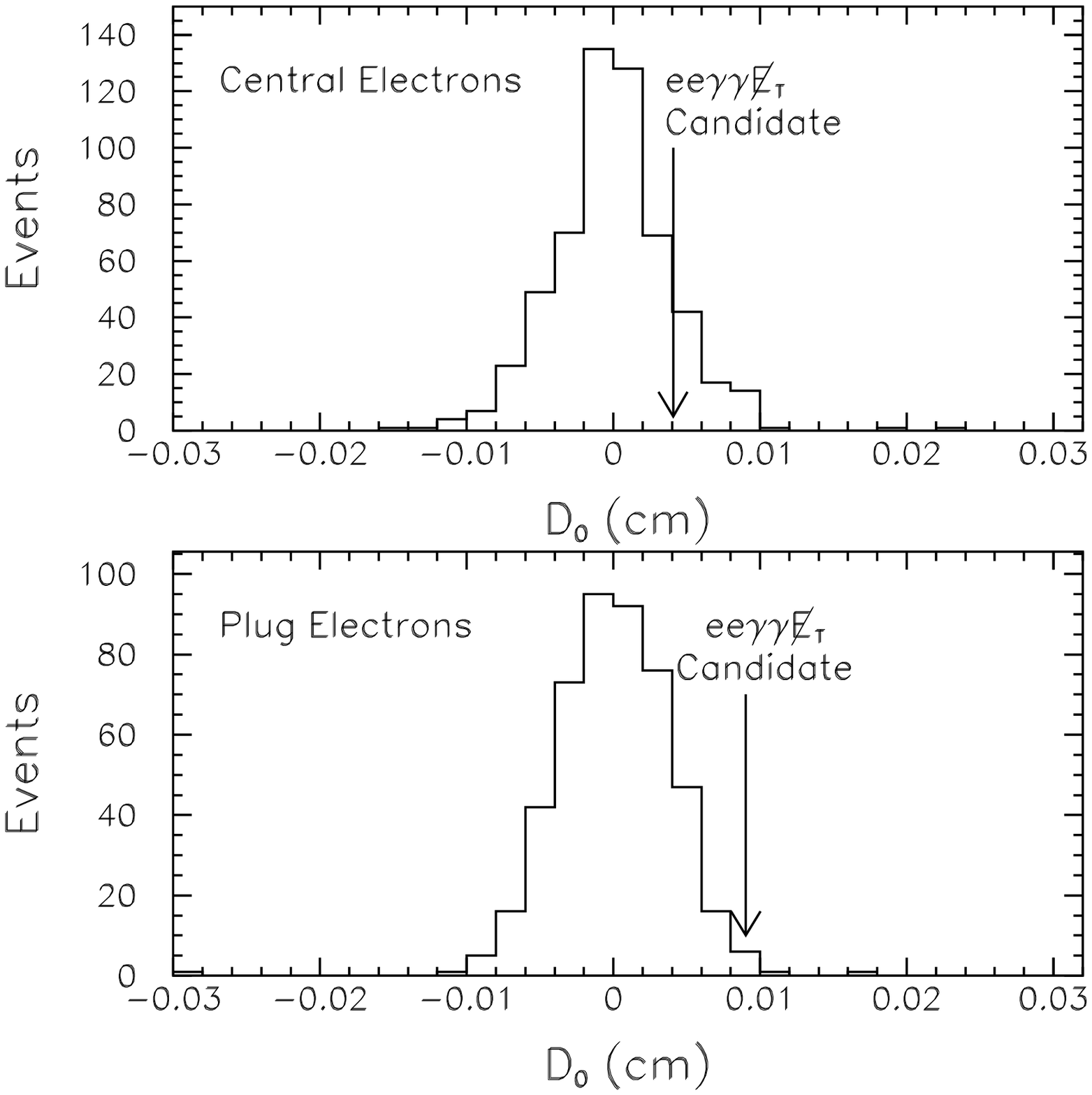}
{The measured impact parameter from the SVX tracker for
electrons from $\zoee$ events. The 
central electron in the $\eeggmett$ candidate event has an
impact parameter of 46$\pm 45\mu$m.
The nearest SVX stub to the $\phi$ of the cluster in the plug electromagnetic
calorimeter has an impact parameter of 90$\pm 45\mu$m. Both are 
consistent with zero within resolution.}
{Impact Parameter}

The tracking information is confusing and 
would be highly unusual for an electron (no
others like it are found in the sample of 1009 
well-measured plug electrons).
Since there is no associated track in the central tracking chamber, 
it is not obvious that the 
stub has anything to do with the cluster in the calorimeter.  
Since there are no 
other large energy clusters in the $\eta-\phi$ region suggested by the stub, 
either this is an SVX or PES failure, or the stub is 
due to a low P$_{\rm T}$ charged particle which is not 
distinguishable in the calorimeter.
Based on the $|\Delta\phi|$ distribution of the
$\zoee$ events, 
the probability that this observation is due to an electron 
is estimated to be less than 0.3\% at 95\%~C.L. 

\subsection{Interpreting the Electromagnetic Cluster}

To summarize, the relevant experimental facts about the electromagnetic
cluster in the plug calorimeter in the $\eeggmett$ candidate event are:

\begin{enumerate}

\item The cluster easily passes all the standard electron identification
selection criteria.

\item There are no SVX clusters in the region 
0.29~rad $< \phi< $ 0.30~rad, with 3 expected.  There is a bad
SVX strip in layer 0 and a gap in the coverage in layer 1 which may lie along
the trajectory and cause clusters to be lost.
There is an SVX stub which is near the expected trajectory,
but is not necessarily correlated with the plug cluster. 
The stub is well measured and appears to be due to 
a charged particle 
traveling at an $\eta$ consistent with the cluster.
However, the  probability for an electron to 
have $|\Delta\phi|>0.03$~rad between the stub and the 
cluster is estimated to be less than 0.3\% at 95\%~C.L. 

\item Assuming the energy and position of the plug cluster are due to the
particle which made this stub, the best-fit 
impact parameter of the SVX stub is 90$\pm 45\mu$m. While 
this is not inconsistent with the prompt electron hypothesis 
(a 2$\sigma$ deviation), the result is highly dependent on the calorimeter
information.
If the calorimeter information is removed from the SVX track finding algorithm,
the best-fit impact parameter becomes $233\pm180\mu$m. 

\item The VTX occupancy indicates that there is at least one 
charged particle traveling 
in the direction of the PEM cluster (this could be the track associated with the
SVX stub).

\end{enumerate}

\noindent Possible sources of the cluster are discussed in the next Sections.

\subsubsection{Two Interactions}

One possibility is that the cluster or the SVX stub comes from a different
$\ppbar$ interaction. As a check, the VTX and
SVX results are investigated using the other vertices in
the event; the results are summarized 
in Table~\ref{SVX and VTX Other Vertex}. The trajectory from a vertex to the
plug passes through the SVX fiducial region for only one other vertex; there
are no stubs associated with it.  The VTX occupancy along the 
trajectories from the other vertices do not indicate a better choice. 
There is no indication that the stub
or cluster is from a different vertex.

\begin{table}[htb]
\centering
\begin{tabular}{c|c|c|c} 
Vertex  & SVX Clusters  & $D_0$                 & VTX Occ       \\
(cm)    & (Exp/Obs)     & ($\mu$m)              & (Exp/Obs)     \\
\hline
20.4    & 3/3           & 90$\pm$45             & 7/7           \\
-8.9    & 3/0           & -                     & 9/3           \\
-38.9   & 0/-           & -                     & 10/7          \\
-33.7   & 0/-           & -                     & 10/4          \\
\end{tabular}
\caption[The VTX and SVX results assuming the plug cluster came from a different
vertex in the event]{The number of expected versus observed hits in 
VTX and SVX detectors 
assuming the cluster in the plug comes from a different
vertex in the event.  
Due to the cluster position in the plug calorimeter, 
only vertices with 
\mbox{-13~cm $< z_{\rm vert} <$ 38~cm} could give three or more
clusters in the SVX. $D_0$ is the impact parameter.
}
\label{SVX and VTX Other Vertex}
\end{table}

\subsubsection{Anomalous Electron Detection}

For example, 
an electron could emit an energetic photon, via 
bremsstrahlung,
while traversing the detector, or the electron could have 
had an elastic 
scattering with a nucleus. 
If the photon emission or collision occurs after the electron leaves the SVX, 
then there should be at least two
final state particles 
and the SVX stub should, by conservation of momentum,  
point to the energy-weighted mean of the energy deposition in the calorimeter
(within the expected resolution and 
bending due to the magnetic field).  No evidence for a 
second cluster is seen in the
calorimeter. If the photon emission occurs
before the electron reaches the SVX, then the initial 
direction of the electron must have been directly toward the
center of the electromagnetic cluster 
(again by conservation of momentum). In this case the SVX
stub is due to the electron going off with low momentum, and 
the electromagnetic cluster is due to the photon. However, 
the impact parameter would be roughly 5 times that 
observed.
A final scenario is that the photon emission or 
collision occurs in the SVX. If this were the case, there 
should be a kink in the trajectory defined by 
the SVX hits and the primary vertex. No such deviation is seen.

\subsubsection{Photonic Interpretation}

The cluster could be a photon with a nearby, but unrelated, charged particle.
Figure~\ref{VTXOCC vs. Etaphi} shows that while the 
occupancy in the VTX has a local maximum at $\phi=0.3$~rad, 
it is above 0.5 for all values of $\phi$.
Thus, even if the SVX stub is
due to an unrelated, low-momentum charged particle that
causes the local maximum in the VTX,
the cluster would, by a side band estimate, still not pass any reasonable
photon VTX occupancy 
requirement.
To estimate the probability that the SVX tracker might find a stub unrelated to
the cluster in the plug, 
the SVX stub-finding algorithm is used. Instead of using the $\phi$
position of the cluster, the $\phi$ is varied between 0 and 2$\pi$ in
increments of 0.01~rad.
A total of 8.4\% of $\phi$ space has a good stub of which
1.8\%  is due to the stub at $\phi$ = 0.265~rad. It is thus not improbable
to find an unrelated stub or high VTX occupancy in this event.

Another way to estimate the probability 
for a photon to have an SVX stub and high
VTX occupancy is to use the central photon sample from \secorchap~2 but with 
the additional requirement of E$_{\rm T}^{\gamma} >$ 20~GeV. 
There are 
268 events in the sample for a total of 536 photons.
Figure~\ref{Photon VTX and SVX} shows the VTX and 
the  $\Delta\phi$ distributions for the photons in the sample. 
A total of 277 photons pass through the
fiducial region of the SVX,
16 have an SVX stub
and 6 have \mbox{$|\Delta\phi|>$ 0.03~rad} ($\approx 2\%$). A total of 
58 of the 536 photons have a VTX occupancy greater than 50\%
($\approx 10\%$). The bottom part of
Figure~\ref{Photon VTX and SVX} shows that most of the 
photons with a stub have low VTX
occupancy. No event with $|\Delta\phi|>$0.03~rad
has a VTX occupancy greater than
0.5 indicating that there is 
no correlation (at low statistics) between large VTX occupancy and large
$|\Delta\phi|$~\cite{Slightly Higher}. 
While the cluster could be due to a photon with a soft track nearby,
it is an unusual example as estimated by the diphoton sample, and the 
hypothesis cannot be proved or excluded. 

\subsubsection{Hadronic $\tau$ decay}

The cluster could be due to the hadronic (1-prong) decay of a 
$\tau$ lepton.
For example, the decay
\mbox{$\tau \rightarrow \pi^+\pi^0\nu_{\tau}$} produces a $\pi^+$ which 
could generate the SVX stub and VTX occupancy, and
a $\pi^0$ which decays via \mbox{$\pi^0 \rightarrow
\gamma\gamma$} and could generate a calorimeter cluster that is 
largely electromagnetic energy and that
passes the remaining electron identification selection criteria.
However, most 
hadronic $\tau$ decays will not shower predominantly in
the electromagnetic calorimeter. The cluster in the event  
deposits roughly 180~GeV in the electromagnetic calorimeter
and only 5~GeV in the hadronic calorimeter.  
The probability that a $\tau$ might fake the electron signature, but
not be from the $\tau\rightarrow e\nu\nu\nu$ decay chain, is estimated using 
a Monte Carlo to simulate a sample of $\tau$'s with a 
one-prong decay (excluding the
electron and muon decays) interacting with the
plug calorimeter~\cite{TAUOLA}.
Unfortunately, the 
detector simulation does not correctly model the VTX
occupancy or the 
$\chi^2$ variables in the calorimeter or strip chambers 
in the plug, each of which provide rejection against the hadronic decays. 
To avoid an underestimate of the probability, the simulated 
cluster is not required 
to pass these requirements. The top part of
Figure~\ref{Tau Fake Plots} shows the
electromagnetic fraction of the energy of clusters 
produced by the $\tau$. 
The rate at which $\tau$ events pass the Had/EM and
calorimeter isolation selection criteria
(corrected for the 50\% one-prong
branching fraction~\cite{PDG}) 
is plotted in the bottom of Figure~\ref{Tau Fake Plots}
as a function of the ratio of reconstructed cluster energy to the
original $\tau$ energy. For most of phase space the fake rate is flat, typically
around 3\%; however at
the end points it rises to almost 10\%.

\twofig{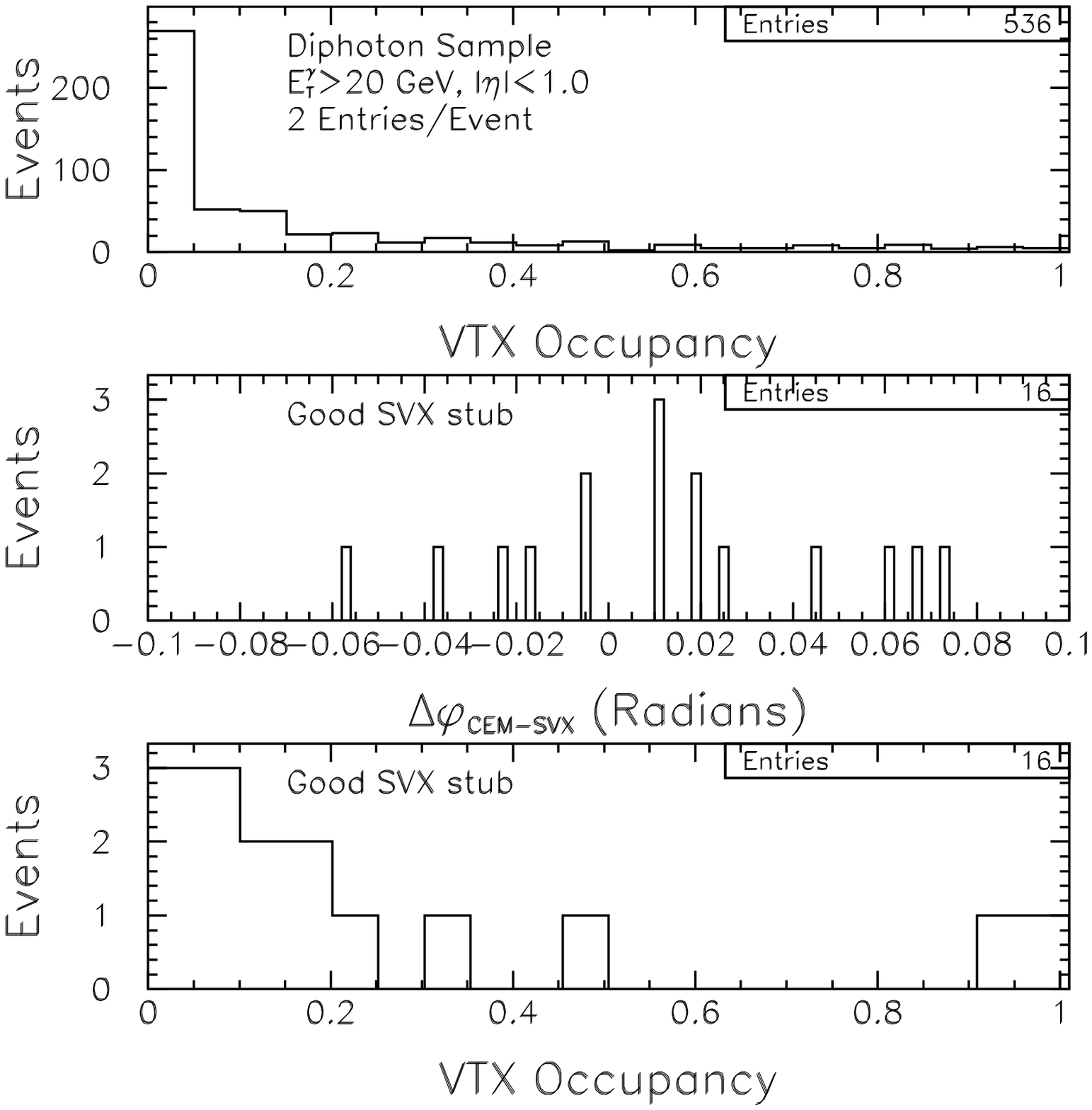}
{The top plot shows 
VTX occupancy (number of hits/number of hits expected) 
for a sample of central photons.
A total of 58 of the 536 photons have a VTX occupancy of greater
than 50\%. The middle plot shows the 
$\Delta\phi$ between the measured photon position and the 
stub found by the SVX tracker for photons with a stub. The 
lower plot shows the VTX occupancy for photons which have an associated 
stub found by the SVX tracker.} 
{Photon VTX and SVX}
{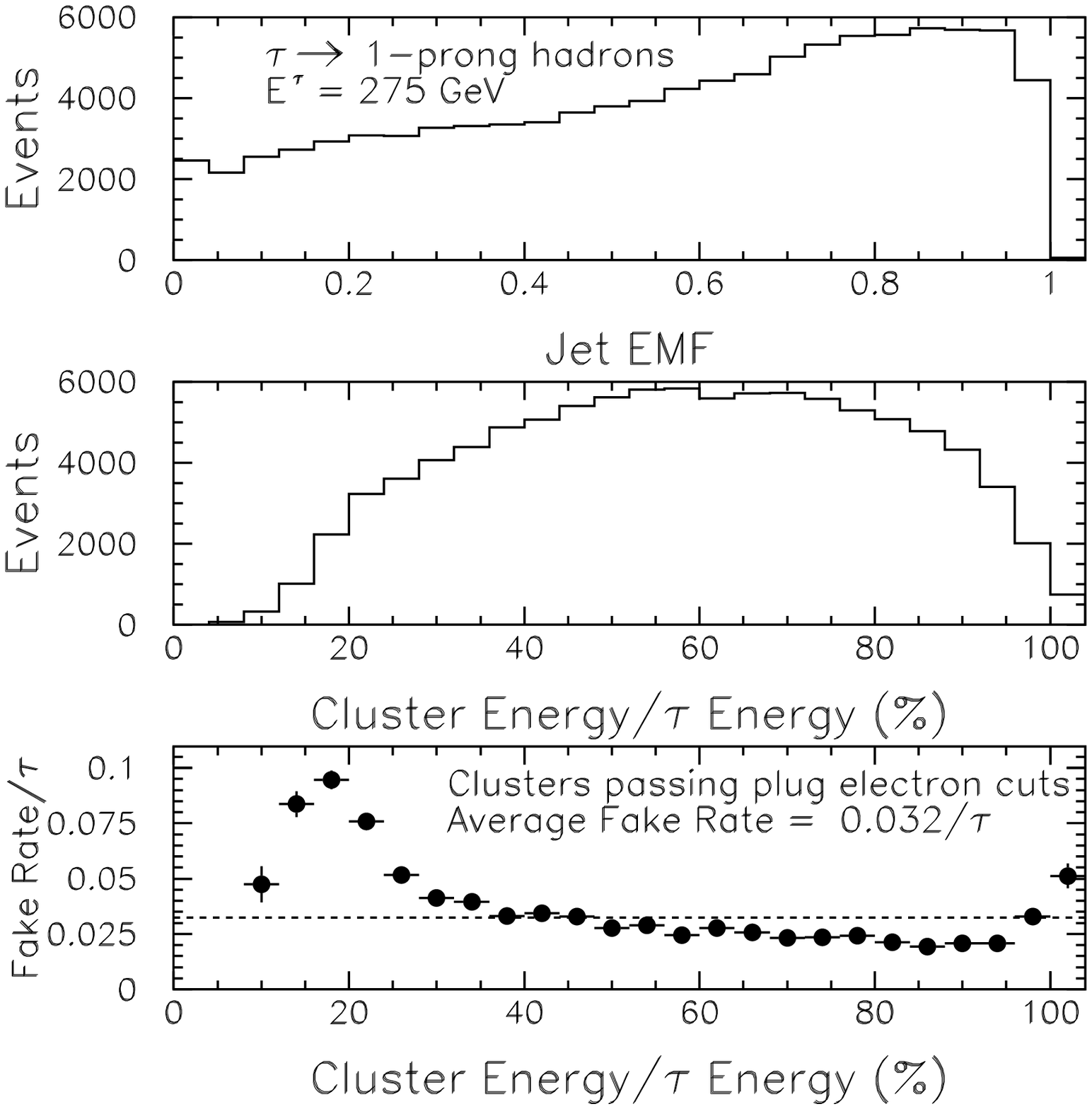}
{A Monte Carlo simulation of one-prong hadronic decays of a
$\tau$ interacting with the plug calorimeter.
The electromagnetic fraction (EMF) of the energy 
deposited in the electromagnetic and hadronic calorimeters 
is shown in the top plot. The measured cluster energy as a fraction of the true
energy of the $\tau$ is shown in the middle plot.  The rate 
at which $\tau$ events pass the Had/EM and isolation selection criteria
(corrected for the 50\% one-prong
branching fraction) as a function of ratio of reconstructed cluster energy to
original $\tau$ energy is shown in the bottom plot.}
{Tau Fake Plots}

While the cluster in the PEM could be due to the  hadronic decay of a $\tau$,
and there is no evidence to the contrary, this would be an unusual example
as estimated by the Monte Carlo. Furthermore, it would in general
increase the \mett\ since only 20\%-100\% of true $\tau$ energy is deposited in
the calorimeter. 
Ignoring additional tracking information from the SVX and VTX, as well as
potential rejection power from the calorimeter, the
probability of a $\tau$ to pass
the electron selection criteria is conservatively estimated to be less 
than a few percent.


\subsubsection{Jet Interpretation}

A jet associated with the event, either as part of the partonic process
or from initial or final state radiation, could fluctuate to 
pass all the electron selection criteria.  The 
rate at which a jet passes the electron selection criteria
is estimated using a method similar to that 
in Ref.~\cite{Marcus}
The fake-rate per jet 
is consistent with being a constant for jets with \mbox{E$_{\rm T} > 25$~GeV}
with a probability of approximately \PEFAKEP/jet.  
Thus, although 
the cluster could be due to a jet which fluctuated to pass the electron
selection criteria it would be an unusual example. 

\subsubsection{Conclusion}
Although the cluster passes all of the standard electron selection
criteria, the tracking information provides 
evidence that the cluster is not due to an
electron.  
The cluster could be interpreted as a photon, 
the hadronic decay of a tau lepton, or
simply as a jet. While all three scenarios are reasonable 
\mbox{{\it a priori}, } and are
consistent with the facts, each is 
unlikely in that this would be an unusual example of any single one of 
them~\cite{Only SM Options}.  
There simply is not enough  information 
to establish the origin of the cluster.  

\subsection{A Study of the Kinematics of the Event}\label{kinematics}

A study of the kinematics of the $\eeggmett$ candidate event is potentially 
useful in
helping understanding its origin.
The energies and momenta are given
in Table~\ref{Kinematics Table}.  There are no additional jets with
E$_{\rm T}^{\rm Corrected}>10$~GeV. \metstatement\
Figure~\ref{Mass ee vs. Met} shows the \mett\ of the system 
if the energy of the cluster in the plug calorimeter were mismeasured. 
For simplicity, the cluster is denoted as $e$, and 
in the plot, M$_{\rm ee}$ is plotted
versus $\mett$ for different correction factors, $C$,  such that E$_{\rm used} =
C \times {\rm E_{measured}}$. The $\mett$ cannot be reduced 
below 25~GeV for any value of $C$. While the value of the $\mett$ is at a local
minimum for \mbox{${\rm M}_{\rm ee} \approx {\rm M}_{\rm Z^0}$,}
where \mbox{$\mettmin = 26.6$~GeV,} it would mean that 
a particle (or particles) with 51~GeV ($\sigma=2$~GeV)
of electromagnetic energy was mismeasured as having 183~GeV of energy. 

\begin{sloppypar}
Table~\ref{Mett Table} lists the masses and transverse momentum 
for various combinations of the
clusters in the event. One of the most interesting combinations is the $e_{\rm
central}\gamma_1$ combination  which has an invariant mass of 91.7~GeV/c$^2$
and a ${\rm P_T}$ of 4.1~GeV.
While this could be a
$Z^0$ where one electron faked the photon signature (there 
is no track or SVX stub pointing at the photon) this would be unusual, as
estimated in Section~\ref{LGG Section}. 
Table \ref{Transverse Masses Table} lists the calculated transverse masses for
various combination of the clusters and the \mett. While 
the $e_{\rm central}\mett$ combination is inconsistent with the decay
of a $W$ via $W\rightarrow e\nu$ (it has 
\mbox{M$_{\rm T} = 4.3$~GeV/c$^2$)} 
the $e_{\rm central}\gamma_2\mett$ combination 
could be the radiative decay of a $W$ via $W\rightarrow e\nu\gamma$ 
\mbox{(M$_{\rm T} = 70.4$~GeV/c$^2$).}  
\end{sloppypar}

\subsection{Conclusions}

\onefig{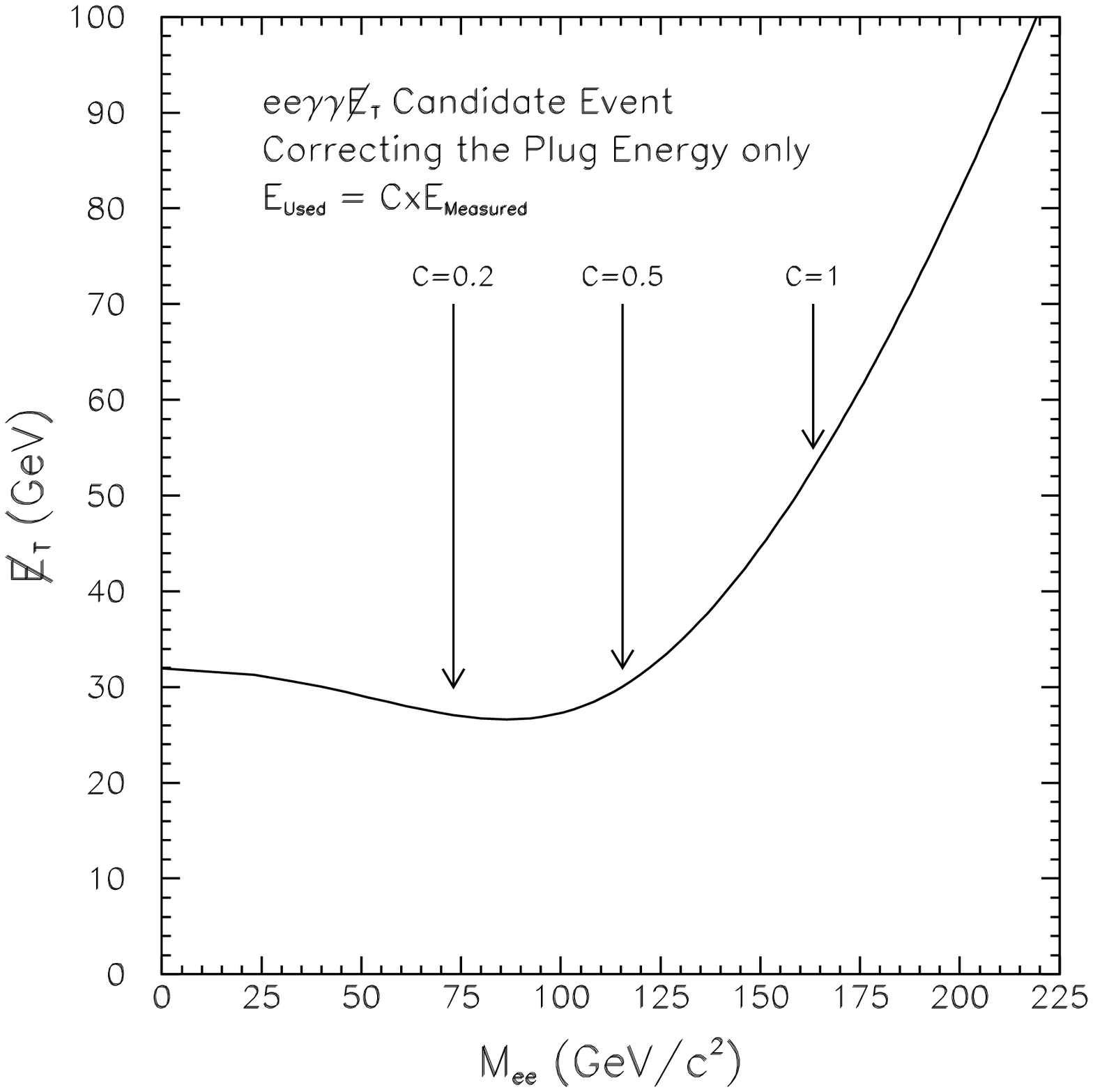}
{The invariant mass of the cluster in the plug calorimeter, here denoted as an
$e$, and the electron in
the central calorimeter (M$_{ee}$) plotted verses 
the $\mett$ as the energy of the
cluster in the plug is varied.}
{Mass ee vs. Met}

The $\eeggmett$ candidate event 
appears to originate from a single $\ppbar$ collision and consists 
of a high quality isolated electron, two isolated 
photons in the central calorimeter, 
significant $\mett$, and an electromagnetic 
cluster in the plug calorimeter. 
While the cluster passes all of the standard electron selection
criteria, further investigation reveals its interpretation is not obvious.   
The tracking chambers indicate that there is a charged particle (or
particles) traveling in the direction of the cluster but not directly at it,
indicating that the cluster might not due to an electron. 
The cluster could be interpreted as a photon, 
the hadronic decay of a tau lepton, or
simply as a jet. While all of three scenarios are reasonable {\it a priori}, 
and are consistent with the facts, each is 
unlikely in that this would be an unusual example of any single one of them.  

\begin{table}[htb]
\centering
\begin{tabular}{l|c|c|c|c|c} 
\multicolumn{6}{c}{ Run 68739, Event 257646} \\
\hline
  & P$_{\rm x} $  & P$_{\rm y}$  & P$_{\rm z}$ & E     & E$_{\rm T}$  \\
  & (GeV/c)       & (GeV/c)      & (GeV/c)     & (GeV) & (GeV)        \\
\hline\hline
 $\gamma_1$                                                                      
   &     32.1(9)
   &    -16.8(5)
   &  -35(1)
   &   50(1)
   &   36(1)
 \\ \hline
 $\gamma_2$                                                                      
   &    -12.9(4)
   &    -29.6(9)
   &    -22.5(7)
   &   39(1)
   &     32.3(9)
 \\ \hline
 $e^-$                                                                           
   &  -34(1)
   &     11.5(3)
   &     21.7(6)
   &   42(1)
   &   36(1)
 \\ \hline
 Plug EM Cluster                                                                 
   &   60(2)
   &     19.0(5)
   & -172(5)
   &  183(5)
   &   63(2)
 \\ \hline
 \mett                                                                           
   &  -54(7)
   &   13(7)
  & ---
   &   ---
   &   55(7)
\\ 
\end{tabular}
\caption[The 4-vectors of the electron and photon candidates
and the missing transverse energy in the $\eeggmett$
candidate event]{
The 4-vectors of the electron and photon candidates
and the missing transverse energy in the $\eeggmett$
candidate event using the primary vertex at $z$=20.4~cm.
The parentheses represent the uncertainty in the last digit and are
as determined in \secorchap~\ref{Resolution Section} 
and~\cite{Resolutions}.
There are no additional jets with E$_{\rm T}^{\rm
Corrected} > 10$~GeV. }
\label{Kinematics Table}
\end{table}

\begin{table}[htb]
\centering
\begin{tabular}{l|r|r|r|r|r}
\multicolumn{1}{l|}{Objects} & 
\multicolumn{1}{c|}{M$^{\rm System}$} &
\multicolumn{1}{c|}{P$_{\rm T}^{\rm System}$} &
\multicolumn{1}{c|}{$\mett$} & 
\multicolumn{1}{c|}{$\phi_{(\mett)}$} & 
\multicolumn{1}{c}{H$_{\rm T}$}\\
\multicolumn{1}{c|}{}        & 
\multicolumn{1}{c|}{(Gev/c$^2$)} &
\multicolumn{1}{c|}{(Gev)} &
\multicolumn{1}{c|}{(GeV)}   & 
\multicolumn{1}{c|}{(deg)} &
\multicolumn{1}{c}{(Gev/c$^2$)} \\
\hline
   $e_{\rm plug}   e_{\rm central}\gamma_1       \gamma_2       
 $ &   232.4 &     48.1 &     52.8 &    167.2 &    221.2\\
  \hline
    $e_{\rm plug}                  \gamma_1       \gamma_2       
 $ &   121.8 &     84.4 &     89.0 &    164.9 &    221.1\\
    $               e_{\rm central}\gamma_1       \gamma_2       
 $ &   121.4 &     38.2 &     32.0 &     73.8 &    137.0\\
    $e_{\rm plug}   e_{\rm central}\gamma_1                      
 $ &   195.6 &     59.7 &     66.9 &    195.6 &    202.8\\
    $e_{\rm plug}   e_{\rm central}               \gamma_2       
 $ &   200.4 &     13.1 &     20.0 &    194.9 &    152.0\\
  \hline
    $e_{\rm plug}   e_{\rm central}                              
 $ &   163.3 &     40.0 &     47.5 &    227.3 &    147.1\\
    $                              \gamma_1       \gamma_2       
 $ &    47.3 &     50.4 &     49.3 &    121.2 &    118.1\\
    $e_{\rm plug}                  \gamma_1                      
 $ &    56.5 &     92.6 &     99.1 &    183.8 &    198.7\\
    $e_{\rm plug}                                 \gamma_2       
 $ &    97.0 &     48.7 &     54.1 &    173.3 &    149.8\\
    $               e_{\rm central}\gamma_1                      
 $ &    91.7 &      5.8 &      4.1 &    166.6 &     76.8\\
    $               e_{\rm central}               \gamma_2       
 $ &    64.1 &     50.7 &     43.4 &     18.6 &    112.1\\
\end{tabular}
\caption[The kinematics of various combinations of the clusters in the
$\eeggmett$ candidate event]
{The kinematics of various combinations of the clusters in the
$\eeggmett$ candidate event. The combination of 
clusters is referred to as a system.
Column 4 ($\mett$) is the transverse imbalance of
that particular sub-system  and takes into account the underlying event.   
The H$_{\rm T}$ is the transverse mass of the
system along with its imbalance.
The cluster in the plug is simply referred to as $e_{\rm Plug}$ for simplicity.
The lowest $\mett$ attainable
by simply removing one electron or photon candidate from the event is 20.0~GeV, 
which occurs by removing $\gamma_1$.
By removing both the central photon and the cluster in the plug the 
$\mett$ becomes 4.1~GeV.}
\label{Mett Table}
\end{table}

\begin{table}[htb]
\centering
\begin{tabular}{l|r}
\multicolumn{1}{l|}{Objects} & 
\multicolumn{1}{c}{M$_{\rm T}$} \\
\multicolumn{1}{c|}{}        & 
\multicolumn{1}{c}{(Gev/c$^2$)} \\
\hline
    $e_{\rm plug}   e_{\rm central}\gamma_1       \gamma_2       \mett          
 $ &   221.1\\
  \hline
    $e_{\rm plug}                  \gamma_1       \gamma_2       \mett          
 $ &   182.0\\
    $               e_{\rm central}\gamma_1       \gamma_2       \mett          
 $ &   141.2\\
    $e_{\rm plug}   e_{\rm central}\gamma_1                      \mett          
 $ &   187.0\\
    $e_{\rm plug}   e_{\rm central}               \gamma_2       \mett          
 $ &   180.4\\
  \hline
    $e_{\rm plug}   e_{\rm central}                              \mett          
 $ &   144.2\\
    $                              \gamma_1       \gamma_2       \mett          
 $ &   111.9\\
    $e_{\rm plug}                  \gamma_1                      \mett          
 $ &   146.2\\
    $e_{\rm plug}                                 \gamma_2       \mett          
 $ &   148.5\\
    $               e_{\rm central}\gamma_1                      \mett          
 $ &   113.2\\
    $               e_{\rm central}               \gamma_2       \mett          
 $ &    70.4\\
  \hline
    $e_{\rm plug}                                                \mett          
 $ &   111.6\\
    $               e_{\rm central}                              \mett          
 $ &     4.3\\
    $                              \gamma_1                      \mett          
 $ &    86.9\\
    $                                             \gamma_2       \mett          
 $ &    52.8\\
\end{tabular}
\caption[The transverse mass for the measured $\mett$ and
various combinations of the electron and photon candidates 
within the $\eeggmett$ candidate event]
{The transverse mass for the measured $\mett$ and
various combinations of the electron and photon candidates 
within the $\eeggmett$ candidate event. 
The cluster in the plug calorimeter, for simplicity, is labeled 
$e_{\rm Plug}$. The
transverse mass of the 
$e_{\rm central}\mett$ and the `$e_{\rm plug}\mett$' candidate pairs 
are 4.3~GeV/c$^2$ and 111.6~GeV/c$^2$ respectively and are thus unlikely to be
from the decay $W\rightarrow e\nu$. However,
the $e_{\rm central}\gamma_2\mett$ combination 
could be due to the radiative decay of a $W$ via $W\rightarrow e\nu\gamma$ 
(M$_{\rm T} = 70.4$~GeV/c$^2$).   }
\label{Transverse Masses Table}
\end{table}


\chapterfive


The \mbox{{\it a posteriori}} estimation of the  
probability of a single event has measure zero. One instead has to  
define an event topology
and estimate the number of events which pass that
set of selection requirements from standard model sources.  
In an attempt to make the requirements similar to the standard 
\mbox{{\it a priori}} criteria used in CDF $W$ and $Z^0$ analyses, 
the $\eeggmett$ event topology is 
defined by the following list of requirements;

\begin{itemize}

\item One isolated electron in the central calorimeter with
E$_{\rm  T} > 25$~GeV

\item \begin{sloppypar} A second isolated electromagnetic 
cluster, in the central or plug calorimeters, 
which passes the electron identification requirements with 
\mbox{E$_{\rm  T} > 25$~GeV.} \end{sloppypar}

\item Two isolated central photons, E$_{\rm  T} > 25$~GeV 

\item $\mett > 25$~GeV \correctedmett

\item An electron-electron invariant mass 
above the mass of the $Z^0$: we use 110~GeV.
\end{itemize}

A subtlety in the topology requirement is that the cluster in the
plug calorimeter is possibly not an electron.  
To take this into account, the possible standard model sources are 
divided into two classes- those in which
the cluster in the plug 
is caused by an electron, and those in which it is not. In 
both cases,
electron candidates are required to pass the standard identification and
isolation requirements described in the previous 
\secorchap~\cite{top,Why Fake}.  Both
photons are required to pass the high-threshold requirements 
in Table~\ref{Event Cuts}. The primary sources are 
standard model $WW\gamma\gamma$ and $t{\bar t}$ production, 
events in which jets fake electrons and/or photons, cosmic ray
interactions, and overlapping events.


\subsection{Standard Model $WW\gamma\gamma$ and $t{\bar t}$ 
production}\label{WWgg}

 The standard model process that is most likely 
to produce the signature directly (assuming the cluster in the plug calorimeter
is due to an electron) is the production and decay of 
$WW\gamma\gamma$ where
\begin{equation}
p{\bar p} \rightarrow 
W^+W^-\gamma\gamma \rightarrow (e^+\nu)(e^-{\bar \nu})\gamma\gamma
\end{equation}
with each $\nu$ leaving the detector and causing \mett.
 To estimate the rate,  the MADGRAPH Monte Carlo~\cite{Mrenna} 
is used to simulate  the process in lowest order.  
The cross-section, $\sigma_{WW\gamma\gamma}$, is estimated
to be  $\sigma_{WW\gamma\gamma} = 0.15\pm 0.05$~fb for two
photons with E$_{\rm T} >$ 10~GeV, and $|\eta|<$~4.0. 
Taking into account the uncertainty on
the cross section, the luminosity (85$\pm$6.8 pb$^{-1}$), 
and differences between
detection efficiencies in the data and in the detector simulation, 
a total of \EWWGGHIGHRATE\  
events is taken as the best estimate
for $WW\gamma\gamma$ producing two electrons, two photons and \mett\ in the
observed topology.


Another source, in which both electrons are real, is
standard model $t{\bar t}$ production and decay. The PYTHIA 
Monte Carlo is used to simulate production, decay, fragmentation and 
the underlying event. Both 
$t$-quarks are forced to decay via
$t\rightarrow Wb$, both $W$'s decay via
$W\rightarrow e\nu$, and the photons are produced from radiation from 
internal fermion lines or are from jets which fake the photon
signature. 
Taking into account the uncertainty on
the $t{\bar t}$ cross section, the luminosity, 
differences between
detection efficiencies in the data and in the detector simulation,  
the extra photon rates and statistical
uncertainties in the sample,
the rate is estimated to be \ETOTTOPEXP\ events.


\subsection{Estimating the Number of Expected Fake Events}
\label{Estimates}

Other processes which contribute to the standard model production rate of
$\eeggmett$ events 
include events with jets that fake either photons or electrons, two 
standard model overlapping  events, or additional objects from
cosmic rays interacting or radiating in the detector.  To estimate the number of
events from these sources, the rate at which a {\it part} of the 
event occurs 
is multiplied by the probability that the rest of constituent
parts
of the event occur in a random event. For example, to estimate the rate
at which $WWjj$ production fakes the event signature, the
rate at which $W^+W^-\rightarrow e^+\nu e^-{\bar \nu}$ 
events occur and pass the $ee\mett$ requirements is multiplied by the
probability that two jets are produced in
association with the $WW$ and both fake the photon requirements.

The number of  observed events in various channels that constitute a
part of the $\eeggmett$ signature are listed in 
Table~\ref{Events Observed Table}. The 
rates at which jets fake the photon and
electron selection criteria
are essentially flat as a function of E$_{\rm T}$ 
for E$_{\rm T}>25~$GeV are also given in Table~\ref{Events Observed Table}.
The datasets used to measure or estimate
these numbers 
are selected
using the standard electron, photon and \mett\ identification requirements
described in \secorchap s~2 and 3.
The second part of the estimate requires a determination 
of the probability of finding an extra object or objects, such as real or fake
photons, in the event. The results for fake \mett, real and fake photons, and
fake electrons are summarized in Table~\ref{Object Fakes Table}.

\begin{table}[htb!]
\centering
\begin{tabular}{l|c}
Type of Event & Observed Number of events \\
\hline
\multicolumn{2}{c}{{\bf $W$-Type Events}}\\ \hline
$e_{\rm central}\mett$                  & \NTOTW        \\ \hline
$e_{\rm central}\mett + \gamma$         & \EGMET             \\ \hline
$e_{\rm central}\mett$ + Central Jet 
                   & \NWCJEVENTS\ Events, \NWCJETS\ jets \\ \hline
$e_{\rm central}\mett$ + Plug Jet       
                   & \NWPJEVENTS\ Events, \NWPJETS\ jets \\ \hline
$e_{\rm plug}\mett$                     & \NTOTPW \\ \hline
$e_{\rm plug}\mett + \gamma$            & \NPGM \\
\hline
\multicolumn{2}{c}{{\bf Photon-Type Events}}\\ \hline
$e_{\rm central}\gamma$                 & \EG            \\ \hline
$e_{\rm plug}\gamma$                    & \NPG            \\ \hline
$e_{\rm central}\gamma\gamma$           &  1            \\ \hline
$e_{\rm plug}\gamma\gamma$              &  0            \\ \hline
$\gamma\gamma$                          & \NGG           \\ \hline
$\gamma\gamma + \mett$ Events                 & 3             \\ \hline
\multicolumn{2}{c}{{\bf Cosmic-Type Events}}\\ \hline
$\gamma + \mett$                        & \NGMET         \\ \hline
\multicolumn{2}{c}{{\bf $Z^0/\gamma^*$-Type Events(CC/CP)}}\\
\hline
 $ee$                    		& $\CCDY/\CPDY           $\\ \hline
 $ee\mett$                    		& $\CCDYMET/\CPDYMET     $\\ \hline
 $ee\gamma$                    		& $\CCDYGAMMA/\CPDYGAMMA $\\ \hline
$\zoee$                                 & $\CCZEE/\CPZEE         $\\ \hline
$\zoee$ + $\mett$	                & $\CCZEEMET/\CPZEEMET   $\\ \hline
$\zoee$ + $\gamma$                      & $\CCZEEGAMMA/\CPZEEGAMMA $\\ \hline
M$_{\rm ee} > 110$~GeV                  & $\CCMEE/\CPMEE          $\\ \hline
M$_{\rm ee} > 110$~GeV + \mett          & $\CCMEEMET/\CPMEEMET    $\\ \hline
M$_{\rm ee} > 110$~GeV + $\gamma$       & $\CCMEEGAMMA/\CPMEEGAMMA $\\ \hline
\hline
\multicolumn{2}{c}{\bf {Other numbers}}\\ \hline
Bunch Crossings                         & \scinotn{3}{12}\\ \hline

Central electron fake rate      & $<$ \CEFAKEP / jet \\ 
                                & (95$\%$ C.L. Upper Limit) \\
\hline

Central photon   fake rate      &  \CPFAKEP / jet \\ \hline

Plug    electron fake rate      & \PEFAKEP / jet \\ 
\end{tabular}
\caption[The number of observed events for the various parts of the $\eeggmett$
signature]{The number of 
observed events for the various parts of the $\eeggmett$
signature, used to calculate fake 
and overlap rates.  The $\zoee$ events
require \mbox{81~GeV $< {\rm M}_{\rm e^+e^-} < 101$~GeV}.
The one event in the
M$_{\rm ee} > 110$~GeV + $\gamma$ category is the $\eeggmett$ event.} 
\label{Events Observed Table}
\end{table}

\begin{table}[htb]
\centering
\begin{tabular}{lll|c}
\multicolumn{3}{l|}{Rate for finding addition objects}    & Rate/Event  \\
\hline
\multicolumn{3}{l|}{Fake $\mett$} &
\METFAKE \\
$\;\;\;R_{\rm  \mettsm}$  & $\approx
\frac{Z^0 \rightarrow {\rm ee\mettsm\; Events}}
     {Z^0 \rightarrow{\rm ee\; Events}}$ & & \\
 & $= \frac{\CCZEEMET + \CPZEEMET}{\CCZEE + \CPZEE}$ & & \\
 &  $\approx$  \METFAKE & & \\ \hline

\multicolumn{3}{l|}{An additional central photon} &
\RADDGAMMA \\
$\;\;\;R_{\rm  \gamma}$  & $\approx
R^{\rm fake}_{\gamma}$ & + $R_{\gamma}^{\rm radiation}$ &    \\
 & $\approx {\it P}^{\rm Central}_{\rm Extra\; Jet}\times 
{\it P}^{\rm fake}_{\gamma}$  &
+ $\frac{\rm ee\gamma\; Events}{\rm ee\; Events}$ &   \\
 & $= \frac{\NWCJETS}{\NTOTW}\times$ (\CPFAKEP) & +
$\frac{\CCDYGAMMA + \CPDYGAMMA}{\CCDY + \CPDY}$ & \\
 & = \CPFAKE & +  \RADGAMMA & \\
 &  $\approx$ \RADDGAMMA  & &   \\ \hline

\multicolumn{3}{l|}{An additional plug electron candidate} &
\PEFAKE \\
 $\;\;\;R^{\rm fake}_{\rm  Plug\; e}$ &
 \multicolumn{2}{l|}{$\approx {\it P}^{\rm Plug}_{\rm Extra\; Jet} \times 
{\it P}_{{\rm Plug}~e}^{\rm fake}$}  &  \\
 & $= \frac{\NWPJETS}{\NTOTW}\times$ (\PEFAKEP)  & &  \\
 &  $\approx$ \PEFAKE &  &  \\ \hline

\multicolumn{3}{l|}{An additional central electron (95$\%$ C.L.)} &
$<$\CEFAKE \\
$\;\;\;R^{\rm fake}_{\rm  Cent\;  e}$ &
\multicolumn{2}{l|}{$\approx {\it P}^{\rm Central}_{\rm Extra\; Jet} 
\times {\it P}^{\rm fake}_{\rm Central\; e}$}  &  (95$\%$ C.L.) \\
 & $= \frac{\NWCJETS}{\NTOTW}\times$ (\CEFAKEP)  & &  \\
 &  $\approx$ \CEFAKE & & \\
\end{tabular}
\caption[The estimated rates for finding fake objects (electrons, photons or 
$\mett$) in an event from various processes]
{The estimated rates for finding fake objects (electrons, photons or 
$\mett$) in an event from various processes. These numbers are estimated using 
the from
the numbers in Table~\protect\ref{Events Observed Table}. 
The rate for finding
an additional central photon is probably an overestimate 
by a factor of two because both
methods include contributions from real photons as well as fakes. } 
\label{Object Fakes Table}
\end{table}

\subsubsection{Events with a Fake Object or Objects}

The number of expected events where part of the event is `real' and
part of the event is `faked' is summarized in 
Table~\ref{Estimated Fake Events Table}. 
Contributions from events with fake
central electrons have not been included 
as the expected rate is negligible compared to the other
sources. A total of \scinotn{3\pm3}{-7} events are expected in the data 
due to fake sources.

\begin{table}[htb]
\centering
\begin{tabular}{c|c|c|c|l}
Real    & Fake          &Fake           & Fake          &
\multicolumn{1}{c}{ Events} \\
Process & Process 1     & Process 2     & Process 3     & 
\multicolumn{1}{c}{in 85~pb$^{-1}$} \\
\hline

$e^+e^- \gamma\mett$ &
$\gamma^{\rm Fake}$ & 
--- & 
--- &  
\\
 
\EEGMET &  
 \CPFAKE    &     
--- &     
--- & 
  \EEGFM
 \\ \hline

$e^+e^- \gamma\gamma$ &
\mettfake & 
--- & 
--- &  
\\
 
\EEGG &  
 \METFAKE    &     
--- &     
--- & 
  \EEGGF
 \\ \hline

$e_{\rm Cent} \gamma\gamma\mett$ &  
$e_{\rm Plug}^{\rm Fake}$ & 
--- & 
--- & 
 \\
         
\EGGMET &  
 \PEFAKE  & 
--- &
--- &
  \EFGGM
 \\ \hline

$ ee\mett$  &  
$\gamma^{\rm Fake}$ & 
$\gamma^{\rm Fake}$ & 
--- &
 \\ 
\NMEEMET &  
 \CPFAKE &
 \CPFAKE &
--- &
  \EEFFM
\\ \hline

$e_{\rm Cent} \gamma\mett$ &
$\gamma^{\rm Fake}$ &
$e_{\rm Plug}^{\rm Fake}$ &
 --- & 
 \\

\EGMET &
 \CPFAKE &
 \PEFAKE &
 --- & 
 \EFGFM \\ \hline
$ee \gamma$ &
$\gamma^{\rm Fake}$ &
\mettfake & 
 --- & 
 \\
\EEG &
 \CPFAKE &
 \METFAKE &
 --- & 
 \EEGFF \\ \hline
$e_{\rm Cent}\gamma \gamma$ &
$e_{\rm Plug}^{\rm Fake}$ &
\mettfake & 
 --- & 
 \\

\EGG &
 \PEFAKE &
 \METFAKE &
 --- & 
 \EFGGF \\
\hline
%
$e_{\rm Cent} \gamma$ & 
$\gamma^{\rm Fake}$ &
$e_{\rm Plug}^{\rm Fake}$ &
\mettfake  & 
 \\

\EG &
 \CPFAKE &
 \PEFAKE &
 \METFAKE &
 \EFGFF  \\
\hline
$e_{\rm Cent} \mett$ & 
$\gamma^{\rm Fake}$ &
$\gamma^{\rm Fake}$ &
$e_{\rm Plug}^{\rm Fake}$ &
 \\

\NTOTW &
 \CPFAKE &
 \CPFAKE &
 \PEFAKE &
 \EFFFM  \\
\hline
$ee$ & 
$\gamma^{\rm Fake}$ &
$\gamma^{\rm Fake}$ &
\mettfake  & 
 \\

\NMEE &
 \CPFAKE &
 \CPFAKE &
 \METFAKE &
 \EEFFF  \\
\hline
%
%
 \multicolumn{3}{l}{Sum}  & \multicolumn{2}{r}{$\approx$ \SUMEEGGMETNOWWGG}\\  
\end{tabular}
\caption[An estimate of the number of 
events passing the $\eeggmett$ selection criteria from
events with fake electrons, photons or $\mett$]
{An estimate of the number of events 
passing the $\eeggmett$ selection criteria from
events with fake electrons, photons or $\mett$.
Individual rates are estimated as being equal 
to the number of observed events (Real
Process) multiplied by 
the rate at which additional objects from fakes (Fake Process)
are observed in the event.
The Real Process rate is taken or estimated from 
Table~\protect\ref{Events Observed Table} and the Fake Process rates are
taken from  Table~\protect\ref{Object Fakes Table}.}
\label{Estimated Fake Events Table}
\end{table}

The `Real Process' rates (Column 1) in Table 
\ref{Estimated Fake Events Table} are derived
from Table~\ref{Events Observed Table} as follows:
\begin{itemize}

\item{\bf $e^+e^- \gamma\mett$:} The \NMEEMET\ events 
from the data with M$_{\rm ee} > 110$~GeV
and \mbox{$\mett > 25$~GeV},  multiplied by 
a factor of \RADGAMMA\ for a real additional 
central photon.

\item{\bf $e^+e^-  \gamma\gamma$:} The \NMEE\ events in the data with 
M$_{\rm ee} > 110$~GeV, multiplied by two factors of \RADGAMMA\ for real
additional central photons.

\item{\bf $e_{\rm Central} \gamma\gamma\mett$:} The \EGMET\
$e_{\rm central}\mett + \gamma$ events, multiplied by a factor of \RADGAMMA\ 
for a real additional central photon.

\item {\bf $e^+e^-\mett$:} The \NMEEMET\ events 
from the data with M$_{\rm ee} > 110$~GeV
and \mbox{$\mett > 25$~GeV}. 

\item{\bf $e_{\rm Central} \gamma\mett$:} The \EGMET\ $e_{\rm Central}
\gamma\mett$ events.

\item{\bf $ee\gamma$:} The \NMEE\ 
$ee$ events in the data with M$_{\rm
ee}>110$~GeV, multiplied by a factor of \RADGAMMA\ for a real additional 
central photon.

\item{\bf $e_{\rm Central} \gamma\gamma$:} The \EG\ $e_{\rm Central}
\gamma$ events, multiplied by a factor of \RADGAMMA\ for a real additional 
central photon.

\item{\bf $e_{\rm Central} \gamma$:} The \EG\ $e_{\rm Central}
\gamma$ events.

\end{itemize}

\subsubsection{Overlapping Events, Including Cosmic Rays}

Events in which two collisions 
occur at the same time, each producing part of the event, can fake
the $\eeggmett$ signature. 
The rate of expected events from each source is estimated to be equal to
the rate at which one part of the event occurs, multiplied by 
the probability of the rest of the
signature occurring in a second overlapping event.  
The probability of getting a particular type of overlapping event is estimated
to be equal to
the number of events with that signature, divided by the total number of 
events studied by the detector during the course of the run (\scinotn{3}{12}).
The total rate sums over all
processes and includes contributions from cosmic rays which leave a photon
in the detector as well as real physics contributions which might occur
in an overlapping event. 

The results for the dominant sources of overlap events are summarized in 
Table~\ref{Overlaps Table}. To take into account the fact
that there are 4 interactions in the $\eeggmett$ candidate event, 
the estimate is multiplied by 
6 to reflect the 6 possible permutations of any two of the four interactions
causing the
signature. Summing all the sources, the total rate due to overlapping events 
is estimated to be \EOVERLAPTOT\ events.

\begin{table}[htb]
\centering
\begin{tabular}{l|l|c@{$\times$}c|c}
Process 1 & Process 2 & \multicolumn{2}{|c|}{Calculation} & Exp. Events \\
\hline
$WW_{high~mass}$  & $\gamma\gamma$           & 
\NMEEMET              & \ovpr{\NGG}     & \OVWWGG      \\ \hline

$W\gamma$          & $W\gamma$                & 
\EGMET               & \ovpr{\EGMET}      & \OVWGWG     \\ \hline

$W\gamma\gamma\goes e_{\rm plug}\gamma\gamma$    
                        & $W\goes e_{\rm cent}\mett$  & 
\PGG & \ovpr{\NTOTW} & \OVPGGW     \\ \hline

$ee\gamma$    & $\gamma\mett$    & 
\EEGTOT             & \ovpr{\NGMET}  & \OVEEGGMET \\ \hline

$W\gamma\gamma \rightarrow e_{\rm cent}\gamma\gamma$ 
                      & $e_{\rm Plug}\mett$   &
\WGG  & \ovpr{\NTOTPW}     & \OVWGGCOS       \\ \hline 

$WW\gamma$              & Cosmic $\rightarrow \gamma$   &
\WWG& \ovpr{\NGMET}  & \OVWWGCOS     \\ \hline
\hline
   \multicolumn{4}{l}{Sum}  & $\approx$ \OVERLAPTOTVER\\ \hline
   \multicolumn{4}{l}{Sum$\times$ 6}  & $\approx$ \OVERLAPTOT\\ 
\end{tabular}
\caption[The number of events with the $\eeggmett$ signature 
due to two overlapping events]
{The number of events with the $\eeggmett$ signature 
due to two overlapping events. 
These include double interactions, 
two separate events occurring in the same bunch crossing,
as well as an event with additional objects from a cosmic ray which 
interacted or
radiated in the detector. The number of expected events is estimated to be
equal to the rate of the `real' part of the event (Process 1)
times the probability of observing a particular type of overlapping event
(Process~2). The probability is defined to
be equal to the number of events with the signature divided by the number of
bunch crossings in the data (\scinotn{3}{12}). To take into account the fact
that 4 interactions are observed in this event the estimate is multiplied by 
6.}
\label{Overlaps Table}
\end{table}

The Process 1 rates (Column 1) in Table 
\ref{Overlaps Table} are derived
as follows:
\begin{itemize}

\item{\bf $e_{\rm plug}\gamma\gamma$:} 
The \NPG\ $e_{\rm plug}\gamma$ events in the data, multiplied by
the fake+real additional photon rate of \RADDGAMMA.

\item{\bf $ee\gamma$:} The \NMEE\ $ee$ events in the data with M$_{\rm
ee}>110$~GeV, multiplied by 
the fake+real additional photon rate of \RADDGAMMA.

\item{\bf $e\gamma\gamma$:} The 
\EG\ $e\gamma$ events in the data, multiplied by 
the fake+real additional photon rate of \RADDGAMMA.

\item{\bf $WW\gamma$:} The \NMEEMET\ $WW$ events in the data, multiplied by
the fake+real additional photon rate of \RADDGAMMA.

\item{\bf $W\gamma\gamma$:} The \EGMET\ $e\gamma\mett$ 
events in the data, multiplied by
the fake+real additional photon rate of \RADDGAMMA.

\item{\bf $e_{\rm Plug}\mett$:} The \NTOTPW\ $e_{\rm plug}\mett$ events in the
data. These events come from both $W$ and cosmic ray production. 

\end{itemize}

\normalsize


\subsection{Total Standard Model Rates}\label{All Plug and Fake Plug E}

The possible standard model sources are 
divided into two classes- those in which the second electromagnetic cluster
passing the electron selection criteria 
is caused by an electron, and those in which it
is not. 
The standard model estimate for the number of events with the $\eeggmett$
signature where the plug cluster is allowed to be due to an electron
is dominated by real $WW\gamma\gamma$ production with a total 
of \WWGGHIGHRATE\  $\eeggmett$ events expected.
standard model 
$t{\bar t}$ production contributes an additional \TOTTOPEXP\ events. 
The total fake rate, split roughly equally between $ee\gamma\gamma$ + fake
\mett, $ee\gamma\mett$ + fake photon
and $e\gamma\gamma\mett$ + fake plug electron, contributes a total of  
\SUMEEGGMETNOWWGG\ events. 
Overlapping events, including cosmic rays 
are estimated to contribute a total of  
\OVERLAPTOT\ events. 
Including the uncertainties, 
the total rate is estimated to be: 
\begin{equation}
N^{\rm Standard~ Plug ~e ~Requirements}_{\rm Expected} = \EEEGGMETTOTRATE\
{\rm events}.
\end{equation}
With the addition of the SVX data and a thorough scrutiny of the plug
cluster, 
there are good indications that the cluster may not
be due to an electron. The total rate where the plug cluster
is not due to an electron is reduced from that above
because the
dominant backgrounds ($WW\gamma\gamma$ and $t{\bar t}$), 
each of which produce two electrons, are removed. 
The total,  dominated by $e\gamma\gamma\mett$ + fake plug electron,
is \EFAKEGGMETTOTRATE. Overlaps are again negligible. Including 
the uncertainties, the total rate is estimated to be: 
\begin{equation}
N^{\rm Plug ~Cluster ~not ~an ~e}_{\rm Expected} = 
\EEFAKEGGMETTOTRATE\ {\rm events}.
\end{equation}

\chaptersix


The low probability of a satisfactory standard model 
explanation for the `$\eeggmett$' candidate event leaves open the possibility of
new physics interpretations.  A number of 
theories put forward to explain the event predict that other
events from related decay modes 
should appear in the $\gamma\gamma + X$ searches. Since there is
no evidence of these events in the central diphoton sample, 
quantitative limits can be set on 
such scenarios. 

\subsection{Anomalous $WW\gamma\gamma$ Production}
\label{Anomalous WWGG Production}

It is improbable  {\it a~priori} that the $\eeggmett$ candidate
event is from standard model $WW\gamma\gamma$
production, as shown in \secorchap~5. However, the event
could be an example of anomalous $WW\gamma\gamma$ 
production~\cite{WWGG and SUSY}.
This hypothesis can be tested quantitatively 
by assuming that the one event was produced at its
mean cross section and using the standard model 
$WW\gamma\gamma$ Monte Carlo to estimate  the mean number
of events in the $WW \rightarrow jjj$ decay channel using the following:
\begin{eqnarray}
N_{\gamma\gamma + jjj}^{\rm Expected} & \approx &
N_{\gamma\gamma + \ell_i\ell_j \mettsm}^{\rm Observed} \\
 & & \times
(\frac{{\rm Rate}~(WW\gamma\gamma \rightarrow \gamma\gamma + jjj)}
   {{\rm Rate}~(WW\gamma\gamma \rightarrow \gamma\gamma + 
                \ell_i\ell_j\mett)})
   \nonumber
\end{eqnarray}
where  the $\gamma\gamma + jjj$ channel is defined as
two photon 
which pass the high-E$_{\rm T}$ diphoton selection criteria (E$_{\rm
T}^{\gamma}>25$~GeV) plus 3 or more 
jets as defined in \secorchap~3 and the $\gamma\gamma + \ell\ell\mett$ channel
is defined as 2 or more
leptons in any combination as defined in \secorchap~3 and 
\mbox{\mett$>$ 25~GeV.}
Only 3 jets are required 
because the acceptance is almost 
cut in half by requiring a fourth jet. 
All leptonic decays ($e, \mu$ or $\tau$) 
of the $WW$ pair is used as a normalization 
rather than just the $ee$ channel to be conservative.

The Monte Carlo estimates
that $WW\gamma\gamma$ production should
produce $\gamma\gamma$ events with 3 or more jets 30 times more often than 
events with two photons, two charged leptons and
$\mett$.  With one $\gamma\gamma\ell\ell\mett$ candidate event and no 
events with 3 or
more jets in the data  (see Table~\ref{found} and 
Figure~\ref{Diphoton NJET Data 25})
anomalous $WW\gamma\gamma$ is excluded as the
source of this event at the 95\% C.L.


\subsection{Supersymmetric Models}\label{SUSY Theory}

Several theories have been proposed 
to explain the $\eeggmett$ candidate event
\cite{Gravitino Reference, Higgsino LSP, Non-SUSY}.
The trademark of many of the supersymmetric 
versions~\cite{Dawson} is that in addition to the
two photons produced, two or more of the 
lightest supersymmetric particles (LSP) are produced in 
every event, either by direct production or by cascade,
and leave the detector causing an energy imbalance.  

In light gravitino scenarios~\cite{Gravitino Reference} 
the gravitino can have a mass on the order of
1~eV and for most of the parameter space the lightest neutralino, \NONE, has
a branching ratio of $\approx$100\% into $\gamma\gravitino$. 
The lifetime of the $N_1$ depends on $M_{\tilde G}$; the decay occurs inside
the detector for a gravitino mass less than approximately 1~KeV.
The $\gravitino$ is
very weakly interacting and escapes the detector leaving an energy imbalance.
These models can produce the $\eeggmett$ signature, for example, via:
\begin{eqnarray}
C_1C_1 & \rightarrow & (\nu{\tilde e})(\nu{\tilde e}) \rightarrow
\nu(e\NONE)\nu(e\NONE) \\
 & & \rightarrow 
e\nu(\gamma\Gravitino)e\nu(\gamma\Gravitino) \rightarrow
\eeggmett \nonumber
\end{eqnarray}

For concreteness, limits on light gravitino scenarios are set using a
gauge-mediated model in the MSSM, 
 hereafter referred to as the minimal gauge-mediated model.
The parameter space in these models is 
spanned by $M_2$, $\tanbeta$ and the sign of $\mu$. To
simulate these models we have used a custom interface~\cite{Kolda} to the 
SPYTHIA Monte Carlo~\cite{SPYTHIA}
which calculates the inputs to SPYTHIA using the full one-loop renormalization
group effects calculated in~\cite{Gravitino Reference}e.
Full simulations are done for a
total of 50 points in parameter space: 
\mbox{$M_2$ = 75,} 100, 125, 150 and 
200~GeV,  \mbox{$\tanbeta$ = 1.1,} 2, 5, 10 and 25 and \mbox{\sgnmu = $\pm$1}.  
For most of parameter space 
${\rm M}_{\CONE} \simeq {\rm M}_{\NTWO} \simeq 2{\rm M}_{\NONE}$ 
and the production cross section 
is dominated by  $C_1N_2$ and $C_1C_1$  production which in turn 
decay to produce two photons and missing $\Et$ in every event. 
Figure~\ref{MGM Kinematics} shows distributions in E$_{\rm T}$ of the photons
and \mett\ (after simulation and the full diphoton and 
$\Delta\phi_{\mettsm-{\rm jet}}$ requirements
of Section~\ref{Resolution Section}) for  \mbox{$M_2$ = 150~GeV,} 
\mbox{$\tanbeta$ = 10}
and \mbox{\sgnmu = 1.}

The \NTGNO\ model of Kane \etal~\cite{Higgsino LSP} 
predicts the $N_2$ to be pure photino, $N_1$ to be pure higgsino and 
the higgsino to be lighter than the photino.
In this case, the \NONE\ is the LSP, the gravitino is too massive to play a
role in the phenomenology and the dominant decay of the \NTWO\ is 
through the one-loop radiative
decay with \mbox{Br(\NTGNO) $\approx$~100\%.} The $\eeggmett$ signature can
be produced, for example, via: 
\begin{equation}\label{Slepton Decay}
{\tilde e_L}{\tilde e_L} \rightarrow
(e\NTWO)(e\NTWO) \rightarrow e(\gamma\NONE)e(\gamma\NONE) \rightarrow
\eeggmett.
\end{equation}

For concreteness, limits are set on a particular point in parameter 
space~\cite{Ambrosanio Table Ref} 
with \mbox{M$_{N_1} = 36.6$~GeV}, \mbox{M$_{N_2} = 64.6$~GeV} and a
total sparticle production cross section of
\KANEXSECTOT~pb.
To provide a normalization point for 
future model-builders to estimate the detector efficiency, 
$\NTWO\NTWO$ production \mbox{($\sigma_{N_2N_2} \approx$ 2~fb)} 
is used to set cross section limits. 
The distributions in E$_{\rm T}$ of the photons
and \mett\ (after simulation and the full diphoton and 
$\Delta\phi_{\mettsm-{\rm jet}}$ requirements
of Section~\ref{Resolution Section}) are shown in Figure~\ref{Kane Kinematics}.

\twofig{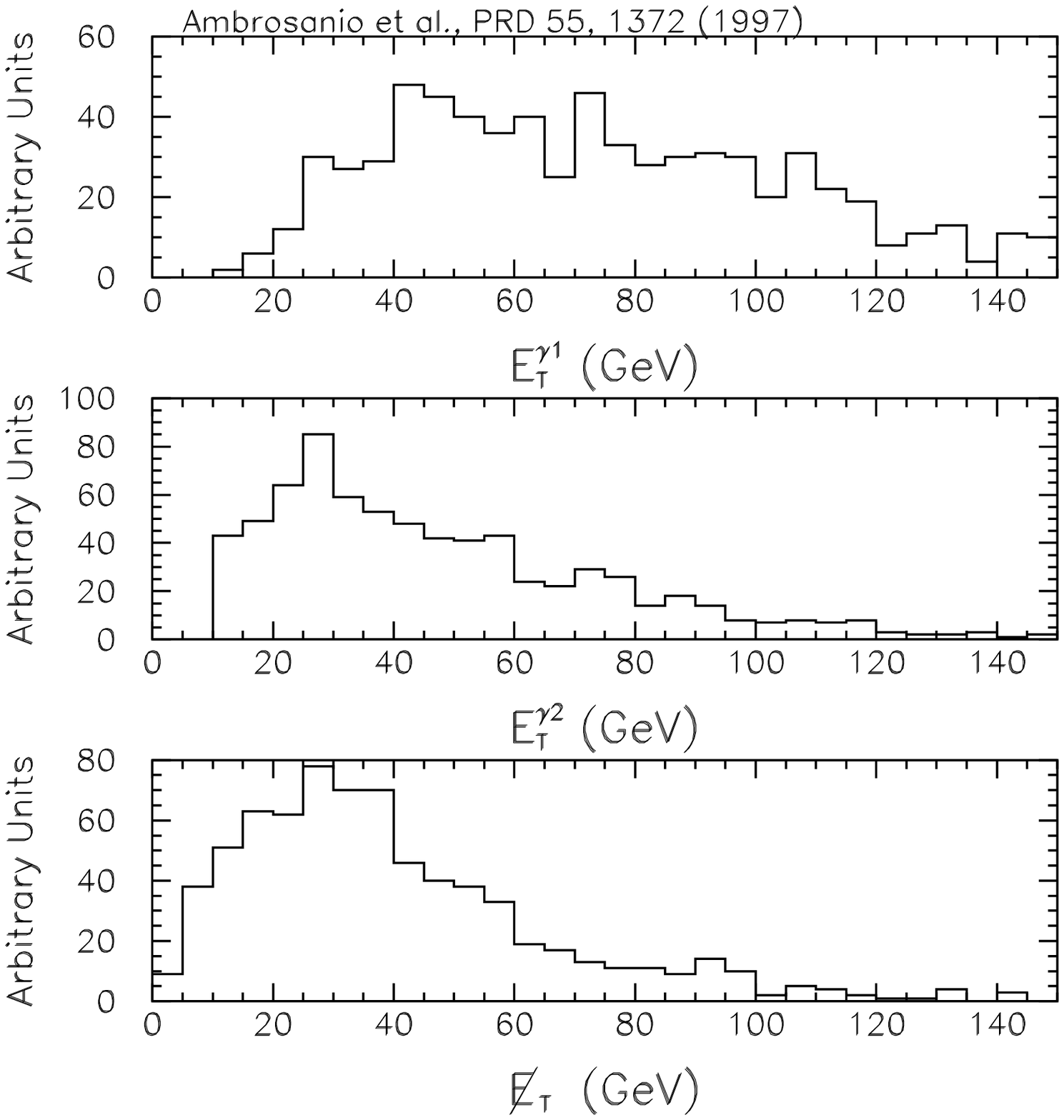}
{The distributions in E$_{\rm T}$ of the photons
and \mett\ in the minimal 
gauge-mediated model with $M_2$ = 150~GeV, \mbox{$\tanbeta$ = 10}
and \mbox{\sgnmu = 1.}
The sample is normalized to 5000 events generated, which correspond
to an integrated luminosity of 7,163~pb$^{-1}$.}
{MGM Kinematics}
{kinematics_plot.ps}
{The distributions in E$_{\rm T}$ of the photons
and \mett\ for \NTWO\NTWO\ production in the \NTGNO\ model with 
M$_{N_2} = 64.6$~GeV, and
M$_{N_1} = 36.6$~GeV.  The sample is normalized to 5000 events 
generated, which correspond to a luminosity of 2,272~fb$^{-1}$.}
{Kane Kinematics}

\subsection{Acceptances}\label{Acceptances}

\begin{sloppypar} 
The acceptance for a given model is determined using the following equation:
\begin{eqnarray}\label{Acc Equation}
Acc & = &  \Sigma 
    A^i_{\rm MC} \times 
    C^i_{\rm (ID~and~Iso)} \times 
    C_{z_{\rm vertex}} \\
 & & \times 
    C_{\rm ETOUT} \times 
    C^i_{\rm Trig} \nonumber 
\end{eqnarray}
where the index, $i$, is for the two different regions (\mbox{12~GeV $< {\rm
E_T^{\gamma_2}} <$22~GeV} and \mbox{${\rm E_T^{\gamma_2}} >$22~GeV})
to take into account the different trigger requirements and photon selections,
$A^i_{\rm MC}$ is the acceptance from the Monte Carlo using the full detector 
simulation for the different regions,
$C^i_{\rm (ID~and~Iso)}$ is the correction for differences between photon
identification and
isolation variables in the data 
and in the detector simulation, $C_{z_{\rm vertex}}$ is the
correction for differences between the
distributions of the interaction point, $z_{\rm vertex}$, 
in the data and that simulated in 
Monte Carlo, $C_{\rm ETOUT}$ is the efficiency of
the energy-out-of-time requirement and $C^i_{\rm Trig}$ is the correction for
the trigger efficiency.  
The corrections (taken from \effref) 
are summarized in Table~\ref{Efficiency Table}.  
\end{sloppypar}

\begin{table}[htb]
\centering
\begin{tabular}{l|c|c}
Correction              & Requirement & Requirement \\
                        & 12~GeV $< {\rm
E_T^{\gamma_2}} <$22~GeV & ${\rm E_T^{\gamma_2}} >$22~GeV \\
\hline
ID and Iso              & $\CORRIDLOW$          & $\CORRIDHIGH$\\

$|z_{\rm vertex}|<$ 60 cm 
                        & $\RATIOVERT$          & $\RATIOVERT$ \\
ETOUT = 0               & $\EFFETOUT$           & $\EFFETOUT$ \\
Trigger                 & $\TRIGLOW$            & $\TRIGHIGH$ \\
\hline
Total Correction        & $\EFFCORLOW$          & $\EFFCORHIGH$ \\
\end{tabular}
\caption[The corrections used to take into account differences between the true
detector response and the detector simulation]
{The corrections used to take into account differences between the true
detector response and the detector simulation. The identification and
isolation requirement corrections are labeled as ID and Iso.
The energy out-of-time requirement correction is labeled as ETOUT.}
\label{Efficiency Table}
\end{table}

\subsection{Results for $\gamma\gamma + \mett$}

A single set of 
requirements, \Etggl\ and \mettgmh, is chosen as it is estimated 
to exclude the maximal
amount of parameter space for the light gravitino model.
The acceptances are typically between 1\% and 10\%. 
Figure~\ref{Acc and NEXP vs. N1} shows the 
acceptance and number of expected events 
versus the \NONE\ mass.

Only one event in the diphoton data sample
passes the requirements (the $\eeggmett$ candidate event). 
Figure~\ref{Contour Limit} shows the 
contour plot of the excluded region in the $\tanbeta$ versus $M_2$ plane.
The shaded regions in 
Figures~\ref{Filled XSEC vs. N1MASS Plot} and \ref{Filled XSEC vs. C1MASS Plot}
show the limits as a function of the $\NONE$ and $\CONE$ masses, 
respectively as the parameters are varied.
The lines show the experimental limit and the theoretically predicted
cross section for the lowest value of the \NONE\ or \CONE\ 
mass which is excluded.  The 
\NONE\ is 
excluded for  
M$_{N_1}<65$~GeV at 95\% C.L. (this occurs at \mbox{$\tanbeta = 5$,} 
\mbox{$\mu> 0$}). The \CONE\ is excluded for
M$_{C_1} < 120$~GeV at 95\% C.L. (this occurs at \mbox{$\tanbeta = 5$,} 
\mbox{$\mu < 0$}). 

The same selection criteria are used for the \NTGNO\ model.
The model is not excluded by the
data as only \KANELOWNEXPTOT\ events from all sparticle production and decay 
are expected to pass the selection criteria. 
The acceptance for $\NTWO\NTWO$ production
is \KANELOWACC\% with a 
95\% C.L. cross section upper limit of \KANELOWXSEC~pb. 
These results, along with the light gravitino results and the results of 
Section~\ref{Chap-Searches}, 
are comparable
to those of LEP~\cite{LEP, LEP II} 
and the D$\O$ collaboration~\cite{D0}.

\twofig{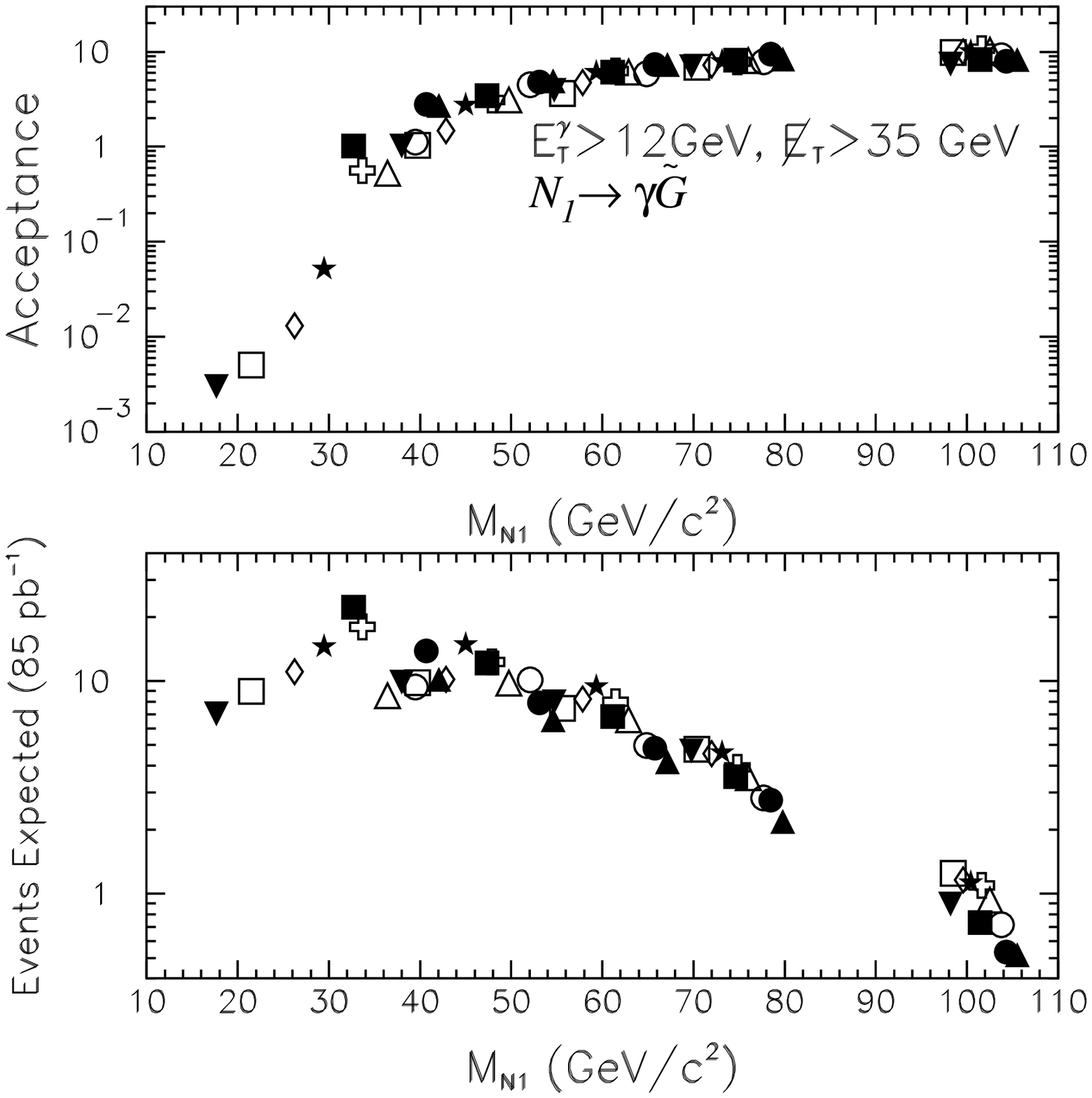}
{The acceptances (upper plot) and the number of expected events (lower plot) 
for various points in parameter space 
plotted versus the \NONE\ mass  in the minimal gauge-mediated
model.}
{Acc and NEXP vs. N1}
{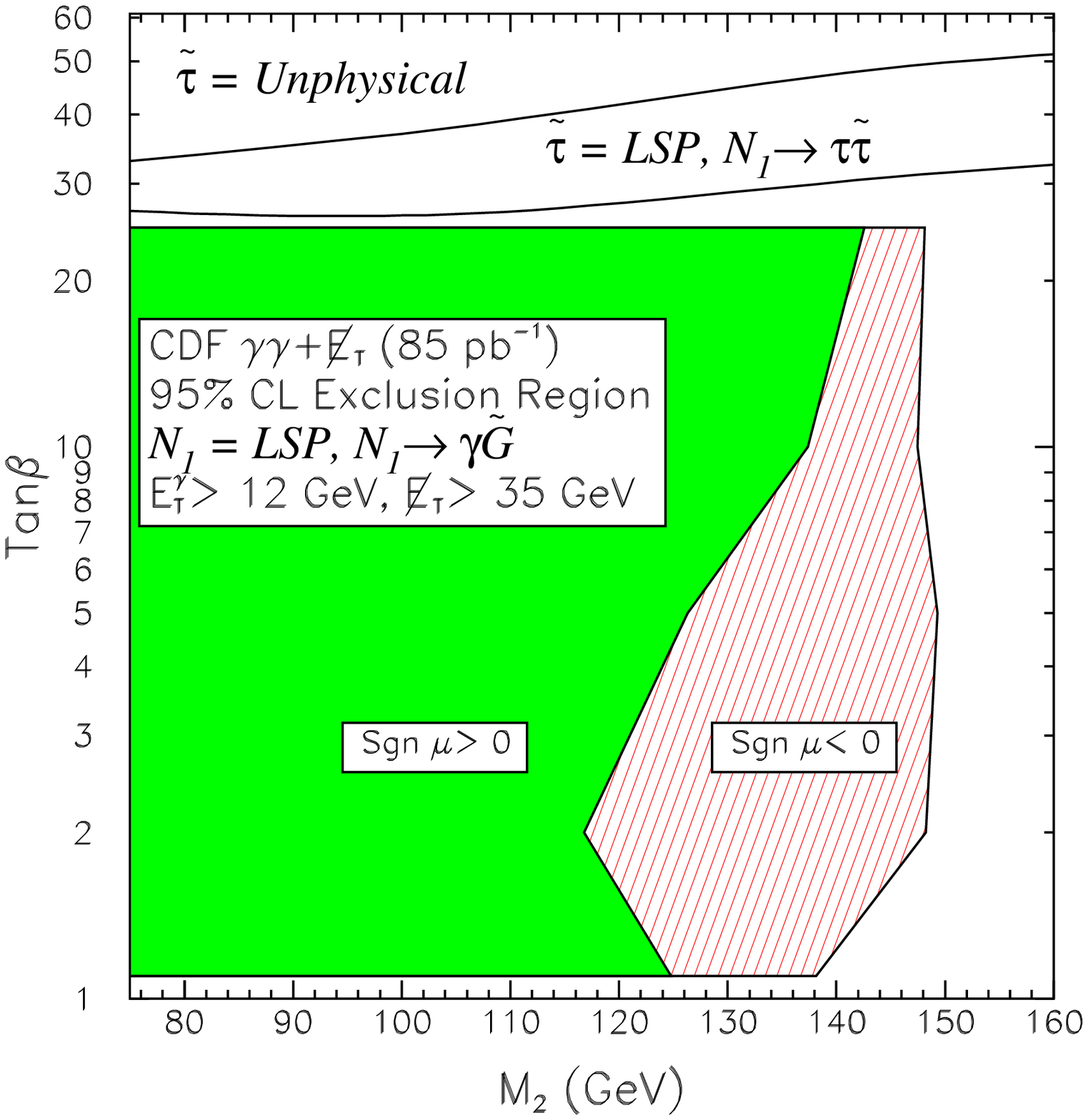}
{The contour plot of the excluded region of the minimal 
gauge-mediated model in the $\tanbeta$
versus $M_2$ plane.}
{Contour Limit}

\twofig{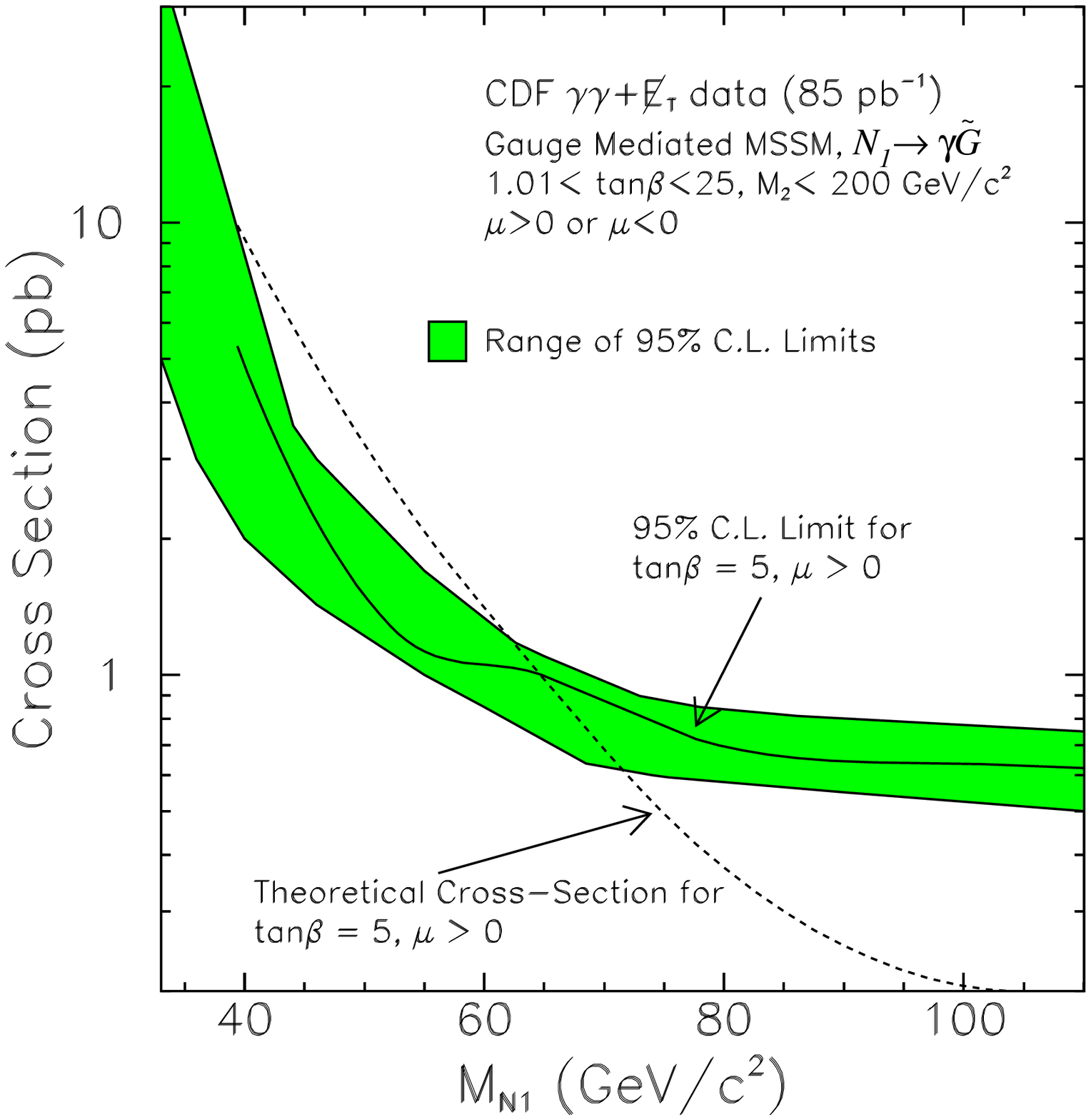}
{The 95\% C.L. cross section upper
limit from the data versus the 
$\NONE$ mass  in the minimal gauge-mediated model.
The shaded region shows the range of cross section limits as the parameters are
varied within the ranges
$1 <\tanbeta <25,$ \mbox{$M_2 <200$~GeV,} and $\mu > 0$ or $\mu <0$.
The lines show the experimental limit (solid line) 
and the theoretically predicted cross
section (dashed line)
for the lowest value of M$_{N_1}$ that is excluded
(M$_{N_1} <65$~GeV at 95\% C.L., for $\tanbeta=5$, \mbox{$\mu > 0$).}}
{Filled XSEC vs. N1MASS Plot}
{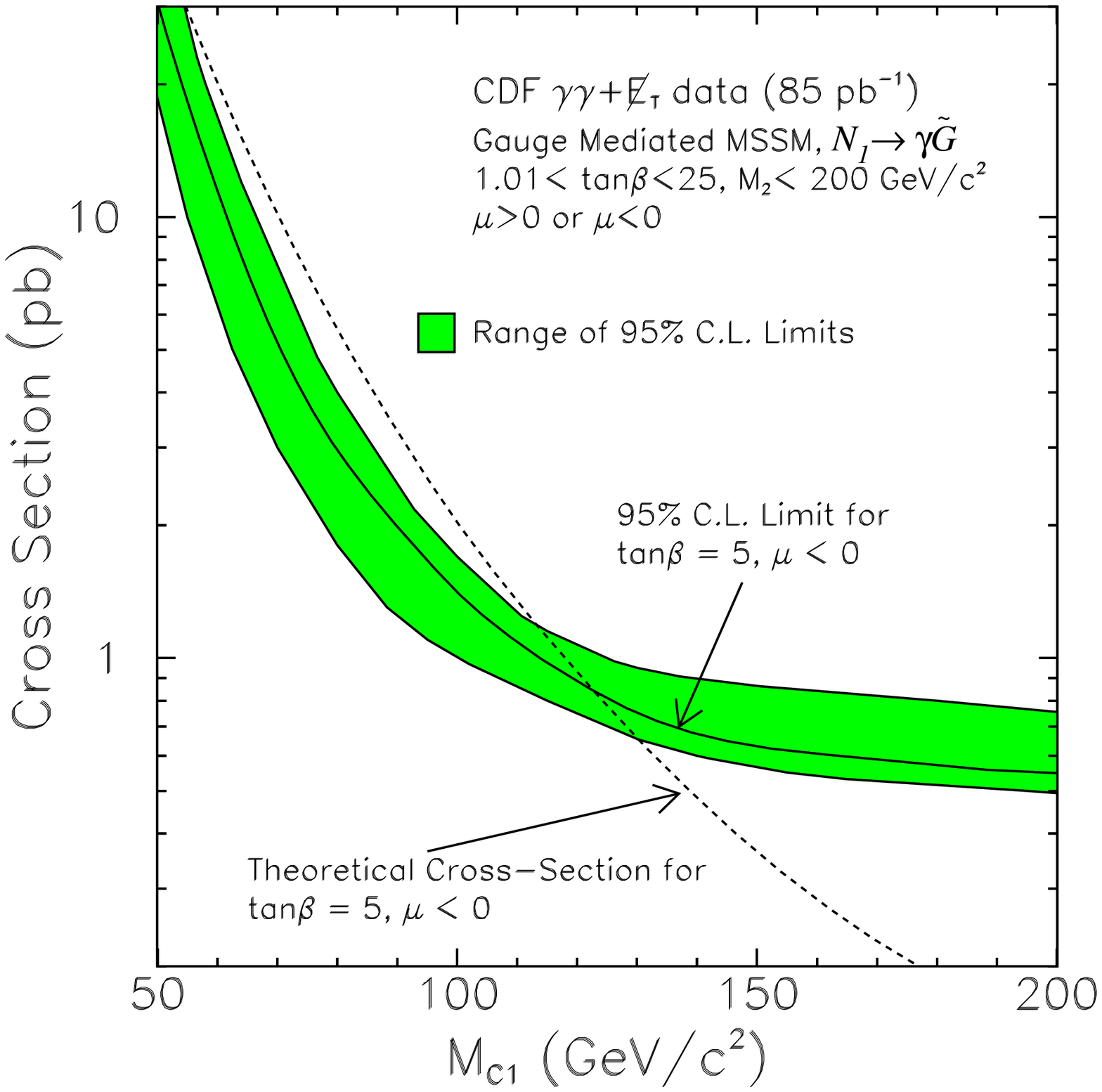}
{The 95\% C.L. cross section upper
limit from the data versus the C$_1$ mass  in the minimal gauge-mediated model.
The shaded region shows the range of cross section limits as the parameters are
varied within the ranges
$1 <\tanbeta <25,$ \mbox{$M_2 <200$~GeV,} and $\mu > 0$ or $\mu <0$.
The lines show the experimental limit (solid line) 
and the theoretically predicted cross
section (dashed line)
for the lowest value of M$_{C_1}$ that is excluded
(M$_{C_1} <120$~GeV at 95\% C.L., for $\tanbeta=5$, \mbox{$\mu <0$).}}
{Filled XSEC vs. C1MASS Plot}

\subsection{Conclusions}\label{Conclusions}
 
The diphoton data set is a good place to search for new physics.
The fact that there are no $\gamma\gamma+\ge 3$ jets in the data excludes 
a model of anomalous $WW\gamma\gamma$ production as the source of the 
$\eeggmett$ candidate at the \mbox{95\% C.L.}  Similarly, 
the diphoton + \mett\ data show no evidence for new physics with the possible
exception of the $\eeggmett$ candidate event. Although, we have some
sensitivity to supersymmetric models with photonic final states, there is
a large amount of parameter space which remains unexplored. More data are
required.    
            
\section {Conclusion}
\label{Chap-Conclusion}

We have  searched a  sample of 85  pb$^{-1}$ of $p{\bar p}$ collisions 
for events
with two central photons and anomalous production of missing transverse energy,
jets,
charged leptons ($e, \mu$, and  $\tau$), $b$-quarks and photons. We find good
agreement  with  standard  model   expectations, with the
possible exception of one
event that sits on the tail of the $\mett$ distribution as well
as  having a  high-$\Et$  central  electron  and a  high-$\Et$  electromagnetic
cluster. 

The $\eeggmett$ candidate event has sparked interest in
the physics community.  
The most probable explanation 
is that this a single $\ppbar$ collision which produced 
a high-$\Et$, isolated electron, two high-$\Et$, isolated
photons, a high-$\Et$ isolated electromagnetic cluster which could be an
electron, photon, tau or jet, and a significant amount of missing
transverse energy.
A conservative estimate predicts that there should be a 
total of \EEEGGMETTOTRATE\ events in the data with the 
$\eeggmett$ signature. If sources which produce a second
electron are excluded the rate drops to \EEFAKEGGMETTOTRATE\ events. 

The $\eeggmett$ candidate event is tantalizing. Perhaps it is a hint of
physics beyond the standard model. 
Then again it may just be one of the rare standard model 
events  that could show up in $10^{12}$ interactions.
Only more data will tell.\\

\Large
\noindent {\bf Acknowledgments}\\
\normalsize

     We thank the Fermilab staff and the technical staffs of the
participating institutions for their contributions.  G.~Kane provided important
theoretical guidance. S.~Mrenna provided
critical help with SPYTHIA and with the $W\gamma, Z\gamma$, and
$WW\gamma\gamma$ calculations.
C.~Kolda provided invaluable assistance in the SUSY modeling.
We are also grateful to G.~Farrar, J.~Rosner, and
F.~Wilczek for helpful conversations. This work was
supported by the U.S. Department of Energy and National Science Foundation,
the Italian Istituto Nazionale di Fisica Nucleare, the Ministry of Science,
Culture, and Education of Japan,
the Natural Sciences and Engineering Research
Council of Canada, the National Science Council of the Republic of China; and
the A.P. Sloan Foundation.

\appendix

\section{Details of the SVX Stub Finding Routines}
\label{App-SVX}

This appendix discusses the Silicon vertex detector (SVX) and the algorithm
for finding stubs for electrons in the plug calorimeter.
The SVX is a single-sided silicon microstrip detector made from 300$\mu$m thick
silicon wafers. 
Three wafers are bonded together to form a 25.5~cm long `ladder' with the
strips running lengthwise to provide $r-\phi$ coordinate measurements.
Four layers of ladders (numbered 0-3) are placed at radii of 2.86~cm,
4.26~cm, 5.69~cm, and 7.87~cm
and arranged in a projective wedge that subtends $30^{\circ}$ in  $\phi$.
Twelve wedges form a ``barrel'';
Two barrels are placed end-to-end along the beam direction to cover the
region 1~cm $<|z|<$ 28~cm.
The SVX tracking results are described using 
three different terms: hit, cluster and stub.  When a charged particle traverses
the SVX it typically deposits energy in 2 or 3 strips per layer.  If enough
energy is deposited in a strip it is referred to as a `hit.' 
A `cluster' finding algorithm joins the 
adjacent hits on the layer 
and determines a mean position with a typical resolution in $\phi$ of 
\mbox{5-15~mrad}.
Joining 3 or more clusters on different layers produces
a `stub' with a typical resolution of 1.5~mrad. 

Typically, the SVX is used to find stubs associated
with CTC tracks. We use a method similar to that in Ref.~\cite{Wasymm} 
which is based only on the calorimeter and
the vertex position information  as an input to the SVX stub finding
algorithms. 
The SVX stub-finding algorithm
searches for SVX clusters in the region of $\approx\pm$100~mrad 
around the $\phi$
of the electron candidate and uses the E$_{\rm T}$ and $\phi$ information in the
fit. 
Any stub found is required to pass the requirements in 
Table~\ref{SVX Requirements} to ensure that it is 
well-measured.

\begin{table}[htb]
\centering
\begin{tabular}{l|l} 
Cut Description         & Requirement \\ \hline
SVX Fiducial        & Trajectory must pass \\
                    & \hspace*{0.5cm} through $\ge$ 3 layers of the SVX \\
                    & 1~cm $<|z_{\rm Trajectory}^{\rm Layer}|< $ 26~cm \\
                    & $\ge$ 3 layers with clusters \\ \hline
Well Measured       & $\chi^2_{\rm SVX} < 2.0$ \\ \hline
Correct $\eta$          & $80 < Q_{\rm Min}^{\rm Central} < 200$ \\
                        & $100 < Q_{\rm Min}^{\rm Plug} < 200$  \\
\end{tabular}
\caption[The SVX stub-finding 
requirements for electrons]{The SVX stub-finding requirements for electrons. 
Any stub
found by the SVX tracker must pass through the fiducial part of the SVX, be
well-measured and be consistent with being from an electron in the central 
or plug
calorimeter.}
\label{SVX Requirements}
\end{table}

Using $\frac{dE}{dX}$ techniques, 
the $\eta$ for the stub can be inferred.
The amount of charge (which is proportional to
the energy deposited in the silicon) collected by the SVX
strips, Q, helps determine the path length of the particle through the strip.
For a given stub, the cluster with the smallest amount of
charge deposited, normalized by the trajectory angle
$Q_{\rm Min}^{\rm Uncorr}\times Sin(\theta) = Q_{\rm Min}$, is a good measure
of the direction of the charged
particle.  We require  $80 < Q_{\rm Min}^{\rm Central} < 200$ and 
                        $100 < Q_{\rm Min}^{\rm Plug} < 200.$

In the $\eeggmett$ candidate event, 
the hypothetical trajectory between the vertex at 20.4~cm and the 
location of the
cluster, as measured in the calorimeter,
passes through the inner three layers of the SVX, and passes
between the two SVX barrels at the radius of the fourth layer, 
as shown in Figure~\ref{Trajectory}. 
The 3-cluster stub which is found 
appears to be well measured, $\chi^2_{\rm SVX}=0.54$ and 
$Q_{\rm Min}=145$.  The  $\Delta\phi$ is measured to be 
$\Delta\phi=$-29~mrad to be compared with an the 
-2.6~mrad expected for a 63~GeV positron.
As a check, the charge deposition in the SVX clusters 
can be studied to infer a best guess for $\eta$.
Using the vertex at 20.4~cm,  $Q_{\rm Min}^{\rm Uncorr} = 422$
and that typically \mbox{$100 < Q_{\rm Min}^{\rm Uncorr} Sin\theta < 200$} 
yields the prediction that 
$1.42 < |\eta| < $2.11. 
Independently, assuming the vertex at 20.4~cm,
the cluster pattern (i.e, only the
inner three layers were hit in a single barrel) implies the range
$-1.9<\eta<-1.61$ or $0.7 < \eta < 0.8$. Both estimates are consistent with 
the cluster at $\eta=-1.72$.



\begin{thebibliography}{99}
\message{BIBLIOGRAPHY}




\bibitem{SM Reference} S.~L.~Glashow, {\it Nucl. Phys} {\bf 22}, 579 (1961); 
S.~Weinberg, \Journal{\PRL}{19}{1264}{1967}; A.~Salam, 
Elementary
Particle Theory: Relativistic Groups and Analyticity (Nobel Symposium No. 8),
edited  by N.~Svartholm (Almqvist and Wiksell, Stockholm, 1968), p. 367.

\bibitem{Gravitino Reference}
a)~P.~Fayet, {\it Phys Rep} {\bf 105} 21 (1984);
b)~M.~Dine, A.~Nelson, Y.~Nir and Y.~Shirman, \Journal{\PRD}{53}{2658}{1996};
c)~S.~Dimopoulos, M.~Dine, S.~Raby and S.~Thomas, 
\Journal{\PRL}{76}{3494}{1996};
d)~S.~Dimopoulos, S.~Thomas and J.~Wells, \Journal{\PRD}{54}{3283}{1996};
e)~K.~Babu, C.~Kolda and F.~Wilczek, \Journal{\PRL}{77}{3070}{1996};
%
f)~J.~Lopez and D.~Nanopoulos,  {\it Mod. Phys. Lett.} A{\bf 10}, 2473 (1996);
%
g)~S.~Ambrosanio, G.~Kane, G.~Kribs, 
S.~Martin and S.~Mrenna, \Journal{\PRD}{54}{5395}{1996}.
h)~J.~Bagger, K.~Matchev,D.~Pierce and R.~Zhang,
 \Journal{\PRD}{55}{3188}{1997},
i)~H.~Baer, M.~Brhlik, C.~Chen, and X.~Tata,
  \Journal{\PRD}{55}{4463}{1997}.

\bibitem{Higgsino LSP}
a)~H.~Haber, G.~Kane  and M.~Quiros, 
{\it Phys Letters} B {\bf 160}, 297 (1985); 
b)~S.~Ambrosanio, G.~Kane, G.~Kribs, S.~Martin and S.~Mrenna, 
\Journal{\PRL}{76}{3498}{1996};
c)~S.~Ambrosanio, G.~Kane, G.~Kribs, S.~Martin and
S.~Mrenna, \Journal{\PRD}{55}{1372}{1997}.
%
%

\bibitem{Non-SUSY}
See for example 
a)~G.~Bhattacharyya and R.~Mohapatra,  \Journal{\PRD}{54}{4204}{1996};
b)~J.~Rosner, \Journal{\PRD}{55}{3143}{1997};
c)~K.~Lane, {\it Phys Letters} B {\bf 357}, 624 (1995).



\bibitem{bluebook} The Collider Detector at Fermilab (CDF), 
A compilation of articles
reprinted from {\em Nuclear Instruments and Methods in Physics Research -- A},
North-Holland (1988).

\bibitem{CDFcoo} At CDF, the  
$z$ (longitudinal) axis is along the proton  beam axis;
$r$ is the transverse coordinate.  Pseudorapidity  ($\eta$) is
$\eta\equiv-$ln(tan($\theta$/2)), where $\theta$ is the polar angle.
Transverse energy is defined as $\Et ={\rm E}\sin(\theta)$. The negative of the
vector sum of the transverse energy is known as 
missing transverse energy or \mett.

\bibitem{Park} S. Park, 10th Topical Workshop on Proton-Antiproton Collider
Physics, R. Raja and J. Yoh, eds., AIP press, May 1995, p. 62.
The electromagnetic cluster at large $\eta$ passes all the standard electron
selection criteria. However, 
there is some indication that this cluster is not from
an electron. This is discussed in \secorchap\ 4.


\bibitem{Toback} D. Toback, Searches for 
New Physics in Diphoton Events in $p{\bar p}$ collisions at
\mbox{$\roots= 1.8$~TeV},
Ph.D. thesis, University of Chicago, 1997; \mbox{F. Abe \etal},
\PrePrintE{9801019} to be published in {\it Phys. Rev. Lett}.

\bibitem{svxnim} D. Amidei \etal, {\it Nucl. Instrum. Methods}, {\bf
A350}, 73 (1994);
P. Azzi \etal, \Journal{\NIMA}{360}{137}{1995}.



\bibitem{photons}
For a more complete
description of the photon identification and determination of backgrounds
(which are expected to be mostly from $\pi^0$'s) in CDF
see \mbox{F. Abe \etal}, \Journal{\PRD}{48}{2998}{1993} and
\mbox{F. Abe \etal}, \Journal{\PRL}{73}{2662}{1994}.
 


\bibitem{WMass} F. Abe \etal, \Journal{\PRD}{52}{4784}{1995}.

\bibitem{R} F. Abe \etal, \Journal{\PRD}{52}{2624}{1995}. A much
larger portion of the plug calorimeter is used to identify electrons than in the
top-quark measurements of Ref.~\protect\cite{top}.

\bibitem{top} F. Abe \etal, \Journal{\PRD}{50}{2966}{1994} and
              F. Abe \etal, \Journal{\PRL}{74}{2627}{1995}.
The $b$-jet identification used here is the SECVTX algorithm only.

\bibitem{Marcus} F. Abe \etal, \Journal{\PRL}{79}{3585}{1997}.
For more detail see M.~Hohlmann,
Observation of Top Quark Pairs in the Dilepton Decay Channel using
Electrons, Muons, and Taus dileptons with hadronically
decaying tau leptons, Ph.D. thesis, University of Chicago (1997).




\bibitem{Doug G} D. A. Glenzinski, Observation of the Top Quark in
Proton-Antiproton Collisions at a Center of Mass Energy of 1.8 TeV, Ph.D.
thesis, The Johns Hopkins University, (1995).




\bibitem{trigger}
D. Amidei    \etal, {\it Nucl. Instrum. Methods}, {\bf A269}, 51 (1988);
T. Carroll   \etal, {\it Nucl. Instrum. Methods}, {\bf A263}, 199 (1988);
G. W. Foster \etal, {\it Nucl. Instrum. Methods}, {\bf A269}, 93 (1988).

\bibitem{Trigger Towers} Trigger towers subtend 0.2 $\times 15\degrees$
in $\eta-\phi$ space.

\bibitem{Wgamma}
D. Benjamin, 10th Topical Workshop on Proton-Antiproton Collider
Physics, R. Raja and J. Yoh, eds., AIP press, May 1995, p. 370.

\bibitem{PYTHIA} H. Bengtsson and T. Sj\"{o}strand, {\it Comput. Phys. Commun.}
{\bf 46}, 43 (1987).


\bibitem{jets}
See F.~Abe \etal, \Journal{\PRL}{45}{1488}{1992}
for a description
of the jet-finding algorithm and the jet energy corrections. Jets are
reconstructed here with a cone in $\eta-\phi$ space of radius 0.4 and are
required to have uncorrected E$_{\rm T} > 10$ GeV to be counted. 

\bibitem{No Met} $e^+e^-$ events with an invariant mass near 
that of the $Z^0$ boson are
selected because the dominant source is 
standard model $\zoee$ production which has a
similar topology to $\gamma\gamma$ events and does not have any 
intrinsic \mett.  While other processes, such as $WW$, $WZ$ and $Z^0 \rightarrow
\tau\tau \rightarrow ee\nu\nu\nu\nu$ can fake the fake this signature, their
contamination rate is less than 1\%.

\bibitem{Why Under} The strip chamber measures, in the $z$ direction, a shower
energy of 52~GeV, to be compared with a 37~GeV measurement from the full
calorimeter.

\bibitem{Huston} 
B.~Bailey, J.~Owens and J.~Ohnemus, \Journal{\PRD}{46}{2018}{1992},
J.~Huston, and J.~Owens, \PrePrintP{9508341}, and 
M.~Drees and T.~Han, \Journal{\PRL}{77}{4142}{1996}.


\bibitem{Ellis} S.~D. Ellis, R.~Kleiss, and W.~J. Stirling, 
{\it Phys. Lett.} B {\bf 154}, 435 (1985).

\bibitem{Hauger} S.~A.~Hauger, Measurement of the $Z^0\rightarrow
e^+e^-$ + N Jet Cross Sections in ${\bar p}p$ collisions at $\sqrt{S}$ = 1.8
TeV, Ph.D. thesis, Duke University, 1995.

\bibitem{Mrenna Lgamma}  S. Mrenna, Private communication. This calculation was
performed using 
MADGRAPH~\protect\cite{MADGRAPH} and PYTHIA~\protect\cite{PYTHIA}. 

\bibitem{MADGRAPH} T.~Stelzer and W.~F.~Long,
{\it Comput.Phys.Commun.} {\bf 81}, 357, (1994).


\bibitem{Method 1} We are using the Method 1 SECVTX $b$-tag
estimate. See Ref.~\protect\cite{top}.

\bibitem{Diboson Back} While there are
three $\ell\gamma\gamma$ events (\Etggl) 
in the data with a background expectation of 0.3$\pm$0.1, the $\eeggmett$
candidate does not look like the background source.
A more appropriate comparison is made by 
removing the $\eeggmett$ candidate event. In that case, 
for both photons above 12~GeV
there are 2 diboson candidates, consistent 
with an expectation of 0.3$\pm$0.1 events, and for both
photons above 25~GeV there are no diboson candidates, consistent 
with an expectation of 0.1$\pm$0.1 events. 




\bibitem{Pathologies} For example, the event could be 
due to two $\ppbar$ collisions occurring at the same time,
each producing part of the event. Similarly, the event could be the overlap of a
$\ppbar$ collision and a cosmic ray interaction. Another 
possibility is that parts of the event
are due to jets which faked the electron and/or photon identification criteria.

\bibitem{CTC Info} The efficiency for finding a track in the CTC falls rapidly 
for $|\eta|>1.4$. The $\eta$ for the cluster is \mbox{$\eta$ = -1.63.} In
addition the inner layers of the CTC  have high occupancy in the event making
finding any evidence of a track impossible.

\bibitem{Other Argument} If the clusters in the event were to come
from any of the other vertices in the event their
E$_{\rm T}$ would significantly change. For
example, the E$_{\rm T}$ of the cluster in the plug calorimeter is, using the
vertex at 20.4~cm, 63~GeV. The E$_{\rm T}$ becomes 72, 83, and 85~GeV for the
vertices at -8.9, -33.7 and -38.9~cm, respectively, 
making each far less likely.


\bibitem{Cosmic Rays} The sample of photons from cosmic rays 
are selected from a
sample of single photon events in which the ratio of \mett/\Etg $> 1.0$.

\bibitem{Kuhlmann Note} 
Since most electrons and photons do not deposit energy 
in the hadronic calorimeters this method is not 
efficient for finding timing information. It is thus not unusual that one of the
three central clusters in the event does not have timing information. 

\bibitem{Resolutions} The
energy resolution of
the central and plug electromagnetic calorimeters are 
$(\frac{\delta E}{E})^{2} =
(\frac{(13.5 \pm 0.7)\%~{\rm GeV}^{1/2}}{\sqrt{{\rm E_T}}})^{2} +
((1.0 \pm 1.0)\%)^2$ and
\mbox{$(\frac{\delta E}{E})^{2} =
(\frac{(28)\%~{\rm GeV}^{1/2}}{\sqrt{E}})^{2} +((2.0 \pm 2.0)\%)^2$,}
respectively. 

\bibitem{top exception} This is not strictly true. The identification of
electrons in the plug calorimeter in the top-quark dilepton analysis
require the presence of a track in the CTC.  However, since
the cluster in the event is in
a region of the detector where the track finding efficiency is approximately 
zero, this requirement is removed. 

\bibitem{Wasymm} 
Q.~Fan and A.~Bodek,
Page 553, Proceedings of VIth International
Conference on
Calorimetry in High Energy Physics, June 8-14, 1996,
Frascati (Rome), Italy.


\bibitem{Large DPHI} There is
only one other electron in the sample
which has \mbox{$|\Delta\phi| > 0.03$,} as shown in the central electron plot in
Figure~\ref{Final Deltaphi}. The stub picked up by
the algorithm is a good 4-cluster
stub which is attached to a soft, nearby track.
The stub associated with the high P$_{\rm T}$ electron CTC track
is a perfectly good 3-cluster stub with
the appropriate $\Delta\phi$. It is not 
selected because the algorithm selects 4-cluster stubs over 3-cluster 
stubs. There are no other such stubs found for
the plug electromagnetic cluster.

\bibitem{Other Problem} The measurement
of the shower in the electromagnetic 
calorimeter would also be unusual for an electron. 
However, a detailed study indicates that the problem is not significant and may
be due to noise in the calorimeter.  

\bibitem{Stuart Private} The amount of energy
deposited in each layer of the SVX is given by a Landau
distribution. The probability of the amount of energy deposited to be below
the requirement to create a cluster is less than 1\%.

\bibitem{Slightly Higher} The photon sample was selected
as `photons' having no nearby tracks (see Table~\ref{Event Cuts}),
the efficiency of these criteria is estimated
to be 95.3\%. Assuming all of the photons rejected from the
sample have a VTX  occupancy of greater than 50\%
and $|\Delta\phi|>$0.03~rad, the true rate at which photons pass these cuts
could be as large as 15\% and 7\%
respectively.

\bibitem{TAUOLA} S. Jadach \etal, TAUOLA version 2.5 (June 1994); Also see
{\it Comput.Phys.Commun.} {\bf 76} 361, (1993) and references therein.

\bibitem{PDG} Particle Data Group, \Journal{\PRD}{54}{1}{1996}. 

\bibitem{Only SM Options} We have considered here only standard model
interpretations. There are exotic possibilities such as a new particle that
decays to $\gamma + \tau$ that are consistent with all the data.


\bibitem{Why Fake}  
The list of electron
requirements does not include
the SVX $\Delta\phi$ cut 
since it is not part of the standard selection criteria. 
The CTC track requirement is removed since it restricts the $\eta$ range in
which electrons can be found.
 Removing these requirements is
conservative as it only
increases the rate at which standard model processes could produce the event.


\bibitem{Mrenna}  S. Mrenna, Private communication. This calculation was
performed using 
MADGRAPH~\protect\cite{MADGRAPH} and PYTHIA~\protect\cite{PYTHIA}. 
Also See Ref.~\protect\cite{Higgsino LSP}b. 
Based  on  experience  with  Diboson  production
there is  about a 30$\%$
uncertainty on the $WW\gamma\gamma$
cross section because of  higher order corrections
and  structure function uncertainties. 


\bibitem{WWGG and SUSY} SUSY scenarios could 
produce such a situation. For example $C^+C^-
\rightarrow W\NONE W\NONE \rightarrow W(\gamma\Gravitino)W(\gamma\Gravitino)$.

\bibitem{Dawson} For an excellent introduction to supersymmetry  see
S.~Dawson, \PrePrintP{9612229} or  M.~Carena,  R.~Culbertson, S.~Eno, 
H.~Frisch and S.~Mrenna, \PrePrintE{9712022}.

\bibitem{Kolda} C. Kolda, Private communication.

\bibitem{SPYTHIA}  S. Mrenna, {\it Comput.Phys.Commun.} {\bf 101}, 232, (1997).

\bibitem{Ambrosanio Table Ref} 
We have used the
model in Appendix B (Table 12) of Ref.~\protect\cite{Higgsino LSP}c
which assumes that the $\eeggmett$ candidate
event is real and
due to ${\tilde e_{L}}{\tilde e_{L}}$ production, and
the ${\tilde t}$ is not light.


\bibitem{LEP}
ALEPH  Collaboration, {\it Phys. Lett. B} {\bf 420} 127 (1997) and CERN-EP/98-0 
submitted to {\it Phys. Lett. B}, 
DELPHI Collaboration, Eur. Phys. J. C {\bf 1}, 1 (1998),
L3     Collaboration, {\it Phys. Lett. B} {\bf 415} 299 (1997),
OPAL   Collaboration, Eur. Phys. J. C {\bf 2}, 607 (1998).


\bibitem{LEP II} Aleph Collaboration CERN-EP/98-077, submitted to {\it
Phys. Lett. B}, OPAL Collaboration, Eur. Phys. J. C {\bf 1}, 31 (1998).

\bibitem{D0} 
S.~Abachi \etal,
\Journal{\PRL}{78}{2070}{1997} and B.~Abbott \etal, 
\Journal{\PRL}{80}{442}{1998}




\end{thebibliography}
\end{document}